\documentclass[showkeys, superscriptaddress, twocolumn, aps,prapplied,longbibliography,10pt]{revtex4-1}

\usepackage[T1]{fontenc}
\usepackage[utf8]{inputenc}
\usepackage{amsmath}
\usepackage{amssymb}
\usepackage{amsthm}
\usepackage{textcomp}
\usepackage{graphicx}
\usepackage[colorlinks=true, citecolor=blue, urlcolor=blue, linkcolor=black]{hyperref}
\usepackage{textgreek}
\usepackage{verbatim}
\usepackage{placeins}
\usepackage{tabularx}

\begin{abstract}

Magnetic skyrmions are chiral spin textures that hold great promise as nanoscale information carriers. Since their first observation at room temperature, progress has been made in their current-induced manipulation, with fast motion reported in stray-field-coupled multilayers. However, the complex spin textures with hybrid chiralities and large power dissipation in these multilayers limit their practical implementation and the fundamental understanding of their dynamics. Here, we report on the current-driven motion of N\'eel skyrmions with diameters in the 100-nm range in an ultrathin Pt/Co/MgO trilayer. We find that these skyrmions can be driven at a speed of 100 m s$^{-1}$ and exhibit a drive-dependent skyrmion Hall effect, which is accounted for by the effect of pinning. Our experiments are well substantiated by an analytical model of the skyrmion dynamics as well as by micromagnetic simulations including material inhomogeneities. This good agreement is enabled by the simple skyrmion spin structure in our system and a thorough characterization of its static and dynamical properties.

\end{abstract}

\begin{document}

\title{Current-Driven Skyrmion Dynamics and Drive-Dependent Skyrmion Hall Effect in an Ultrathin Film}

\author{Rom\'eo Juge}
\affiliation{Univ. Grenoble Alpes, CNRS, CEA, Grenoble INP, IRIG-Spintec, Grenoble, France}

\author{Soong-Geun Je}
\affiliation{Univ. Grenoble Alpes, CNRS, CEA, Grenoble INP, IRIG-Spintec, Grenoble, France}

\author{Dayane de Souza Chaves}
\affiliation{Univ. Grenoble Alpes, CNRS, Institut N\'eel, 38000 Grenoble, France}

\author{Liliana D. Buda-Prejbeanu}
\affiliation{Univ. Grenoble Alpes, CNRS, CEA, Grenoble INP, IRIG-Spintec, Grenoble, France}

\author{Jos\'e Pe\~na-Garcia}
\affiliation{Univ. Grenoble Alpes, CNRS, Institut N\'eel, 38000 Grenoble, France}

\author{Jayshankar Nath}
\affiliation{Univ. Grenoble Alpes, CNRS, CEA, Grenoble INP, IRIG-Spintec, Grenoble, France}

\author{Ioan Mihai Miron}
\affiliation{Univ. Grenoble Alpes, CNRS, CEA, Grenoble INP, IRIG-Spintec, Grenoble, France}

\author{Kumari Gaurav Rana}
\affiliation{Univ. Grenoble Alpes, CNRS, CEA, Grenoble INP, IRIG-Spintec, Grenoble, France}

\author{Lucia Aballe}
\affiliation{ALBA Synchrotron Light Facility, 08290 Cerdanyola del Vall\`es, Barcelona, Spain}

\author{Michael Foerster}
\affiliation{ALBA Synchrotron Light Facility, 08290 Cerdanyola del Vall\`es, Barcelona, Spain}

\author{Francesca Genuzio}
\affiliation{Elettra-Sincrotrone S.C.p.A., 34149 Basovizza, Trieste, Italy}

\author{Tevfik Onur Mente\c{s}}
\affiliation{Elettra-Sincrotrone S.C.p.A., 34149 Basovizza, Trieste, Italy}

\author{Andrea Locatelli}
\affiliation{Elettra-Sincrotrone S.C.p.A., 34149 Basovizza, Trieste, Italy}

\author{Francesco Maccherozzi}
\affiliation{Diamond Light Source Ltd., Didcot OX11 0DE, United Kingdom}

\author{Sarnjeet S. Dhesi}
\affiliation{Diamond Light Source Ltd., Didcot OX11 0DE, United Kingdom}

\author{Mohamed Belmeguenai}
\affiliation{Laboratoire des Sciences des Proc\'ed\'es et des Mat\'eriaux, CNRS, Universit\'e Paris 13, 93430 Villetaneuse, France}

\author{Yves Roussign\'e}
\affiliation{Laboratoire des Sciences des Proc\'ed\'es et des Mat\'eriaux, CNRS, Universit\'e Paris 13, 93430 Villetaneuse, France}

\author{St\'ephane Auffret}
\affiliation{Univ. Grenoble Alpes, CNRS, CEA, Grenoble INP, IRIG-Spintec, Grenoble, France}

\author{Stefania Pizzini}
\affiliation{Univ. Grenoble Alpes, CNRS, Institut N\'eel, 38000 Grenoble, France}

\author{Gilles Gaudin}
\affiliation{Univ. Grenoble Alpes, CNRS, CEA, Grenoble INP, IRIG-Spintec, Grenoble, France}

\author{Jan Vogel}
\affiliation{Univ. Grenoble Alpes, CNRS, Institut N\'eel, 38000 Grenoble, France}

\author{Olivier Boulle}
\email{Corresponding author: olivier.boulle@cea.fr}
\affiliation{Univ. Grenoble Alpes, CNRS, CEA, Grenoble INP, IRIG-Spintec, Grenoble, France}

\maketitle

\section{Introduction}

Magnetic skyrmions are fascinating spin textures that have recently attracted considerable attention. Their peculiar topology and nanometer size confer on them quasiparticlelike properties that, combined with their ability to be moved by an electrical current, make them promising candidates for the storage and manipulation of information. They are envisaged to be information carriers in
racetrack memories \cite{fertSkyrmionsTrack2013} and logic devices \cite{zhangMagneticSkyrmionLogic2015} combining high density, thermal stability, and high data flow. Among the different classes of materials hosting skyrmions, sputtered multilayered stacks comprising ultrathin heavy-metal (HM)/ferromagnet (FM) layers combine several interesting features. Their structural-inversion asymmetry along with the large spin-orbit coupling of the HM lead to a large interfacial Dzyaloshinskii-Moriya interaction (DMI), a key ingredient in the skyrmion stabilization that ensures
their homochirality and their Néel nature \cite{bogdanovStabilityVortexlikeStructures1999}. In addition, the current-induced spin-orbit torques (SOTs) in these systems are expected to provide an efficient way to drive the skyrmions \cite{sampaioNucleationStabilityCurrentinduced2013}. In recent years, a concerted effort has led to the observation of stable magnetic skyrmions under ambient conditions in these structures \cite{jiangBlowingMagneticSkyrmion2015, moreau-luchaireAdditiveinterfacialchiral2016, boulleRoomtemperatureChiralMagnetic2016, wooObservationRoomtemperatureMagnetic2016, yuRoomTemperatureCreationSpin2016, soumyanarayananTunableroomtemperaturemagnetic2017, wooSpinorbitTorquedrivenSkyrmion2017, hrabecCurrentinducedSkyrmionGeneration2017, carettaFastCurrentdrivenDomain2018} as well as their current-driven dynamics \cite{jiangBlowingMagneticSkyrmion2015, wooObservationRoomtemperatureMagnetic2016, hrabecCurrentinducedSkyrmionGeneration2017, wooCurrentdrivenDynamicsInhibition2018}. Their topological properties affect their dynamics in a nontrivial manner. Notably, due to their quantized topological charge, magnetic skyrmions are subject to the so-called skyrmion Hall effect (SkHE) \cite{jiangDirectObservationSkyrmion2017, litziusSkyrmionHallEffect2017}, which leads to their deviation from the trajectory imposed by the current. This dynamical signature of their topology needs to be characterized and taken into account for future applications. Numerous studies of skyrmion dynamics have been focused on multilayers with a large number of repetitions \cite{wooObservationRoomtemperatureMagnetic2016, litziusSkyrmionHallEffect2017, wooCurrentdrivenDynamicsInhibition2018}. In particular, fast motion (100 m s$^{-1}$) of small skyrmions (100 nm) has been reported in [Pt/CoFeB/MgO]$_{15}$ \cite{wooObservationRoomtemperatureMagnetic2016, litziusSkyrmionHallEffect2017}. However, in these stray-field-coupled multilayers, the large dipolar fields can outweigh the DMI and stabilize twisted spin structures with a nonuniform chirality across the film thickness \cite{legrandHybridchiraldomain2018, dovzhenkoMagnetostaticTwistsRoomtemperature2018, liAnatomySkyrmionicTextures}. This leads to a complex current-driven dynamics due to layer-dependent SOTs and topologies \cite{lemeshTwistedDomainWalls2018, legrandModelingShapeAxisymmetric2018}. Notably, these hybrid textures may not conserve their topological charge during motion at high current density and therefore may not be skyrmions which cannot be described by existing models \cite{lemeshTwistedDomainWalls2018}. Furthermore, for a given current density, a larger Joule dissipation is also expected, since it scales with the total film thickness. Therefore, the demonstration of fast current-driven skyrmion motion in nanometer-thick film systems is a prerequisite for the development of low-power skyrmion-based devices and for a simple understanding of their dynamics.

In this work, we report on the observation of the current-driven motion of isolated magnetic skyrmions in an ultrathin Pt/Co/MgO film. Previously, we have shown that this system exhibits a large interfacial DMI and hosts homochiral Néel skyrmions at room temperature \cite{boulleRoomtemperatureChiralMagnetic2016}. Here, we find that these skyrmions, with diameters in the 100-nm range, move at a velocity up to 100 m s$^{-1}$ and that the SkHE is markedly drive dependent. These observations are complemented by a detailed characterization of the static and dynamical properties of the stack. These measurements and the simple skyrmion spin structure in our system allow us to approach a quantitative comparison with the prediction of the Thiele equation at high current density. Real-scale micromagnetic simulations including material inhomogeneities reproduce the observed drive-dependence of the skyrmion mobility and the SkHE, allowing the identification of different dynamic regimes. The simulations reveal that the drive-dependence of the SkHE is not accounted for by the fieldlike SOT (FL-SOT), as previously claimed \cite{litziusSkyrmionHallEffect2017} but by pinning, which greatly impacts the skyrmion dynamics at low current density.
 
\section{Results and discussion}

\subsection{Experimental results}

\begin{figure}[h]
\includegraphics[width=0.45\textwidth]{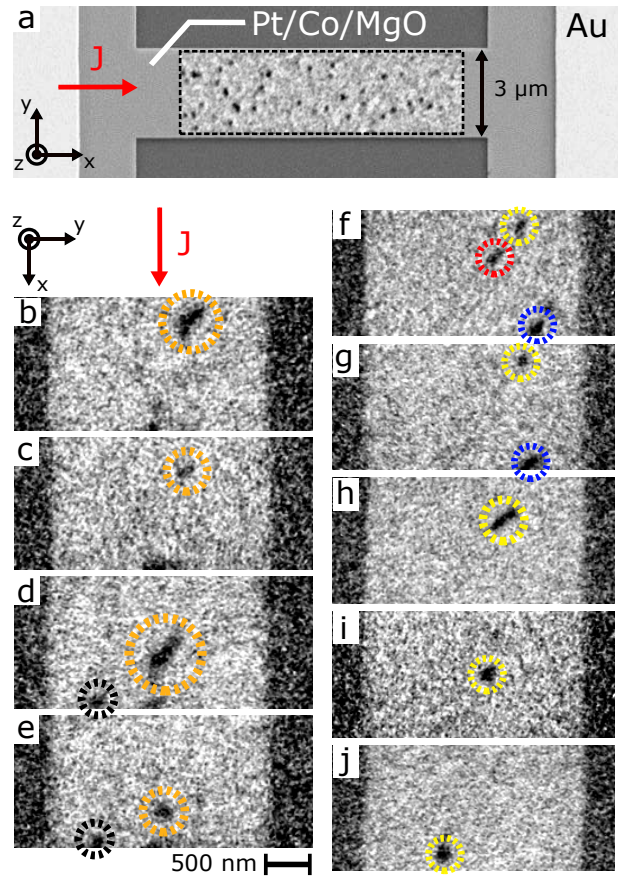}
\centering
\caption{\textbf{a.} A scanning electron microscopy (SEM) image of the device consisting of a 3-\textmu{}m-wide Pt/Co/MgO track contacted with Ti/Au pads for the current injection. A XCMD-PEEM image (black dashed rectangle) showing isolated skyrmions in the track is superimposed on the SEM image. \textbf{b-e.} and \textbf{f-j.} Two distinct sequences of images. Each image was acquired after a single-11 ns current pulse. In both cases, the charge current is flowing along $+\boldsymbol{\hat{x}}$ with $J=5.6\times{}10^{11}$ A m$^{-2}$. The applied magnetic field is $\mu_0H_z\approx{}-5$ mT.}
\label{skyrmion motion EXP}
\end{figure}

We study the current-driven dynamics of magnetic skyrmions in a sputter-deposited Ta(3){\slash}Pt(3){\slash}Co(0.9){\slash}MgO(0.9){\slash}Ta(2) film (thicknesses in nanometers) at room temperature, using the high-spatial-resolution photoemission electron microscopy combined with x-ray magnetic circular dichroism (XMCD-PEEM). The sample is patterned into 3-{\textmu}m-wide tracks with injection pads consisting of Ti(10 nm)/Au(100 nm), as represented in Fig. \ref{skyrmion motion EXP}.a. An external out-of-plane magnetic field $\mu_0H_z\approx{}-5$ mT is
applied on the initially demagnetized configuration in order to stabilize skyrmions with a core magnetized along $+\boldsymbol{\hat{z}}$. In Fig. \ref{skyrmion motion EXP}.b-e and Fig. \ref{skyrmion motion EXP}.f-j, two sequences of images display the characteristic current-driven motion of the skyrmions. Between each acquisition, a single 11-ns current pulse of amplitude $J=5.6\times{}10^{11}$ A m$^{-2}$ is injected along $+\boldsymbol{\hat{x}}$. Overall, some skyrmions exhibit a net motion of several hundreds of nanometers along the current direction, that is, against the electron flow (yellow and orange circles). The same directionality is also observed for skyrmions with the opposite core polarity, when $H_z>0$. In addition, these skyrmions experience a net motion perpendicular to the current direction, which is the signature of the SkHE. This kind of motion is consistent with the current-driven dynamics of left-handed N\'eel skyrmions
governed by the spin Hall effect (SHE), as discussed hereafter. However, it also exhibits a stochastic character, with events of nucleation (black circle) and annihilation (red and blue circles). In addition, the sequences b-e and f-j reveal that the skyrmions displacements are not identical for each current pulse. Finally, some skyrmions appear to be distorted after a current pulse and sometimes acquire an elongated shape. This stochastic behavior can be attributed to the presence of pinning sites obstructing the skyrmion motion \cite{kimCurrentdrivenSkyrmionDynamics2017, legrandRoomTemperatureCurrentInducedGeneration2017}. This local disorder can also explain the large dispersion observed in the skyrmion size and shape \cite{zeisslerPinningHysteresisField2017, jugeMagneticSkyrmionsConfined2018, grossSkyrmionMorphologyUltrathin2018} with a measured average effective diameter $d_{Sk}=156\pm{}45$ nm (see Sec. 1.4 within the Supplementary Material).

\begin{figure}[t!]
\includegraphics[width=0.46\textwidth]{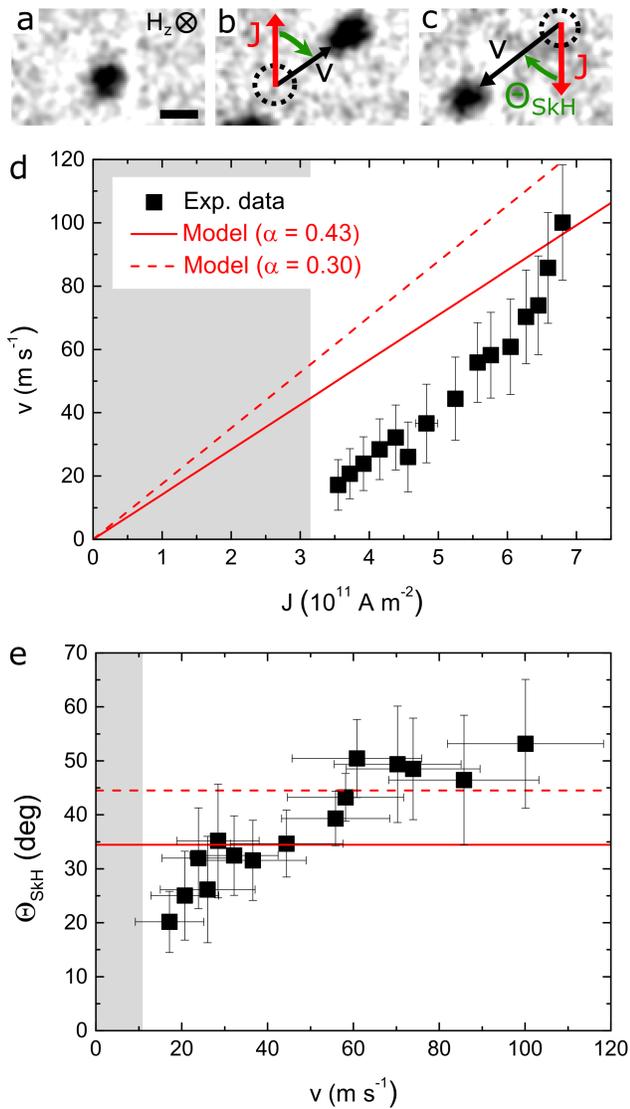}
\centering
\caption{\textbf{a-c.} Sequence of XMCD-PEEM images showing a skyrmion after 2 consecutive 8-ns current pulses with opposite polarities (scale bar is 200 nm). The skyrmion Hall angle (SkHA) $\Theta{}_{SkH}$ is defined as the angle between the current and the skyrmion-motion directions. \textbf{d.} The skyrmion velocity as a function of the current density. \textbf{e.} The SkHA as a function of the skyrmion velocity. The red solid lines and red dashed lines are the analytical skyrmion velocity and SkHA calculated from Eqs. (\ref{equationVELOCITY}) and (\ref{equationSkHA}) using the parameters given in Table \ref{table param} for $\alpha=0.43$ and $\alpha=0.30$, respectively. The error bars denote the sum of the systematic measurement error and the standard deviation. The shaded areas highlight the regime for which no significant displacement was observed after a single current pulse.}
\label{velocity SkHA EXP}
\end{figure}

Despite this irregular motion, systematic measurements of the current-induced displacements averaged over a large number of skyrmions allow us to extract an average skyrmion velocity and skyrmion Hall angle (SkHA) --- namely, the angle $\Theta_{SkH}$ between the current and the skyrmion motion directions (see Fig. \ref{velocity SkHA EXP}.a-c). Fig. \ref{velocity SkHA EXP}.d displays the dependence of the skyrmion velocity with the current density. The velocity is calculated as the total displacement divided by the measured pulse width (8 ns and 11 ns in our experiments) and then averaged over multiple events at the same current density. The thermal drift between each acquisition is corrected from larger images using the right-angled corner of the track (see Fig. \ref{skyrmion motion EXP}.a). Note that below a certain current density, highlighted by the shaded area in Fig. \ref{velocity SkHA EXP}.d, no significant displacement is observed after a single current pulse. Above $J=3.5\times{}10^{11}$ A m$^{-2}$, the skyrmion velocity increases monotonically with the current density and reaches 100 m$\cdot$s$^{-1}$ for $J=6.8\times{}10^{11}$ A m$^{-2}$. This value is comparable to the largest velocities recorded for skyrmions of similar size in stray-field-coupled multilayers \cite{wooObservationRoomtemperatureMagnetic2016}. However, the use of a single repetition allows to lower the power dissipation by one order of magnitude. Furthermore, as previously mentioned, the current injection also leads to a motion perpendicular to the current direction. The direction of deflection with respect to the current direction is unchanged when reversing the current direction (see Fig. \ref{velocity SkHA EXP}.a-c), which is characteristic of the SkHE. The measurements (Fig. \ref{velocity SkHA EXP}.e, black squares) reveal that the SkHA depends on the skyrmion velocity and exhibits a monotonic increase up to approximately 50{\textdegree}.

\begin{table*}[ht]
\def\arraystretch{1.7}
\setlength{\tabcolsep}{3pt}
\caption{A summary of the parameters used in the analytical model and the simulations. The saturation magnetisation $M_s$ and uniaxial anisotropy $K_u$ obtained from SQUID magnetometry measurements. The DMI parameter $D$ measured using Brillouin light scattering (BLS) spectroscopy. The damping parameter $\alpha$ extracted from the DW mobility in the field-driven steady-flow regime. The exchange constant $A$ extracted from micromagnetic simulations and BLS measurements. The DW width parameter $\Delta$ deduced from micromagnetic simulations. The average skyrmion radius $R$ measured at $\mu_0H_z\approx{}-5$ mT. The DL-SOT and FL-SOT effective fields per unit current density, $C_{DL}$ and $C_{FL}$ respectively (given in mT per $10^{11}$ A m$^{-2}$), extracted from harmonic Hall voltage measurements. The current density is calculated from the measured current transmitted through the device, assuming that it flows uniformly in Pt(3){\slash}Co(0.9), owing to the larger resistivity of the Ta(3) under-layer. Additional details can be found in the Supplementary Material.} 
\vspace{0.05 in}
\begin{tabularx}{\textwidth}{c c c c c c c c c}
\hline
$M_s$ (MA m$^{-1}$) & $K_u$ (MJ m$^{-3}$)& $|D|$ (mJ m$^{-2}$)& $\alpha$ & $A$ (pJ m$^{-1}$) & $\Delta$ (nm) & $R$ (nm)& $C_{DL}$ (T A$^{-1}$ m$^2$) & $C_{FL}$ (T A$^{-1}$ m$^2$)  \\ [1ex]\hline\hline
$1.42\pm0.05$  & $1.34\pm0.12$  & $1.27\pm0.04$ & $0.43\pm0.08$ & $16\pm6$ & 11.5 & $78\pm23$ & $2.1$ & $0.9$ \\ \hline
\end{tabularx}
\label{table param}
\end{table*}

In the following, we discuss and interpret these results in the light of the Thiele model as well as micromagnetic simulations. For this purpose, we realize a detailed characterization of the static and transport properties of the Pt/Co/MgO stack to approach a quantitative comparison with the experimental results. Details of the different measurements are given in the Supplementary Material and the parameters extracted are summarized in Table \ref{table param}. 

\subsection{Comparison with the analytical model}

The experimental results are first compared with the prediction of the Thiele equation \cite{thieleSteadyStateMotionMagnetic1973}, which captures the
dynamics of magnetic skyrmions well. Under the assumption that the skyrmions behave as rigid spin textures, the steady-state velocity $\boldsymbol{v}$ results from an equilibrium between different forces acting on the skyrmion:

\begin{equation}
\boldsymbol{F}_{DL}+\boldsymbol{G}\times\boldsymbol{v}-\alpha\boldsymbol{\mathcal{D}}\cdot\boldsymbol{v}=\boldsymbol{0}
\label{thiele equation}
\end{equation}

Here, $\boldsymbol{F}_{DL}$ is the force due to the current injection. In the case of HM/FM ultrathin films, the force results from the action of the so-called dampinglike spin-orbit torque (DL-SOT): the charge current flowing in the Pt layer is converted into a spin current due to the SHE. The spin accumulation at the Pt/Co interface leads to a torque acting on the Co magnetization. In the case of Pt/Co/MgO, since the spin Hall angle (SHA) is positive for Pt \cite{nguyenSpinTorqueStudy2016} and skyrmions are of N\'eel type and exhibit a left-handed chirality \cite{boulleRoomtemperatureChiralMagnetic2016}, the symmetries of the DL-SOT impose that the resulting force drives the skyrmion in the current direction (\textit{i.e.} opposite to the electron flow) \cite{hrabecCurrentinducedSkyrmionGeneration2017}. In addition, the topology of the skyrmion leads to a gyrotropic force, $\boldsymbol{G}\times\boldsymbol{v}$, where $\boldsymbol{G}=G\boldsymbol{\hat{z}}$ is the gyrovector, whose sign depends only on the topological charge $N_{Sk}$, with $G\propto{}N_{Sk}$ \cite{thieleSteadyStateMotionMagnetic1973, thieleApplicationsGyrocouplingVector1974}. For instance, in Fig. \ref{velocity SkHA EXP}.a-c, $\boldsymbol{G}$ points in the $-\boldsymbol{\hat{z}}$ direction. This second term thus describes the SkHE.  The third term describes the dissipative force with $\boldsymbol{\mathcal{D}}$ the dissipative tensor and $\alpha$ the Gilbert damping parameter. Note that the effect of the FL-SOT is neglected. Indeed, the action of the FL-SOT is equivalent to that of an external magnetic field applied in a direction perpendicular to the current direction and a spatially homogeneous magnetic field does not result in a force on a rigid skyrmion in Thiele’s approach. This approximation is justified in the next section by micromagnetic simulations. Although simple, this equation captures most of the physical ingredients that drive the skyrmion dynamics and is particularly relevant in our ultrathin films with a homogeneous homochiral N\'eel spin configuration across the film thickness. Assuming that the skyrmion has a radial 360{\textdegree} Bloch wall profile \footnote{In a previous work, we observed using XMCD-PEEM, that the skyrmion profile in Pt/Co/MgO was fitted well by a 360{\textdegree} Bloch wall profile \cite{boulleRoomtemperatureChiralMagnetic2016}}, a rotational symmetry, and that its radius $R$ is much larger than the domain wall (DW) width parameter $\Delta$ ($R\gg\Delta$), the skyrmion velocity and the SkHA can be written as follows \cite{tomaselloStrategyDesignSkyrmion2014, hrabecCurrentinducedSkyrmionGeneration2017}:

\begin{align}
\label{equationVELOCITY}
v &= \frac{\gamma\pi}{4}\frac{R}{\sqrt{\left(\frac{\alpha{}R}{2\Delta}\right)^2+1}}C_{DL}J \\
\label{equationSkHA}
\tan\Theta_{SkH} &= \frac{2\Delta{}}{\alpha{}R}
\end{align}

Here, $\gamma$ is the gyromagnetic ratio, $\alpha$ the Gilbert damping parameter, $J$ the current density flowing in the Pt layer and $C_{DL}$ the effective field (per unit current density) associated with the DL-SOT. We use the parameters listed in Table \ref{table param}. The analytical solution for the velocity is plotted in Fig. \ref{velocity SkHA EXP}.d (red solid line): it reveals a relatively good agreement with the experimental data at high current density, considering that the velocity depends critically on all the aforementioned parameters. Nevertheless, we observe that the experimental values are smaller than the one predicted by the Thiele equation at low current density. A similar observation has been pointed out in the study of the current-driven dynamics of DWs in Pt/Co/AlO$_x$ ultrathin films \cite{mironFastCurrentinducedDomainwall2011}. It can be accounted for by the effect of pinning on the skyrmion dynamics, not taken into account in this simple model, which is expected to lead to a thermally activated regime at low drive current, characterized by smaller velocities. This indicates that the skyrmion dynamics in our experiments is in a depinning regime and this points at the existence of a flow regime for the largest current densities injected.

With regard to the SkHA, we find a value $\Theta_{SkH}=$ 34{\textdegree} $\pm$ 8{\textdegree} using Eq. (\ref{equationSkHA}). The uncertainty includes the uncertainty on the skyrmion radius. Note that this value is independent of the skyrmion velocity or the applied current, in sharp contrast with our observations as well as previous experimental studies \cite{jiangDirectObservationSkyrmion2017, litziusSkyrmionHallEffect2017}. This disagreement with the experiments can also be explained by the presence of pinning sites \cite{reichhardtQuantizedTransportSkyrmion2015, reichhardtCollectiveTransportProperties2015, reichhardtNoiseFluctuationsDrive2016, kimCurrentdrivenSkyrmionDynamics2017, legrandRoomTemperatureCurrentInducedGeneration2017}. The calculated SkHA (\textit{i.e.} the steady-flow SkHA) is found to be smaller than the maximum measured values (see Fig. \ref{velocity SkHA EXP}.e, red solid line). However, as detailed in the next section, one expects the SkHA to tend toward the calculated value for large skyrmion velocities. Note that the damping is a key parameter in the calculation of the skyrmion velocity and the SkHA. Here, we use the value $\alpha=0.43$. It is extracted from field-driven DW dynamics measurements in a Pt{\slash}Co(0.63 nm){\slash}MgO (see Sec. 1.5 within the Supplementary Material). This value is in good agreement with previous DW dynamics experiments in ultrathin Pt/Co/Pt \cite{metaxasCreepFlowRegimes2007} and Pt/Co/AlO$_x$ \cite{mironFastCurrentinducedDomainwall2011, phamVeryLargeDomain2016} films. To explain such large damping values, besides the contributions of surface roughness \cite{minEffectsDisorderInternal2010} and spin pumping into the Pt layer \cite{beaujourMagnetizationDampingUltrathin2006}, the presence of Rashba spin-orbit coupling is also predicted to enhance the damping \cite{kimPredictionGiantSpin2012}. This additional contribution can be expected in our stack considering the non-negligible FL-SOT (see Table \ref{table param}) and can lead to a significant reduction of the DW and skyrmion velocity as well as the SkHA \cite{kimRoleNonlinearAnisotropic2015}. We note that the damping parameter is measured in a stack with a slightly thinner Co layer, which is required to stabilize larger domains and drive DWs with an external field. In a thicker Co layer, the different interfacial contributions to the damping mentioned above are expected to decrease. The simplest approximation to account for the change in the Co thickness is to assume that $\alpha\sim{}1/t$, leading to $\alpha=0.30$ for the Pt{\slash}Co(0.9nm){\slash}MgO film used to study the skyrmion dynamics. Accounting for this correction leads to an enhancement of the calculated skyrmion velocity and of the SkHA $\Theta_{SkH}=$ 45{\textdegree} $\pm$ 8{\textdegree} (see Fig. \ref{velocity SkHA EXP}.d and e, red dashed lines) that provides a better agreement with the experimental results, as it will be justified in the next section.

\subsection{Micromagnetic simulations}

\begin{figure}[hbt!]
\includegraphics[width=0.5\textwidth]{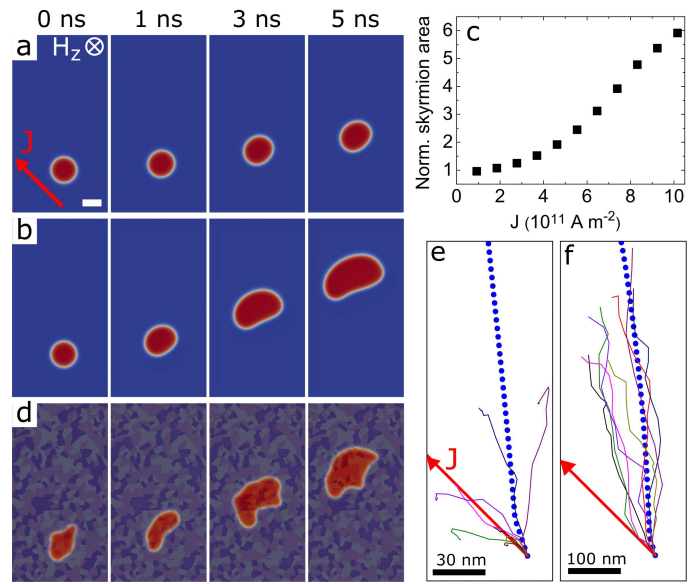}
%\centering
\caption{\textbf{a-b.} Sequences of snapshots showing the current-driven dynamics of a skyrmion in an ideal disorder-free film with \textbf{a} $J=2.9\times{}10^{11}$ A m$^{-2}$ and \textbf{b} $J=6.7\times{}10^{11}$ A m$^{-2}$ (scale bar is 100 nm). \textbf{c.} Evolution of the skyrmion size as a function of the applied current density (at 5 ns). The size is defined as the area for which $m_z>0$ (in red) normalised by the area at rest. \textbf{d.} A sequence of snapshots in a disordered film. The dark grains are region of higher anisotropy and the bright grains those of lower anisotropy. \textbf{e-f.} A summary of the trajectories recorded for different grain distributions for \textbf{e} $J=2.9\times10^{11}$ A m$^{-2}$ and \textbf{f} $6.7\times10^{11}$ A m$^{-2}$. The blue dotted lines indicate the skyrmion trajectories in the disorder-free scenario. The position of deformed skyrmions is defined as that of their barycenter. The simulations were performed for the experimental parameters listed in Table \ref{table param} with the applied field $\mu_0H_z=-5.4$ mT.}
\label{skyrmion motion SIMU}
\end{figure}

To go beyond the assumptions and limitations of the analytical model, we performed micromagnetic simulations using the parameters listed in Table \ref{table param}. Both DL-SOT and FL-SOT were taken into account. These torques are given respectively by $\boldsymbol{T}_{DL}=-\gamma_0C_{DL}J\boldsymbol{m}\times[(\boldsymbol{\hat{z}}\times\boldsymbol{\hat{j}})\times\boldsymbol{m}]$ and $\boldsymbol{T}_{FL}=-\gamma_0C_{FL}J\boldsymbol{m}\times(\boldsymbol{\hat{z}}\times\boldsymbol{\hat{j}})$, with $\gamma_0=\mu_0\gamma$, $\boldsymbol{m}$ the reduced magnetisation vector, $\boldsymbol{\hat{j}}$ the unit vector in the current direction and $C_{DL}$ (resp. $C_{FL}$) the measured effective field per unit current density associated with the DL-SOT (FL-SOT) \cite{khvalkovskiyMatchingDomainwallConfiguration2013}. The simulations were carried out using the MICRO3D code \cite{budaMicromagneticSimulationsMagnetisation2002}. The geometry consists of a $1040\times560\times0.9$ nm$^3$ track with a cell size of $3.2\times3.2\times0.45$ nm$^3$. An out-of-plane external field $\mu_0H_z=-5.4$ mT is applied which sets the skyrmion radius to 78.8 nm ---  that is the experimental one --- and thermal fluctuations are neglected. Fig. \ref{skyrmion motion SIMU}.a and b show snapshots of the skyrmion dynamics during the application of a current with $J=2.9\times{}10^{11}$ A m$^{-2}$ and $6.7\times{}10^{11}$ A m$^{-2}$ respectively. As indicated by the red arrow, the current is tilted by 45{\textdegree} with respect to the track direction so that the skyrmions move along the track. The simulations reveal that the skyrmion experiences an expansion accompanied by a deformation that becomes more pronounced with increasing driving current. This effect is purely dynamical: when the current is turned off, the skyrmion shrinks back to its original size and recovers its rotational symmetry. In addition, we find that this deformation is independent of the FL-SOT and is therefore only due to the DL-SOT (see Sec. 2.2 within the Supplementary Material). To highlight this effect, we calculated the effective skyrmion size, defined as the total area for which $m_z>0$, normalized by the skyrmion area at rest. Fig. \ref{skyrmion motion SIMU}.c displays its evolution with the current density. It shows that the skyrmion size increases significantly, in contradiction with the rigid-core assumption of the Thiele model. We note that the simulated skyrmion size matches that of the observed skyrmions and it is significantly larger than the one considered in most studies \cite{sampaioNucleationStabilityCurrentinduced2013, kimCurrentdrivenSkyrmionDynamics2017, legrandRoomTemperatureCurrentInducedGeneration2017, litziusSkyrmionHallEffect2017}, which could explain why this dynamical effect had not been emphasized before.

\begin{figure}[h]
\includegraphics[width=0.45\textwidth]{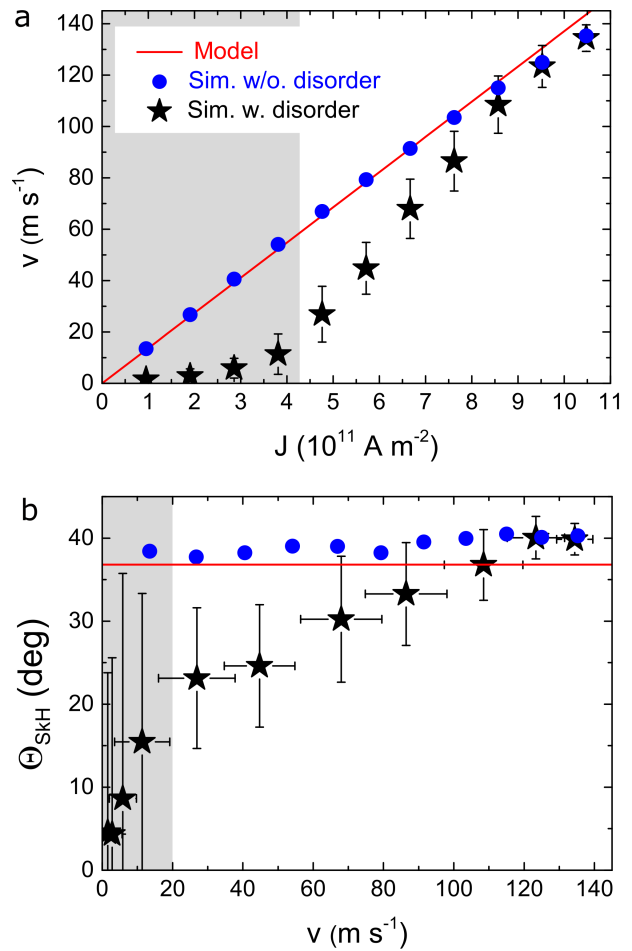}
\centering
\caption{\textbf{a.} The skyrmion velocity as a function of the current density in the case of a disorder-free film (blue circles) and in the case of a disordered film (black stars). \textbf{b.} The SkHA $\Theta_{SkH}$ as a function of the skyrmion velocity. The analytical solutions (red solid lines) are calculated  from \textbf{a} Eq. (\ref{equationVELOCITY}) and \textbf{b} Eq. (\ref{equationSkHA}) using the parameters given in Table \ref{table param}. To mimic the experimental conditions, the velocity and the SkHA are calculated from the total skyrmion displacement within an 8-ns time window. The error bars denote the standard deviation. The shaded areas highlight the pinned regime.}
\label{velocity SkHA SIMU}
\end{figure}

To quantify the dynamics of a noncircular skyrmion, we define its position as that of its barycenter. The obtained skyrmion velocity and SkHA are plotted as blue dots in Fig. \ref{velocity SkHA SIMU}.a and b respectively. Remarkably, the simulations are in very good agreement with the simple prediction of the analytical model (red line). This highlights the relevance of discussing our experimental results in the light of the Thiele model. In addition, it emphasizes that the effect of the FL-SOT, which is not taken into account in the model, is not significant in our system. Surprisingly, the DL-SOT-induced expansion and/or deformation does not significantly alter the skyrmion dynamics in the range of current density considered, in spite of the large increase in the skyrmion size (Fig. \ref{skyrmion motion SIMU}.c). Therefore, this effect cannot account for the evolution of the skyrmion velocity and SkHA observed experimentally (Fig. \ref{velocity SkHA EXP}). Furthermore, mechanisms involving a sizeable FL-SOT have been put forward to explain the drive-dependence of the SkHA of slightly deformed skyrmions \cite{litziusSkyrmionHallEffect2017} or breathing skyrmions \cite{tomaselloMicromagneticUnderstandingSkyrmion2018}. However, our micromagnetic simulations including both SOTs with realistic relative amplitudes (up to $C_{FL}=C_{DL}$)  show no significant effect of the FL-SOT (see Sec. 2.2 within the Supplementary Material). Note finally that the Oersted field may also impact the measured SkHA as it generates a force on the skyrmion perpendicular to the current direction. For skyrmions with a core magnetised along $+\boldsymbol{\hat{z}}$, which corresponds to the condition of most of our experiments, this force is opposite to the gyrotropic force \cite{hrabecCurrentinducedSkyrmionGeneration2017}. Thus, besides being of negligible amplitude, it may only reduce the measured SkHA and therefore cannot explain the pronounced drive-dependence observed experimentally (Fig. \ref{velocity SkHA EXP}.e). 

Another explanation to account for our experimental results is the presence of pinning sites in the material \cite{reichhardtCollectiveTransportProperties2015, reichhardtNoiseFluctuationsDrive2016, kimCurrentdrivenSkyrmionDynamics2017, legrandRoomTemperatureCurrentInducedGeneration2017}. To include such an effect, we introduce disorder, modeled by a distribution of grains with different anisotropies \cite{votoEffectsGrainSize2016, garcia-sanchezSkyrmionbasedSpintorqueNanooscillator2016}. The average grain size is 30 nm, a value in line with previous estimations in similar systems \cite{legrandRoomTemperatureCurrentInducedGeneration2017, zeisslerPinningHysteresisField2017, grossSkyrmionMorphologyUltrathin2018}, and the anisotropy is varied randomly between $\left(1\pm0.05\right)K_u$. This approach was shown to provide a good qualitative agreement with experimental observations of the skyrmions statics under applied field \cite{zeisslerPinningHysteresisField2017, jugeMagneticSkyrmionsConfined2018, grossSkyrmionMorphologyUltrathin2018, kimCurrentdrivenSkyrmionDynamics2017, legrandRoomTemperatureCurrentInducedGeneration2017}. Fig. \ref{skyrmion motion SIMU}.d displays the motion of a skyrmion for a given disorder configuration and for $J=6.7\times{}10^{11}$ A m$^{-2}$. It reveals that the skyrmion experiences an additional dynamical deformation due to the local variation of anisotropy that creates minima for the DW energy. This makes it difficult to draw a distinction between the two contributions, namely that of the DL-SOT and that of the disorder. The skyrmion at rest is also distorted, since it relaxes on an inhomogeneous energy landscape, explaining the (static) deformation of some skyrmions in Fig. \ref{skyrmion motion EXP} \cite{zeisslerPinningHysteresisField2017, grossSkyrmionMorphologyUltrathin2018, jugeMagneticSkyrmionsConfined2018}. Fig. \ref{skyrmion motion SIMU}.e and f display the skyrmion trajectories recorded for different grain distributions and for $J=2.9\times{}10^{11}$ A m$^{-2}$ and $J=6.7\times{}10^{11}$ A m$^{-2}$, respectively. The blue dotted lines represent the trajectories in the absence of disorder, which deviate slightly from linearity due to the skyrmion deformation. The first plot reveals a stochastic dynamics, with a large dispersion in the direction of motion. In some cases, the skyrmions are stopped on a strong pinning site, preventing further motion. For larger current densities, the number of such events decreases and the trajectory approaches that of the ideal case. This is emphasized in Fig. \ref{velocity SkHA SIMU}.a, which shows the skyrmion velocity as a function of the current density in the ideal case (blue dots) and the disordered case (black stars).

The disorder introduces different regimes in the skyrmion dynamics: at low driving current, the average velocity, calculated from the total displacement after a computation time of 8 ns, is close to zero. This pinned regime is highlighted by the shaded areas in Fig. \ref{velocity SkHA SIMU}. At higher current density, the force due to the current becomes large compared to the pinning force and the velocity increases and finally reaches the disorder-free velocity, defining the flow regime. This behavior is analogous to that of current-driven DWs \cite{mironFastCurrentinducedDomainwall2011} and points out that care should be taken when interpreting the regime of the skyrmion dynamics. In the pinned regime, a reduction in the average SkHA is observed. Upon increasing the driving force, the SkHA increases monotonically and converges to the disorder-free SkHA in the flow regime. These results can be understood as follows: the pinning can be viewed as a force acting opposite to the ideal skyrmion motion direction \cite{iwasakiUniversalCurrentvelocityRelation2013}, thus lowering its velocity. The gyrotropic response to this force will make the skyrmion deviate from its initial trajectory in a direction transverse to this pinning force, according to the second term of Eq. (\ref{thiele equation}). This effect, referred to as extrinsic SkHE \cite{kimCurrentdrivenSkyrmionDynamics2017}, prevails at low current density and becomes negligible as the force due to the current becomes sufficiently large as compared to the pinning force.

These results are in excellent agreement with our observations and strongly suggest that the observed drive-dependence of the skyrmion velocity and that of the SkHA is largely due to pinning. Note that only the nonshaded areas in Fig. \ref{velocity SkHA SIMU} are to be compared with the experimental results in Fig. \ref{velocity SkHA EXP}. Although the critical current densities and skyrmion velocities for which the depinning and flow regimes are reached depend on the choice of disorder parameters \cite{kimCurrentdrivenSkyrmionDynamics2017, legrandRoomTemperatureCurrentInducedGeneration2017, salimathCurrentdrivenSkyrmionDepinning2019}, this approach constitutes a good qualitative description of the pinning effect on the current-driven skyrmion dynamics. Qualitatively, the micromagnetic simulations reveal that the disorder-free velocity and SkHA constitute upper bounds for the velocity and SkHA in disordered systems. This supports the aforementioned assumption of a lower damping in our film, which would increase the disorder-free velocity and the SkHA (Fig. \ref{velocity SkHA EXP}, red dashed lines), leading to an excellent agreement with the simulations. Therefore, it confirms that the observed skyrmion dynamics are in a depinning regime and this suggests that the flow regime is reached for the largest current densities.

\section{Conclusions}

We have studied the current-driven dynamics of small magnetic skyrmions in an ultrathin Pt{\slash}Co{\slash}MgO film at room temperature. We observe that isolated magnetic skyrmions exhibit fast motion (100 m s$^{-1}$) and a drive-dependent SkHE. Supported by a detailed characterization of the film properties and the SOTs governing the skyrmion dynamics, we show that these observations can be well accounted for by Thiele’s model at high current density, which is particularly relevant in an ultrathin film hosting skyrmions with a well defined chirality. Micromagnetic simulations including the measured DL-SOT and FL-SOT as well as material inhomogeneities reproduce the different regimes of the drive-dependent skyrmion velocity and SkHA. This, in turn, allows us to rule out the impact of the FL-SOT on the skyrmion dynamics and to identify pinning as the dominant effect responsible for the drive dependence of the SkHE. Our results shed light on the current-driven skyrmion dynamics in ultrathin films, which paves the way for future experimental investigations for the purpose of developing low-power skyrmion-based applications.  

\section*{ACKNOWLEDGEMENT}

We thank Nikita Strelkov for his contribution to the simulations and Jordi Prat for his support at the CIRCE beam line. We acknowledge the help from Olivier Fruchart and Simon Le Denmat for performing magnetic force microscopy with low-moment tips. The authors acknowledge the support of the Agence Nationale de la Recherche, Project ANR-17-CE24-0045 (SKYLOGIC), and the support of the DARPA TEE program through Grant No. MIPR HR0011831554. The XMCD-PEEM experiments were performed at (1) the CIRCE beam line at the ALBA synchrotron \cite{AballeALBAspectroscopicLEEMPEEM2015, FoersterCustomsampleenvironments2016}, with CALIPSOplus funding (Grant No. 730872), (2) the Nanospectroscopy beam line at the Elettra synchrotron \cite{mentesCathodeLensSpectromicroscopy2014}, with funding from the European Community’s Horizon 2020 Framework Programme under Grant Agreement 730872, and (3) the the beam line I06 at Diamond Light Source under proposal number SI-I4035.

\nocite{apsrev41Control}
\bibliographystyle{apsrev4-1}
\bibliography{main_bbl}

\newcommand{\noopsort}[1]{}
\begin{thebibliography}{51}%
\makeatletter
\providecommand \@ifxundefined [1]{%
 \@ifx{#1\undefined}
}%
\providecommand \@ifnum [1]{%
 \ifnum #1\expandafter \@firstoftwo
 \else \expandafter \@secondoftwo
 \fi
}%
\providecommand \@ifx [1]{%
 \ifx #1\expandafter \@firstoftwo
 \else \expandafter \@secondoftwo
 \fi
}%
\providecommand \natexlab [1]{#1}%
\providecommand \enquote  [1]{``#1''}%
\providecommand \bibnamefont  [1]{#1}%
\providecommand \bibfnamefont [1]{#1}%
\providecommand \citenamefont [1]{#1}%
\providecommand \href@noop [0]{\@secondoftwo}%
\providecommand \href [0]{\begingroup \@sanitize@url \@href}%
\providecommand \@href[1]{\@@startlink{#1}\@@href}%
\providecommand \@@href[1]{\endgroup#1\@@endlink}%
\providecommand \@sanitize@url [0]{\catcode `\\12\catcode `\$12\catcode
  `\&12\catcode `\#12\catcode `\^12\catcode `\_12\catcode `\%12\relax}%
\providecommand \@@startlink[1]{}%
\providecommand \@@endlink[0]{}%
\providecommand \url  [0]{\begingroup\@sanitize@url \@url }%
\providecommand \@url [1]{\endgroup\@href {#1}{\urlprefix }}%
\providecommand \urlprefix  [0]{URL }%
\providecommand \Eprint [0]{\href }%
\providecommand \doibase [0]{http://dx.doi.org/}%
\providecommand \selectlanguage [0]{\@gobble}%
\providecommand \bibinfo  [0]{\@secondoftwo}%
\providecommand \bibfield  [0]{\@secondoftwo}%
\providecommand \translation [1]{[#1]}%
\providecommand \BibitemOpen [0]{}%
\providecommand \bibitemStop [0]{}%
\providecommand \bibitemNoStop [0]{.\EOS\space}%
\providecommand \EOS [0]{\spacefactor3000\relax}%
\providecommand \BibitemShut  [1]{\csname bibitem#1\endcsname}%
\let\auto@bib@innerbib\@empty
%</preamble>
\bibitem [{\citenamefont {Fert}\ \emph {et~al.}(2013)\citenamefont {Fert},
  \citenamefont {Cros},\ and\ \citenamefont
  {Sampaio}}]{fertSkyrmionsTrack2013}%
  \BibitemOpen
  \bibfield  {author} {\bibinfo {author} {\bibfnamefont {A.}~\bibnamefont
  {Fert}}, \bibinfo {author} {\bibfnamefont {V.}~\bibnamefont {Cros}}, \ and\
  \bibinfo {author} {\bibfnamefont {J.}~\bibnamefont {Sampaio}},\ }\bibfield
  {title} {\enquote {\bibinfo {title} {Skyrmions on the track},}\ }\href
  {\doibase 10.1038/nnano.2013.29} {\bibfield  {journal} {\bibinfo  {journal}
  {Nat. Nanotech.}\ }\textbf {\bibinfo {volume} {8}},\ \bibinfo {pages} {152}
  (\bibinfo {year} {2013})}\BibitemShut {NoStop}%
\bibitem [{\citenamefont {Zhang}\ \emph {et~al.}(2015)\citenamefont {Zhang},
  \citenamefont {Ezawa},\ and\ \citenamefont
  {Zhou}}]{zhangMagneticSkyrmionLogic2015}%
  \BibitemOpen
  \bibfield  {author} {\bibinfo {author} {\bibfnamefont {X.}~\bibnamefont
  {Zhang}}, \bibinfo {author} {\bibfnamefont {M.}~\bibnamefont {Ezawa}}, \ and\
  \bibinfo {author} {\bibfnamefont {Y.}~\bibnamefont {Zhou}},\ }\bibfield
  {title} {\enquote {\bibinfo {title} {Magnetic skyrmion logic gates:
  Conversion, duplication and merging of skyrmions},}\ }\href {\doibase
  10.1038/srep09400} {\bibfield  {journal} {\bibinfo  {journal} {Sci. Rep.}\
  }\textbf {\bibinfo {volume} {5}},\ \bibinfo {pages} {9400} (\bibinfo {year}
  {2015})}\BibitemShut {NoStop}%
\bibitem [{\citenamefont {Bogdanov}\ and\ \citenamefont
  {Hubert}(1999)}]{bogdanovStabilityVortexlikeStructures1999}%
  \BibitemOpen
  \bibfield  {author} {\bibinfo {author} {\bibfnamefont {A.}~\bibnamefont
  {Bogdanov}}\ and\ \bibinfo {author} {\bibfnamefont {A.}~\bibnamefont
  {Hubert}},\ }\bibfield  {title} {\enquote {\bibinfo {title} {The stability of
  vortex-like structures in uniaxial ferromagnets},}\ }\href {\doibase
  10.1016/S0304-8853(98)01038-5} {\bibfield  {journal} {\bibinfo  {journal} {J.
  Magn. Magn. Mater.}\ }\textbf {\bibinfo {volume} {195}},\ \bibinfo {pages}
  {182} (\bibinfo {year} {1999})}\BibitemShut {NoStop}%
\bibitem [{\citenamefont {Sampaio}\ \emph {et~al.}(2013)\citenamefont
  {Sampaio}, \citenamefont {Cros}, \citenamefont {Rohart}, \citenamefont
  {Thiaville},\ and\ \citenamefont
  {Fert}}]{sampaioNucleationStabilityCurrentinduced2013}%
  \BibitemOpen
  \bibfield  {author} {\bibinfo {author} {\bibfnamefont {J.}~\bibnamefont
  {Sampaio}}, \bibinfo {author} {\bibfnamefont {V.}~\bibnamefont {Cros}},
  \bibinfo {author} {\bibfnamefont {S.}~\bibnamefont {Rohart}}, \bibinfo
  {author} {\bibfnamefont {A.}~\bibnamefont {Thiaville}}, \ and\ \bibinfo
  {author} {\bibfnamefont {A.}~\bibnamefont {Fert}},\ }\bibfield  {title}
  {\enquote {\bibinfo {title} {Nucleation, stability and current-induced motion
  of isolated magnetic skyrmions in nanostructures},}\ }\href {\doibase
  10.1038/nnano.2013.210} {\bibfield  {journal} {\bibinfo  {journal} {Nat.
  Nanotech.}\ }\textbf {\bibinfo {volume} {8}},\ \bibinfo {pages} {839}
  (\bibinfo {year} {2013})}\BibitemShut {NoStop}%
\bibitem [{\citenamefont {Jiang}\ \emph {et~al.}(2015)\citenamefont {Jiang},
  \citenamefont {Upadhyaya}, \citenamefont {Zhang}, \citenamefont {Yu},
  \citenamefont {Jungfleisch}, \citenamefont {Fradin}, \citenamefont {Pearson},
  \citenamefont {Tserkovnyak}, \citenamefont {Wang}, \citenamefont {Heinonen},
  \citenamefont {{\noopsort{velthuis}}te Velthuis},\ and\ \citenamefont
  {Hoffmann}}]{jiangBlowingMagneticSkyrmion2015}%
  \BibitemOpen
  \bibfield  {author} {\bibinfo {author} {\bibfnamefont {W.}~\bibnamefont
  {Jiang}}, \bibinfo {author} {\bibfnamefont {P.}~\bibnamefont {Upadhyaya}},
  \bibinfo {author} {\bibfnamefont {W.}~\bibnamefont {Zhang}}, \bibinfo
  {author} {\bibfnamefont {G.}~\bibnamefont {Yu}}, \bibinfo {author}
  {\bibfnamefont {M.~B.}\ \bibnamefont {Jungfleisch}}, \bibinfo {author}
  {\bibfnamefont {F.~Y.}\ \bibnamefont {Fradin}}, \bibinfo {author}
  {\bibfnamefont {J.~E.}\ \bibnamefont {Pearson}}, \bibinfo {author}
  {\bibfnamefont {Y.}~\bibnamefont {Tserkovnyak}}, \bibinfo {author}
  {\bibfnamefont {K.~L.}\ \bibnamefont {Wang}}, \bibinfo {author}
  {\bibfnamefont {O.}~\bibnamefont {Heinonen}}, \bibinfo {author}
  {\bibfnamefont {S.~G.~E.}\ \bibnamefont {{\noopsort{velthuis}}te Velthuis}},
  \ and\ \bibinfo {author} {\bibfnamefont {A.}~\bibnamefont {Hoffmann}},\
  }\bibfield  {title} {\enquote {\bibinfo {title} {Blowing magnetic skyrmion
  bubbles},}\ }\href {\doibase 10.1126/science.aaa1442} {\bibfield  {journal}
  {\bibinfo  {journal} {Science}\ }\textbf {\bibinfo {volume} {349}},\ \bibinfo
  {pages} {283} (\bibinfo {year} {2015})}\BibitemShut {NoStop}%
\bibitem [{\citenamefont {{Moreau-Luchaire}}\ \emph {et~al.}(2016)\citenamefont
  {{Moreau-Luchaire}}, \citenamefont {Moutafis}, \citenamefont {Reyren},
  \citenamefont {Sampaio}, \citenamefont {Vaz}, \citenamefont {Horne},
  \citenamefont {Bouzehouane}, \citenamefont {Garcia}, \citenamefont
  {Deranlot}, \citenamefont {Warnicke}, \citenamefont {Wohlh{\"u}ter},
  \citenamefont {George}, \citenamefont {Weigand}, \citenamefont {Raabe},
  \citenamefont {Cros},\ and\ \citenamefont
  {Fert}}]{moreau-luchaireAdditiveinterfacialchiral2016}%
  \BibitemOpen
  \bibfield  {author} {\bibinfo {author} {\bibfnamefont {C.}~\bibnamefont
  {{Moreau-Luchaire}}}, \bibinfo {author} {\bibfnamefont {C.}~\bibnamefont
  {Moutafis}}, \bibinfo {author} {\bibfnamefont {N.}~\bibnamefont {Reyren}},
  \bibinfo {author} {\bibfnamefont {J.}~\bibnamefont {Sampaio}}, \bibinfo
  {author} {\bibfnamefont {C.~a.~F.}\ \bibnamefont {Vaz}}, \bibinfo {author}
  {\bibfnamefont {N.~V.}\ \bibnamefont {Horne}}, \bibinfo {author}
  {\bibfnamefont {K.}~\bibnamefont {Bouzehouane}}, \bibinfo {author}
  {\bibfnamefont {K.}~\bibnamefont {Garcia}}, \bibinfo {author} {\bibfnamefont
  {C.}~\bibnamefont {Deranlot}}, \bibinfo {author} {\bibfnamefont
  {P.}~\bibnamefont {Warnicke}}, \bibinfo {author} {\bibfnamefont
  {P.}~\bibnamefont {Wohlh{\"u}ter}}, \bibinfo {author} {\bibfnamefont {J.-M.}\
  \bibnamefont {George}}, \bibinfo {author} {\bibfnamefont {M.}~\bibnamefont
  {Weigand}}, \bibinfo {author} {\bibfnamefont {J.}~\bibnamefont {Raabe}},
  \bibinfo {author} {\bibfnamefont {V.}~\bibnamefont {Cros}}, \ and\ \bibinfo
  {author} {\bibfnamefont {A.}~\bibnamefont {Fert}},\ }\bibfield  {title}
  {\enquote {\bibinfo {title} {Additive interfacial chiral interaction in
  multilayers for stabilization of small individual skyrmions at room
  temperature},}\ }\href {\doibase 10.1038/nnano.2015.313} {\bibfield
  {journal} {\bibinfo  {journal} {Nat. Nanotech.}\ }\textbf {\bibinfo {volume}
  {11}},\ \bibinfo {pages} {444} (\bibinfo {year} {2016})}\BibitemShut
  {NoStop}%
\bibitem [{\citenamefont {Boulle}\ \emph {et~al.}(2016)\citenamefont {Boulle},
  \citenamefont {Vogel}, \citenamefont {Yang}, \citenamefont {Pizzini},
  \citenamefont {{\noopsort{souza chaves}}{de Souza Chaves}}, \citenamefont
  {Locatelli}, \citenamefont {Mente{\c s}}, \citenamefont {Sala}, \citenamefont
  {{Buda-Prejbeanu}}, \citenamefont {Klein}, \citenamefont {Belmeguenai},
  \citenamefont {Roussign{\'e}}, \citenamefont {Stashkevich}, \citenamefont
  {Ch{\'e}rif}, \citenamefont {Aballe}, \citenamefont {Foerster}, \citenamefont
  {Chshiev}, \citenamefont {Auffret}, \citenamefont {Miron},\ and\
  \citenamefont {Gaudin}}]{boulleRoomtemperatureChiralMagnetic2016}%
  \BibitemOpen
  \bibfield  {author} {\bibinfo {author} {\bibfnamefont {O.}~\bibnamefont
  {Boulle}}, \bibinfo {author} {\bibfnamefont {J.}~\bibnamefont {Vogel}},
  \bibinfo {author} {\bibfnamefont {H.}~\bibnamefont {Yang}}, \bibinfo {author}
  {\bibfnamefont {S.}~\bibnamefont {Pizzini}}, \bibinfo {author} {\bibfnamefont
  {D.}~\bibnamefont {{\noopsort{souza chaves}}{de Souza Chaves}}}, \bibinfo
  {author} {\bibfnamefont {A.}~\bibnamefont {Locatelli}}, \bibinfo {author}
  {\bibfnamefont {T.~O.}\ \bibnamefont {Mente{\c s}}}, \bibinfo {author}
  {\bibfnamefont {A.}~\bibnamefont {Sala}}, \bibinfo {author} {\bibfnamefont
  {L.~D.}\ \bibnamefont {{Buda-Prejbeanu}}}, \bibinfo {author} {\bibfnamefont
  {O.}~\bibnamefont {Klein}}, \bibinfo {author} {\bibfnamefont
  {M.}~\bibnamefont {Belmeguenai}}, \bibinfo {author} {\bibfnamefont
  {Y.}~\bibnamefont {Roussign{\'e}}}, \bibinfo {author} {\bibfnamefont
  {A.}~\bibnamefont {Stashkevich}}, \bibinfo {author} {\bibfnamefont {S.~M.}\
  \bibnamefont {Ch{\'e}rif}}, \bibinfo {author} {\bibfnamefont
  {L.}~\bibnamefont {Aballe}}, \bibinfo {author} {\bibfnamefont
  {M.}~\bibnamefont {Foerster}}, \bibinfo {author} {\bibfnamefont
  {M.}~\bibnamefont {Chshiev}}, \bibinfo {author} {\bibfnamefont
  {S.}~\bibnamefont {Auffret}}, \bibinfo {author} {\bibfnamefont {I.~M.}\
  \bibnamefont {Miron}}, \ and\ \bibinfo {author} {\bibfnamefont
  {G.}~\bibnamefont {Gaudin}},\ }\bibfield  {title} {\enquote {\bibinfo {title}
  {Room-temperature chiral magnetic skyrmions in ultrathin magnetic
  nanostructures},}\ }\href {\doibase 10.1038/nnano.2015.315} {\bibfield
  {journal} {\bibinfo  {journal} {Nat. Nanotech.}\ }\textbf {\bibinfo {volume}
  {11}},\ \bibinfo {pages} {449} (\bibinfo {year} {2016})}\BibitemShut
  {NoStop}%
\bibitem [{\citenamefont {Woo}\ \emph {et~al.}(2016)\citenamefont {Woo},
  \citenamefont {Litzius}, \citenamefont {Kr{\"u}ger}, \citenamefont {Im},
  \citenamefont {Caretta}, \citenamefont {Richter}, \citenamefont {Mann},
  \citenamefont {Krone}, \citenamefont {Reeve}, \citenamefont {Weigand},
  \citenamefont {Agrawal}, \citenamefont {Lemesh}, \citenamefont {Mawass},
  \citenamefont {Fischer}, \citenamefont {Kl{\"a}ui},\ and\ \citenamefont
  {Beach}}]{wooObservationRoomtemperatureMagnetic2016}%
  \BibitemOpen
  \bibfield  {author} {\bibinfo {author} {\bibfnamefont {S.}~\bibnamefont
  {Woo}}, \bibinfo {author} {\bibfnamefont {K.}~\bibnamefont {Litzius}},
  \bibinfo {author} {\bibfnamefont {B.}~\bibnamefont {Kr{\"u}ger}}, \bibinfo
  {author} {\bibfnamefont {M.-Y.}\ \bibnamefont {Im}}, \bibinfo {author}
  {\bibfnamefont {L.}~\bibnamefont {Caretta}}, \bibinfo {author} {\bibfnamefont
  {K.}~\bibnamefont {Richter}}, \bibinfo {author} {\bibfnamefont
  {M.}~\bibnamefont {Mann}}, \bibinfo {author} {\bibfnamefont {A.}~\bibnamefont
  {Krone}}, \bibinfo {author} {\bibfnamefont {R.~M.}\ \bibnamefont {Reeve}},
  \bibinfo {author} {\bibfnamefont {M.}~\bibnamefont {Weigand}}, \bibinfo
  {author} {\bibfnamefont {P.}~\bibnamefont {Agrawal}}, \bibinfo {author}
  {\bibfnamefont {I.}~\bibnamefont {Lemesh}}, \bibinfo {author} {\bibfnamefont
  {M.-A.}\ \bibnamefont {Mawass}}, \bibinfo {author} {\bibfnamefont
  {P.}~\bibnamefont {Fischer}}, \bibinfo {author} {\bibfnamefont
  {M.}~\bibnamefont {Kl{\"a}ui}}, \ and\ \bibinfo {author} {\bibfnamefont
  {G.~S.~D.}\ \bibnamefont {Beach}},\ }\bibfield  {title} {\enquote {\bibinfo
  {title} {Observation of room-temperature magnetic skyrmions and their
  current-driven dynamics in ultrathin metallic ferromagnets},}\ }\href
  {\doibase 10.1038/nmat4593} {\bibfield  {journal} {\bibinfo  {journal} {Nat.
  Mater.}\ }\textbf {\bibinfo {volume} {15}},\ \bibinfo {pages} {501} (\bibinfo
  {year} {2016})}\BibitemShut {NoStop}%
\bibitem [{\citenamefont {Yu}\ \emph {et~al.}(2016)\citenamefont {Yu},
  \citenamefont {Upadhyaya}, \citenamefont {Li}, \citenamefont {Li},
  \citenamefont {Kim}, \citenamefont {Fan}, \citenamefont {Wong}, \citenamefont
  {Tserkovnyak}, \citenamefont {Amiri},\ and\ \citenamefont
  {Wang}}]{yuRoomTemperatureCreationSpin2016}%
  \BibitemOpen
  \bibfield  {author} {\bibinfo {author} {\bibfnamefont {G.}~\bibnamefont
  {Yu}}, \bibinfo {author} {\bibfnamefont {P.}~\bibnamefont {Upadhyaya}},
  \bibinfo {author} {\bibfnamefont {X.}~\bibnamefont {Li}}, \bibinfo {author}
  {\bibfnamefont {W.}~\bibnamefont {Li}}, \bibinfo {author} {\bibfnamefont
  {S.~K.}\ \bibnamefont {Kim}}, \bibinfo {author} {\bibfnamefont
  {Y.}~\bibnamefont {Fan}}, \bibinfo {author} {\bibfnamefont {K.~L.}\
  \bibnamefont {Wong}}, \bibinfo {author} {\bibfnamefont {Y.}~\bibnamefont
  {Tserkovnyak}}, \bibinfo {author} {\bibfnamefont {P.~K.}\ \bibnamefont
  {Amiri}}, \ and\ \bibinfo {author} {\bibfnamefont {K.~L.}\ \bibnamefont
  {Wang}},\ }\bibfield  {title} {\enquote {\bibinfo {title} {Room-{{Temperature
  Creation}} and {{Spin}}\textendash{{Orbit Torque Manipulation}} of
  {{Skyrmions}} in {{Thin Films}} with {{Engineered Asymmetry}}},}\ }\href
  {\doibase 10.1021/acs.nanolett.5b05257} {\bibfield  {journal} {\bibinfo
  {journal} {Nano Lett.}\ }\textbf {\bibinfo {volume} {16}},\ \bibinfo {pages}
  {1981} (\bibinfo {year} {2016})}\BibitemShut {NoStop}%
\bibitem [{\citenamefont {Soumyanarayanan}\ \emph {et~al.}(2017)\citenamefont
  {Soumyanarayanan}, \citenamefont {Raju}, \citenamefont {Oyarce},
  \citenamefont {Tan}, \citenamefont {Im}, \citenamefont {Petrovi{\'c}},
  \citenamefont {Ho}, \citenamefont {Khoo}, \citenamefont {Tran}, \citenamefont
  {Gan}, \citenamefont {Ernult},\ and\ \citenamefont
  {Panagopoulos}}]{soumyanarayananTunableroomtemperaturemagnetic2017}%
  \BibitemOpen
  \bibfield  {author} {\bibinfo {author} {\bibfnamefont {A.}~\bibnamefont
  {Soumyanarayanan}}, \bibinfo {author} {\bibfnamefont {M.}~\bibnamefont
  {Raju}}, \bibinfo {author} {\bibfnamefont {A.~L.~G.}\ \bibnamefont {Oyarce}},
  \bibinfo {author} {\bibfnamefont {A.~K.~C.}\ \bibnamefont {Tan}}, \bibinfo
  {author} {\bibfnamefont {M.-Y.}\ \bibnamefont {Im}}, \bibinfo {author}
  {\bibfnamefont {A.~P.}\ \bibnamefont {Petrovi{\'c}}}, \bibinfo {author}
  {\bibfnamefont {P.}~\bibnamefont {Ho}}, \bibinfo {author} {\bibfnamefont
  {K.~H.}\ \bibnamefont {Khoo}}, \bibinfo {author} {\bibfnamefont
  {M.}~\bibnamefont {Tran}}, \bibinfo {author} {\bibfnamefont {C.~K.}\
  \bibnamefont {Gan}}, \bibinfo {author} {\bibfnamefont {F.}~\bibnamefont
  {Ernult}}, \ and\ \bibinfo {author} {\bibfnamefont {C.}~\bibnamefont
  {Panagopoulos}},\ }\bibfield  {title} {\enquote {\bibinfo {title} {Tunable
  room-temperature magnetic skyrmions in {{Ir}}/{{Fe}}/{{Co}}/{{Pt}}
  multilayers},}\ }\href {\doibase 10.1038/nmat4934} {\bibfield  {journal}
  {\bibinfo  {journal} {Nat. Mater.}\ }\textbf {\bibinfo {volume} {16}},\
  \bibinfo {pages} {898} (\bibinfo {year} {2017})}\BibitemShut {NoStop}%
\bibitem [{\citenamefont {Woo}\ \emph {et~al.}(2017)\citenamefont {Woo},
  \citenamefont {Song}, \citenamefont {Han}, \citenamefont {Jung},
  \citenamefont {Im}, \citenamefont {Lee}, \citenamefont {Song}, \citenamefont
  {Fischer}, \citenamefont {Hong}, \citenamefont {Choi}, \citenamefont {Min},
  \citenamefont {Koo},\ and\ \citenamefont
  {Chang}}]{wooSpinorbitTorquedrivenSkyrmion2017}%
  \BibitemOpen
  \bibfield  {author} {\bibinfo {author} {\bibfnamefont {S.}~\bibnamefont
  {Woo}}, \bibinfo {author} {\bibfnamefont {K.~M.}\ \bibnamefont {Song}},
  \bibinfo {author} {\bibfnamefont {H.-S.}\ \bibnamefont {Han}}, \bibinfo
  {author} {\bibfnamefont {M.-S.}\ \bibnamefont {Jung}}, \bibinfo {author}
  {\bibfnamefont {M.-Y.}\ \bibnamefont {Im}}, \bibinfo {author} {\bibfnamefont
  {K.-S.}\ \bibnamefont {Lee}}, \bibinfo {author} {\bibfnamefont {K.~S.}\
  \bibnamefont {Song}}, \bibinfo {author} {\bibfnamefont {P.}~\bibnamefont
  {Fischer}}, \bibinfo {author} {\bibfnamefont {J.-I.}\ \bibnamefont {Hong}},
  \bibinfo {author} {\bibfnamefont {J.~W.}\ \bibnamefont {Choi}}, \bibinfo
  {author} {\bibfnamefont {B.-C.}\ \bibnamefont {Min}}, \bibinfo {author}
  {\bibfnamefont {H.~C.}\ \bibnamefont {Koo}}, \ and\ \bibinfo {author}
  {\bibfnamefont {J.}~\bibnamefont {Chang}},\ }\bibfield  {title} {\enquote
  {\bibinfo {title} {Spin-orbit torque-driven skyrmion dynamics revealed by
  time-resolved {{X}}-ray microscopy},}\ }\href {\doibase 10.1038/ncomms15573}
  {\bibfield  {journal} {\bibinfo  {journal} {Nat. Commun.}\ }\textbf {\bibinfo
  {volume} {8}},\ \bibinfo {pages} {15573} (\bibinfo {year}
  {2017})}\BibitemShut {NoStop}%
\bibitem [{\citenamefont {Hrabec}\ \emph {et~al.}(2017)\citenamefont {Hrabec},
  \citenamefont {Sampaio}, \citenamefont {Belmeguenai}, \citenamefont {Gross},
  \citenamefont {Weil}, \citenamefont {Ch{\'e}rif}, \citenamefont
  {Stashkevich}, \citenamefont {Jacques}, \citenamefont {Thiaville},\ and\
  \citenamefont {Rohart}}]{hrabecCurrentinducedSkyrmionGeneration2017}%
  \BibitemOpen
  \bibfield  {author} {\bibinfo {author} {\bibfnamefont {A.}~\bibnamefont
  {Hrabec}}, \bibinfo {author} {\bibfnamefont {J.}~\bibnamefont {Sampaio}},
  \bibinfo {author} {\bibfnamefont {M.}~\bibnamefont {Belmeguenai}}, \bibinfo
  {author} {\bibfnamefont {I.}~\bibnamefont {Gross}}, \bibinfo {author}
  {\bibfnamefont {R.}~\bibnamefont {Weil}}, \bibinfo {author} {\bibfnamefont
  {S.~M.}\ \bibnamefont {Ch{\'e}rif}}, \bibinfo {author} {\bibfnamefont
  {A.}~\bibnamefont {Stashkevich}}, \bibinfo {author} {\bibfnamefont
  {V.}~\bibnamefont {Jacques}}, \bibinfo {author} {\bibfnamefont
  {A.}~\bibnamefont {Thiaville}}, \ and\ \bibinfo {author} {\bibfnamefont
  {S.}~\bibnamefont {Rohart}},\ }\bibfield  {title} {\enquote {\bibinfo {title}
  {Current-induced skyrmion generation and dynamics in symmetric bilayers},}\
  }\href {\doibase 10.1038/ncomms15765} {\bibfield  {journal} {\bibinfo
  {journal} {Nat. Commun.}\ }\textbf {\bibinfo {volume} {8}},\ \bibinfo {pages}
  {15765} (\bibinfo {year} {2017})}\BibitemShut {NoStop}%
\bibitem [{\citenamefont {Caretta}\ \emph {et~al.}(2018)\citenamefont
  {Caretta}, \citenamefont {Mann}, \citenamefont {B{\"u}ttner}, \citenamefont
  {Ueda}, \citenamefont {Pfau}, \citenamefont {G{\"u}nther}, \citenamefont
  {Hessing}, \citenamefont {Churikova}, \citenamefont {Klose}, \citenamefont
  {Schneider}, \citenamefont {Engel}, \citenamefont {Marcus}, \citenamefont
  {Bono}, \citenamefont {Bagschik}, \citenamefont {Eisebitt},\ and\
  \citenamefont {Beach}}]{carettaFastCurrentdrivenDomain2018}%
  \BibitemOpen
  \bibfield  {author} {\bibinfo {author} {\bibfnamefont {L.}~\bibnamefont
  {Caretta}}, \bibinfo {author} {\bibfnamefont {M.}~\bibnamefont {Mann}},
  \bibinfo {author} {\bibfnamefont {F.}~\bibnamefont {B{\"u}ttner}}, \bibinfo
  {author} {\bibfnamefont {K.}~\bibnamefont {Ueda}}, \bibinfo {author}
  {\bibfnamefont {B.}~\bibnamefont {Pfau}}, \bibinfo {author} {\bibfnamefont
  {C.~M.}\ \bibnamefont {G{\"u}nther}}, \bibinfo {author} {\bibfnamefont
  {P.}~\bibnamefont {Hessing}}, \bibinfo {author} {\bibfnamefont
  {A.}~\bibnamefont {Churikova}}, \bibinfo {author} {\bibfnamefont
  {C.}~\bibnamefont {Klose}}, \bibinfo {author} {\bibfnamefont
  {M.}~\bibnamefont {Schneider}}, \bibinfo {author} {\bibfnamefont
  {D.}~\bibnamefont {Engel}}, \bibinfo {author} {\bibfnamefont
  {C.}~\bibnamefont {Marcus}}, \bibinfo {author} {\bibfnamefont
  {D.}~\bibnamefont {Bono}}, \bibinfo {author} {\bibfnamefont {K.}~\bibnamefont
  {Bagschik}}, \bibinfo {author} {\bibfnamefont {S.}~\bibnamefont {Eisebitt}},
  \ and\ \bibinfo {author} {\bibfnamefont {G.~S.~D.}\ \bibnamefont {Beach}},\
  }\bibfield  {title} {\enquote {\bibinfo {title} {Fast current-driven domain
  walls and small skyrmions in a compensated ferrimagnet},}\ }\href {\doibase
  10.1038/s41565-018-0255-3} {\bibfield  {journal} {\bibinfo  {journal} {Nat.
  Nanotech.}\ }\textbf {\bibinfo {volume} {13}},\ \bibinfo {pages} {1154}
  (\bibinfo {year} {2018})}\BibitemShut {NoStop}%
\bibitem [{\citenamefont {Woo}\ \emph {et~al.}(2018)\citenamefont {Woo},
  \citenamefont {Song}, \citenamefont {Zhang}, \citenamefont {Zhou},
  \citenamefont {Ezawa}, \citenamefont {Liu}, \citenamefont {Finizio},
  \citenamefont {Raabe}, \citenamefont {Lee}, \citenamefont {Kim},
  \citenamefont {Park}, \citenamefont {Kim}, \citenamefont {Kim}, \citenamefont
  {Lee}, \citenamefont {Lee}, \citenamefont {Choi}, \citenamefont {Min},
  \citenamefont {Koo},\ and\ \citenamefont
  {Chang}}]{wooCurrentdrivenDynamicsInhibition2018}%
  \BibitemOpen
  \bibfield  {author} {\bibinfo {author} {\bibfnamefont {S.}~\bibnamefont
  {Woo}}, \bibinfo {author} {\bibfnamefont {K.~M.}\ \bibnamefont {Song}},
  \bibinfo {author} {\bibfnamefont {X.}~\bibnamefont {Zhang}}, \bibinfo
  {author} {\bibfnamefont {Y.}~\bibnamefont {Zhou}}, \bibinfo {author}
  {\bibfnamefont {M.}~\bibnamefont {Ezawa}}, \bibinfo {author} {\bibfnamefont
  {X.}~\bibnamefont {Liu}}, \bibinfo {author} {\bibfnamefont {S.}~\bibnamefont
  {Finizio}}, \bibinfo {author} {\bibfnamefont {J.}~\bibnamefont {Raabe}},
  \bibinfo {author} {\bibfnamefont {N.~J.}\ \bibnamefont {Lee}}, \bibinfo
  {author} {\bibfnamefont {S.-I.}\ \bibnamefont {Kim}}, \bibinfo {author}
  {\bibfnamefont {S.-Y.}\ \bibnamefont {Park}}, \bibinfo {author}
  {\bibfnamefont {Y.}~\bibnamefont {Kim}}, \bibinfo {author} {\bibfnamefont
  {J.-Y.}\ \bibnamefont {Kim}}, \bibinfo {author} {\bibfnamefont
  {D.}~\bibnamefont {Lee}}, \bibinfo {author} {\bibfnamefont {O.}~\bibnamefont
  {Lee}}, \bibinfo {author} {\bibfnamefont {J.~W.}\ \bibnamefont {Choi}},
  \bibinfo {author} {\bibfnamefont {B.-C.}\ \bibnamefont {Min}}, \bibinfo
  {author} {\bibfnamefont {H.~C.}\ \bibnamefont {Koo}}, \ and\ \bibinfo
  {author} {\bibfnamefont {J.}~\bibnamefont {Chang}},\ }\bibfield  {title}
  {\enquote {\bibinfo {title} {Current-driven dynamics and inhibition of the
  skyrmion {{Hall}} effect of ferrimagnetic skyrmions in {{GdFeCo}} films},}\
  }\href {\doibase 10.1038/s41467-018-03378-7} {\bibfield  {journal} {\bibinfo
  {journal} {Nat. Commun.}\ }\textbf {\bibinfo {volume} {9}},\ \bibinfo {pages}
  {959} (\bibinfo {year} {2018})}\BibitemShut {NoStop}%
\bibitem [{\citenamefont {Jiang}\ \emph {et~al.}(2017)\citenamefont {Jiang},
  \citenamefont {Zhang}, \citenamefont {Yu}, \citenamefont {Zhang},
  \citenamefont {Wang}, \citenamefont {Benjamin~Jungfleisch}, \citenamefont
  {Pearson}, \citenamefont {Cheng}, \citenamefont {Heinonen}, \citenamefont
  {Wang}, \citenamefont {Zhou}, \citenamefont {Hoffmann},\ and\ \citenamefont
  {{\noopsort{velthuis}}{te Velthuis}}}]{jiangDirectObservationSkyrmion2017}%
  \BibitemOpen
  \bibfield  {author} {\bibinfo {author} {\bibfnamefont {W.}~\bibnamefont
  {Jiang}}, \bibinfo {author} {\bibfnamefont {X.}~\bibnamefont {Zhang}},
  \bibinfo {author} {\bibfnamefont {G.}~\bibnamefont {Yu}}, \bibinfo {author}
  {\bibfnamefont {W.}~\bibnamefont {Zhang}}, \bibinfo {author} {\bibfnamefont
  {X.}~\bibnamefont {Wang}}, \bibinfo {author} {\bibfnamefont {M.}~\bibnamefont
  {Benjamin~Jungfleisch}}, \bibinfo {author} {\bibfnamefont {J.~E.}\
  \bibnamefont {Pearson}}, \bibinfo {author} {\bibfnamefont {X.}~\bibnamefont
  {Cheng}}, \bibinfo {author} {\bibfnamefont {O.}~\bibnamefont {Heinonen}},
  \bibinfo {author} {\bibfnamefont {K.~L.}\ \bibnamefont {Wang}}, \bibinfo
  {author} {\bibfnamefont {Y.}~\bibnamefont {Zhou}}, \bibinfo {author}
  {\bibfnamefont {A.}~\bibnamefont {Hoffmann}}, \ and\ \bibinfo {author}
  {\bibfnamefont {S.~G.~E.}\ \bibnamefont {{\noopsort{velthuis}}{te
  Velthuis}}},\ }\bibfield  {title} {\enquote {\bibinfo {title} {Direct
  observation of the skyrmion {{Hall}} effect},}\ }\href {\doibase
  10.1038/nphys3883} {\bibfield  {journal} {\bibinfo  {journal} {Nat. Phys.}\
  }\textbf {\bibinfo {volume} {13}},\ \bibinfo {pages} {162} (\bibinfo {year}
  {2017})}\BibitemShut {NoStop}%
\bibitem [{\citenamefont {Litzius}\ \emph {et~al.}(2017)\citenamefont
  {Litzius}, \citenamefont {Lemesh}, \citenamefont {Kr{\"u}ger}, \citenamefont
  {Bassirian}, \citenamefont {Caretta}, \citenamefont {Richter}, \citenamefont
  {B{\"u}ttner}, \citenamefont {Sato}, \citenamefont {Tretiakov}, \citenamefont
  {F{\"o}rster}, \citenamefont {Reeve}, \citenamefont {Weigand}, \citenamefont
  {Bykova}, \citenamefont {Stoll}, \citenamefont {Sch{\"u}tz}, \citenamefont
  {Beach},\ and\ \citenamefont {Kl{\"a}ui}}]{litziusSkyrmionHallEffect2017}%
  \BibitemOpen
  \bibfield  {author} {\bibinfo {author} {\bibfnamefont {K.}~\bibnamefont
  {Litzius}}, \bibinfo {author} {\bibfnamefont {I.}~\bibnamefont {Lemesh}},
  \bibinfo {author} {\bibfnamefont {B.}~\bibnamefont {Kr{\"u}ger}}, \bibinfo
  {author} {\bibfnamefont {P.}~\bibnamefont {Bassirian}}, \bibinfo {author}
  {\bibfnamefont {L.}~\bibnamefont {Caretta}}, \bibinfo {author} {\bibfnamefont
  {K.}~\bibnamefont {Richter}}, \bibinfo {author} {\bibfnamefont
  {F.}~\bibnamefont {B{\"u}ttner}}, \bibinfo {author} {\bibfnamefont
  {K.}~\bibnamefont {Sato}}, \bibinfo {author} {\bibfnamefont {O.~A.}\
  \bibnamefont {Tretiakov}}, \bibinfo {author} {\bibfnamefont {J.}~\bibnamefont
  {F{\"o}rster}}, \bibinfo {author} {\bibfnamefont {R.~M.}\ \bibnamefont
  {Reeve}}, \bibinfo {author} {\bibfnamefont {M.}~\bibnamefont {Weigand}},
  \bibinfo {author} {\bibfnamefont {I.}~\bibnamefont {Bykova}}, \bibinfo
  {author} {\bibfnamefont {H.}~\bibnamefont {Stoll}}, \bibinfo {author}
  {\bibfnamefont {G.}~\bibnamefont {Sch{\"u}tz}}, \bibinfo {author}
  {\bibfnamefont {G.~S.~D.}\ \bibnamefont {Beach}}, \ and\ \bibinfo {author}
  {\bibfnamefont {M.}~\bibnamefont {Kl{\"a}ui}},\ }\bibfield  {title} {\enquote
  {\bibinfo {title} {Skyrmion {{Hall}} effect revealed by direct time-resolved
  {{X}}-ray microscopy},}\ }\href {\doibase 10.1038/nphys4000} {\bibfield
  {journal} {\bibinfo  {journal} {Nat. Phys.}\ }\textbf {\bibinfo {volume}
  {13}},\ \bibinfo {pages} {170} (\bibinfo {year} {2017})}\BibitemShut
  {NoStop}%
\bibitem [{\citenamefont {Legrand}\ \emph
  {et~al.}(2018{\natexlab{a}})\citenamefont {Legrand}, \citenamefont
  {Chauleau}, \citenamefont {Maccariello}, \citenamefont {Reyren},
  \citenamefont {Collin}, \citenamefont {Bouzehouane}, \citenamefont {Jaouen},
  \citenamefont {Cros},\ and\ \citenamefont
  {Fert}}]{legrandHybridchiraldomain2018}%
  \BibitemOpen
  \bibfield  {author} {\bibinfo {author} {\bibfnamefont {W.}~\bibnamefont
  {Legrand}}, \bibinfo {author} {\bibfnamefont {J.-Y.}\ \bibnamefont
  {Chauleau}}, \bibinfo {author} {\bibfnamefont {D.}~\bibnamefont
  {Maccariello}}, \bibinfo {author} {\bibfnamefont {N.}~\bibnamefont {Reyren}},
  \bibinfo {author} {\bibfnamefont {S.}~\bibnamefont {Collin}}, \bibinfo
  {author} {\bibfnamefont {K.}~\bibnamefont {Bouzehouane}}, \bibinfo {author}
  {\bibfnamefont {N.}~\bibnamefont {Jaouen}}, \bibinfo {author} {\bibfnamefont
  {V.}~\bibnamefont {Cros}}, \ and\ \bibinfo {author} {\bibfnamefont
  {A.}~\bibnamefont {Fert}},\ }\bibfield  {title} {\enquote {\bibinfo {title}
  {Hybrid chiral domain walls and skyrmions in magnetic multilayers},}\ }\href
  {\doibase 10.1126/sciadv.aat0415} {\bibfield  {journal} {\bibinfo  {journal}
  {Sci. Adv.}\ }\textbf {\bibinfo {volume} {4}},\ \bibinfo {pages} {eaat0415}
  (\bibinfo {year} {2018}{\natexlab{a}})}\BibitemShut {NoStop}%
\bibitem [{\citenamefont {Dovzhenko}\ \emph {et~al.}(2018)\citenamefont
  {Dovzhenko}, \citenamefont {Casola}, \citenamefont {Schlotter}, \citenamefont
  {Zhou}, \citenamefont {B{\"u}ttner}, \citenamefont {Walsworth}, \citenamefont
  {Beach},\ and\ \citenamefont
  {Yacoby}}]{dovzhenkoMagnetostaticTwistsRoomtemperature2018}%
  \BibitemOpen
  \bibfield  {author} {\bibinfo {author} {\bibfnamefont {Y.}~\bibnamefont
  {Dovzhenko}}, \bibinfo {author} {\bibfnamefont {F.}~\bibnamefont {Casola}},
  \bibinfo {author} {\bibfnamefont {S.}~\bibnamefont {Schlotter}}, \bibinfo
  {author} {\bibfnamefont {T.~X.}\ \bibnamefont {Zhou}}, \bibinfo {author}
  {\bibfnamefont {F.}~\bibnamefont {B{\"u}ttner}}, \bibinfo {author}
  {\bibfnamefont {R.~L.}\ \bibnamefont {Walsworth}}, \bibinfo {author}
  {\bibfnamefont {G.~S.~D.}\ \bibnamefont {Beach}}, \ and\ \bibinfo {author}
  {\bibfnamefont {A.}~\bibnamefont {Yacoby}},\ }\bibfield  {title} {\enquote
  {\bibinfo {title} {Magnetostatic twists in room-temperature skyrmions
  explored by nitrogen-vacancy center spin texture reconstruction},}\ }\href
  {\doibase 10.1038/s41467-018-05158-9} {\bibfield  {journal} {\bibinfo
  {journal} {Nat. Commun.}\ }\textbf {\bibinfo {volume} {9}},\ \bibinfo {pages}
  {2712} (\bibinfo {year} {2018})}\BibitemShut {NoStop}%
\bibitem [{\citenamefont {Li}\ \emph {et~al.}(2019)\citenamefont {Li},
  \citenamefont {Bykova}, \citenamefont {Zhang}, \citenamefont {Yu},
  \citenamefont {Tomasello}, \citenamefont {Carpentieri}, \citenamefont {Liu},
  \citenamefont {Guang}, \citenamefont {Gr{\"a}fe}, \citenamefont {Weigand},
  \citenamefont {Burn}, \citenamefont {{\noopsort{laan}}van~der Laan},
  \citenamefont {Hesjedal}, \citenamefont {Yan}, \citenamefont {Feng},
  \citenamefont {Wan}, \citenamefont {Wei}, \citenamefont {Wang}, \citenamefont
  {Zhang}, \citenamefont {Xu}, \citenamefont {Guo}, \citenamefont {Wei},
  \citenamefont {Finocchio}, \citenamefont {Han},\ and\ \citenamefont
  {Sch{\"u}tz}}]{liAnatomySkyrmionicTextures}%
  \BibitemOpen
  \bibfield  {author} {\bibinfo {author} {\bibfnamefont {W.}~\bibnamefont
  {Li}}, \bibinfo {author} {\bibfnamefont {I.}~\bibnamefont {Bykova}}, \bibinfo
  {author} {\bibfnamefont {S.}~\bibnamefont {Zhang}}, \bibinfo {author}
  {\bibfnamefont {G.}~\bibnamefont {Yu}}, \bibinfo {author} {\bibfnamefont
  {R.}~\bibnamefont {Tomasello}}, \bibinfo {author} {\bibfnamefont
  {M.}~\bibnamefont {Carpentieri}}, \bibinfo {author} {\bibfnamefont
  {Y.}~\bibnamefont {Liu}}, \bibinfo {author} {\bibfnamefont {Y.}~\bibnamefont
  {Guang}}, \bibinfo {author} {\bibfnamefont {J.}~\bibnamefont {Gr{\"a}fe}},
  \bibinfo {author} {\bibfnamefont {M.}~\bibnamefont {Weigand}}, \bibinfo
  {author} {\bibfnamefont {D.~M.}\ \bibnamefont {Burn}}, \bibinfo {author}
  {\bibfnamefont {G.}~\bibnamefont {{\noopsort{laan}}van~der Laan}}, \bibinfo
  {author} {\bibfnamefont {T.}~\bibnamefont {Hesjedal}}, \bibinfo {author}
  {\bibfnamefont {Z.}~\bibnamefont {Yan}}, \bibinfo {author} {\bibfnamefont
  {J.}~\bibnamefont {Feng}}, \bibinfo {author} {\bibfnamefont {C.}~\bibnamefont
  {Wan}}, \bibinfo {author} {\bibfnamefont {J.}~\bibnamefont {Wei}}, \bibinfo
  {author} {\bibfnamefont {X.}~\bibnamefont {Wang}}, \bibinfo {author}
  {\bibfnamefont {X.}~\bibnamefont {Zhang}}, \bibinfo {author} {\bibfnamefont
  {H.}~\bibnamefont {Xu}}, \bibinfo {author} {\bibfnamefont {C.}~\bibnamefont
  {Guo}}, \bibinfo {author} {\bibfnamefont {H.}~\bibnamefont {Wei}}, \bibinfo
  {author} {\bibfnamefont {G.}~\bibnamefont {Finocchio}}, \bibinfo {author}
  {\bibfnamefont {X.}~\bibnamefont {Han}}, \ and\ \bibinfo {author}
  {\bibfnamefont {G.}~\bibnamefont {Sch{\"u}tz}},\ }\bibfield  {title}
  {\enquote {\bibinfo {title} {Anatomy of {{Skyrmionic Textures}} in {{Magnetic
  Multilayers}}},}\ }\href {\doibase 10.1002/adma.201807683} {\bibfield
  {journal} {\bibinfo  {journal} {Adv. Mater.}\ }\textbf {\bibinfo {volume}
  {31}},\ \bibinfo {pages} {1807683} (\bibinfo {year} {2019})}\BibitemShut
  {NoStop}%
\bibitem [{\citenamefont {Lemesh}\ and\ \citenamefont
  {Beach}(2018)}]{lemeshTwistedDomainWalls2018}%
  \BibitemOpen
  \bibfield  {author} {\bibinfo {author} {\bibfnamefont {I.}~\bibnamefont
  {Lemesh}}\ and\ \bibinfo {author} {\bibfnamefont {G.~S.~D.}\ \bibnamefont
  {Beach}},\ }\bibfield  {title} {\enquote {\bibinfo {title} {Twisted domain
  walls and skyrmions in perpendicularly magnetized multilayers},}\ }\href
  {\doibase 10.1103/PhysRevB.98.104402} {\bibfield  {journal} {\bibinfo
  {journal} {Phys. Rev. B}\ }\textbf {\bibinfo {volume} {98}},\ \bibinfo
  {pages} {104402} (\bibinfo {year} {2018})}\BibitemShut {NoStop}%
\bibitem [{\citenamefont {Legrand}\ \emph
  {et~al.}(2018{\natexlab{b}})\citenamefont {Legrand}, \citenamefont
  {Ronceray}, \citenamefont {Reyren}, \citenamefont {Maccariello},
  \citenamefont {Cros},\ and\ \citenamefont
  {Fert}}]{legrandModelingShapeAxisymmetric2018}%
  \BibitemOpen
  \bibfield  {author} {\bibinfo {author} {\bibfnamefont {W.}~\bibnamefont
  {Legrand}}, \bibinfo {author} {\bibfnamefont {N.}~\bibnamefont {Ronceray}},
  \bibinfo {author} {\bibfnamefont {N.}~\bibnamefont {Reyren}}, \bibinfo
  {author} {\bibfnamefont {D.}~\bibnamefont {Maccariello}}, \bibinfo {author}
  {\bibfnamefont {V.}~\bibnamefont {Cros}}, \ and\ \bibinfo {author}
  {\bibfnamefont {A.}~\bibnamefont {Fert}},\ }\bibfield  {title} {\enquote
  {\bibinfo {title} {Modeling the {{Shape}} of {{Axisymmetric Skyrmions}} in
  {{Magnetic Multilayers}}},}\ }\href {\doibase
  10.1103/PhysRevApplied.10.064042} {\bibfield  {journal} {\bibinfo  {journal}
  {Phys. Rev. Appl.}\ }\textbf {\bibinfo {volume} {10}},\ \bibinfo {pages}
  {064042} (\bibinfo {year} {2018}{\natexlab{b}})}\BibitemShut {NoStop}%
\bibitem [{\citenamefont {Kim}\ and\ \citenamefont
  {Yoo}(2017)}]{kimCurrentdrivenSkyrmionDynamics2017}%
  \BibitemOpen
  \bibfield  {author} {\bibinfo {author} {\bibfnamefont {J.-V.}\ \bibnamefont
  {Kim}}\ and\ \bibinfo {author} {\bibfnamefont {M.-W.}\ \bibnamefont {Yoo}},\
  }\bibfield  {title} {\enquote {\bibinfo {title} {Current-driven skyrmion
  dynamics in disordered films},}\ }\href {\doibase 10.1063/1.4979316}
  {\bibfield  {journal} {\bibinfo  {journal} {Appl. Phys. Lett.}\ }\textbf
  {\bibinfo {volume} {110}},\ \bibinfo {pages} {132404} (\bibinfo {year}
  {2017})}\BibitemShut {NoStop}%
\bibitem [{\citenamefont {Legrand}\ \emph {et~al.}(2017)\citenamefont
  {Legrand}, \citenamefont {Maccariello}, \citenamefont {Reyren}, \citenamefont
  {Garcia}, \citenamefont {Moutafis}, \citenamefont {{Moreau-Luchaire}},
  \citenamefont {Collin}, \citenamefont {Bouzehouane}, \citenamefont {Cros},\
  and\ \citenamefont
  {Fert}}]{legrandRoomTemperatureCurrentInducedGeneration2017}%
  \BibitemOpen
  \bibfield  {author} {\bibinfo {author} {\bibfnamefont {W.}~\bibnamefont
  {Legrand}}, \bibinfo {author} {\bibfnamefont {D.}~\bibnamefont
  {Maccariello}}, \bibinfo {author} {\bibfnamefont {N.}~\bibnamefont {Reyren}},
  \bibinfo {author} {\bibfnamefont {K.}~\bibnamefont {Garcia}}, \bibinfo
  {author} {\bibfnamefont {C.}~\bibnamefont {Moutafis}}, \bibinfo {author}
  {\bibfnamefont {C.}~\bibnamefont {{Moreau-Luchaire}}}, \bibinfo {author}
  {\bibfnamefont {S.}~\bibnamefont {Collin}}, \bibinfo {author} {\bibfnamefont
  {K.}~\bibnamefont {Bouzehouane}}, \bibinfo {author} {\bibfnamefont
  {V.}~\bibnamefont {Cros}}, \ and\ \bibinfo {author} {\bibfnamefont
  {A.}~\bibnamefont {Fert}},\ }\bibfield  {title} {\enquote {\bibinfo {title}
  {Room-{{Temperature Current}}-{{Induced Generation}} and {{Motion}} of
  sub-100 nm {{Skyrmions}}},}\ }\href {\doibase 10.1021/acs.nanolett.7b00649}
  {\bibfield  {journal} {\bibinfo  {journal} {Nano Lett.}\ }\textbf {\bibinfo
  {volume} {17}},\ \bibinfo {pages} {2703} (\bibinfo {year}
  {2017})}\BibitemShut {NoStop}%
\bibitem [{\citenamefont {Zeissler}\ \emph {et~al.}(2017)\citenamefont
  {Zeissler}, \citenamefont {Mruczkiewicz}, \citenamefont {Finizio},
  \citenamefont {Raabe}, \citenamefont {Shepley}, \citenamefont {Sadovnikov},
  \citenamefont {Nikitov}, \citenamefont {Fallon}, \citenamefont {McFadzean},
  \citenamefont {McVitie}, \citenamefont {Moore}, \citenamefont {Burnell},\
  and\ \citenamefont {Marrows}}]{zeisslerPinningHysteresisField2017}%
  \BibitemOpen
  \bibfield  {author} {\bibinfo {author} {\bibfnamefont {K.}~\bibnamefont
  {Zeissler}}, \bibinfo {author} {\bibfnamefont {M.}~\bibnamefont
  {Mruczkiewicz}}, \bibinfo {author} {\bibfnamefont {S.}~\bibnamefont
  {Finizio}}, \bibinfo {author} {\bibfnamefont {J.}~\bibnamefont {Raabe}},
  \bibinfo {author} {\bibfnamefont {P.~M.}\ \bibnamefont {Shepley}}, \bibinfo
  {author} {\bibfnamefont {A.~V.}\ \bibnamefont {Sadovnikov}}, \bibinfo
  {author} {\bibfnamefont {S.~A.}\ \bibnamefont {Nikitov}}, \bibinfo {author}
  {\bibfnamefont {K.}~\bibnamefont {Fallon}}, \bibinfo {author} {\bibfnamefont
  {S.}~\bibnamefont {McFadzean}}, \bibinfo {author} {\bibfnamefont
  {S.}~\bibnamefont {McVitie}}, \bibinfo {author} {\bibfnamefont {T.~A.}\
  \bibnamefont {Moore}}, \bibinfo {author} {\bibfnamefont {G.}~\bibnamefont
  {Burnell}}, \ and\ \bibinfo {author} {\bibfnamefont {C.~H.}\ \bibnamefont
  {Marrows}},\ }\bibfield  {title} {\enquote {\bibinfo {title} {Pinning and
  hysteresis in the field dependent diameter evolution of skyrmions in
  {{Pt}}/{{Co}}/{{Ir}} superlattice stacks},}\ }\href {\doibase
  10.1038/s41598-017-15262-3} {\bibfield  {journal} {\bibinfo  {journal} {Sci.
  Rep.}\ }\textbf {\bibinfo {volume} {7}},\ \bibinfo {pages} {15125} (\bibinfo
  {year} {2017})}\BibitemShut {NoStop}%
\bibitem [{\citenamefont {Juge}\ \emph {et~al.}(2018)\citenamefont {Juge},
  \citenamefont {Je}, \citenamefont {{\noopsort{souza chaves}}{de Souza
  Chaves}}, \citenamefont {Pizzini}, \citenamefont {{Buda-Prejbeanu}},
  \citenamefont {Aballe}, \citenamefont {Foerster}, \citenamefont {Locatelli},
  \citenamefont {Mente{\c s}}, \citenamefont {Sala}, \citenamefont
  {Maccherozzi}, \citenamefont {Dhesi}, \citenamefont {Auffret}, \citenamefont
  {Gautier}, \citenamefont {Gaudin}, \citenamefont {Vogel},\ and\ \citenamefont
  {Boulle}}]{jugeMagneticSkyrmionsConfined2018}%
  \BibitemOpen
  \bibfield  {author} {\bibinfo {author} {\bibfnamefont {R.}~\bibnamefont
  {Juge}}, \bibinfo {author} {\bibfnamefont {S.-G.}\ \bibnamefont {Je}},
  \bibinfo {author} {\bibfnamefont {D.}~\bibnamefont {{\noopsort{souza
  chaves}}{de Souza Chaves}}}, \bibinfo {author} {\bibfnamefont
  {S.}~\bibnamefont {Pizzini}}, \bibinfo {author} {\bibfnamefont {L.~D.}\
  \bibnamefont {{Buda-Prejbeanu}}}, \bibinfo {author} {\bibfnamefont
  {L.}~\bibnamefont {Aballe}}, \bibinfo {author} {\bibfnamefont
  {M.}~\bibnamefont {Foerster}}, \bibinfo {author} {\bibfnamefont
  {A.}~\bibnamefont {Locatelli}}, \bibinfo {author} {\bibfnamefont {T.~O.}\
  \bibnamefont {Mente{\c s}}}, \bibinfo {author} {\bibfnamefont
  {A.}~\bibnamefont {Sala}}, \bibinfo {author} {\bibfnamefont {F.}~\bibnamefont
  {Maccherozzi}}, \bibinfo {author} {\bibfnamefont {S.~S.}\ \bibnamefont
  {Dhesi}}, \bibinfo {author} {\bibfnamefont {S.}~\bibnamefont {Auffret}},
  \bibinfo {author} {\bibfnamefont {E.}~\bibnamefont {Gautier}}, \bibinfo
  {author} {\bibfnamefont {G.}~\bibnamefont {Gaudin}}, \bibinfo {author}
  {\bibfnamefont {J.}~\bibnamefont {Vogel}}, \ and\ \bibinfo {author}
  {\bibfnamefont {O.}~\bibnamefont {Boulle}},\ }\bibfield  {title} {\enquote
  {\bibinfo {title} {Magnetic skyrmions in confined geometries: {{Effect}} of
  the magnetic field and the disorder},}\ }\href {\doibase
  10.1016/j.jmmm.2017.10.030} {\bibfield  {journal} {\bibinfo  {journal} {J.
  Magn. Magn. Mater.}\ }\textbf {\bibinfo {volume} {455}},\ \bibinfo {pages}
  {3} (\bibinfo {year} {2018})}\BibitemShut {NoStop}%
\bibitem [{\citenamefont {Gross}\ \emph {et~al.}(2018)\citenamefont {Gross},
  \citenamefont {Akhtar}, \citenamefont {Hrabec}, \citenamefont {Sampaio},
  \citenamefont {Mart{\'i}nez}, \citenamefont {Chouaieb}, \citenamefont
  {Shields}, \citenamefont {Maletinsky}, \citenamefont {Thiaville},
  \citenamefont {Rohart},\ and\ \citenamefont
  {Jacques}}]{grossSkyrmionMorphologyUltrathin2018}%
  \BibitemOpen
  \bibfield  {author} {\bibinfo {author} {\bibfnamefont {I.}~\bibnamefont
  {Gross}}, \bibinfo {author} {\bibfnamefont {W.}~\bibnamefont {Akhtar}},
  \bibinfo {author} {\bibfnamefont {A.}~\bibnamefont {Hrabec}}, \bibinfo
  {author} {\bibfnamefont {J.}~\bibnamefont {Sampaio}}, \bibinfo {author}
  {\bibfnamefont {L.~J.}\ \bibnamefont {Mart{\'i}nez}}, \bibinfo {author}
  {\bibfnamefont {S.}~\bibnamefont {Chouaieb}}, \bibinfo {author}
  {\bibfnamefont {B.~J.}\ \bibnamefont {Shields}}, \bibinfo {author}
  {\bibfnamefont {P.}~\bibnamefont {Maletinsky}}, \bibinfo {author}
  {\bibfnamefont {A.}~\bibnamefont {Thiaville}}, \bibinfo {author}
  {\bibfnamefont {S.}~\bibnamefont {Rohart}}, \ and\ \bibinfo {author}
  {\bibfnamefont {V.}~\bibnamefont {Jacques}},\ }\bibfield  {title} {\enquote
  {\bibinfo {title} {Skyrmion morphology in ultrathin magnetic films},}\ }\href
  {\doibase 10.1103/PhysRevMaterials.2.024406} {\bibfield  {journal} {\bibinfo
  {journal} {Phys. Rev. Mater.}\ }\textbf {\bibinfo {volume} {2}},\ \bibinfo
  {pages} {024406} (\bibinfo {year} {2018})}\BibitemShut {NoStop}%
\bibitem [{\citenamefont {Thiele}(1973)}]{thieleSteadyStateMotionMagnetic1973}%
  \BibitemOpen
  \bibfield  {author} {\bibinfo {author} {\bibfnamefont {A.~A.}\ \bibnamefont
  {Thiele}},\ }\bibfield  {title} {\enquote {\bibinfo {title} {Steady-{{State
  Motion}} of {{Magnetic Domains}}},}\ }\href {\doibase
  10.1103/PhysRevLett.30.230} {\bibfield  {journal} {\bibinfo  {journal} {Phys.
  Rev. Lett.}\ }\textbf {\bibinfo {volume} {30}},\ \bibinfo {pages} {230}
  (\bibinfo {year} {1973})}\BibitemShut {NoStop}%
\bibitem [{\citenamefont {Nguyen}\ \emph {et~al.}(2016)\citenamefont {Nguyen},
  \citenamefont {Ralph},\ and\ \citenamefont
  {Buhrman}}]{nguyenSpinTorqueStudy2016}%
  \BibitemOpen
  \bibfield  {author} {\bibinfo {author} {\bibfnamefont {M.-H.}\ \bibnamefont
  {Nguyen}}, \bibinfo {author} {\bibfnamefont {D.~C.}\ \bibnamefont {Ralph}}, \
  and\ \bibinfo {author} {\bibfnamefont {R.~A.}\ \bibnamefont {Buhrman}},\
  }\bibfield  {title} {\enquote {\bibinfo {title} {Spin {{Torque Study}} of the
  {{Spin Hall Conductivity}} and {{Spin Diffusion Length}} in {{Platinum Thin
  Films}} with {{Varying Resistivity}}},}\ }\href {\doibase
  10.1103/PhysRevLett.116.126601} {\bibfield  {journal} {\bibinfo  {journal}
  {Phys. Rev. Lett.}\ }\textbf {\bibinfo {volume} {116}},\ \bibinfo {pages}
  {126601} (\bibinfo {year} {2016})}\BibitemShut {NoStop}%
\bibitem [{\citenamefont
  {Thiele}(1974)}]{thieleApplicationsGyrocouplingVector1974}%
  \BibitemOpen
  \bibfield  {author} {\bibinfo {author} {\bibfnamefont {A.~A.}\ \bibnamefont
  {Thiele}},\ }\bibfield  {title} {\enquote {\bibinfo {title} {Applications of
  the gyrocoupling vector and dissipation dyadic in the dynamics of magnetic
  domains},}\ }\href {\doibase 10.1063/1.1662989} {\bibfield  {journal}
  {\bibinfo  {journal} {J. Appl. Phys.}\ }\textbf {\bibinfo {volume} {45}},\
  \bibinfo {pages} {377} (\bibinfo {year} {1974})}\BibitemShut {NoStop}%
\bibitem [{Note1()}]{Note1}%
  \BibitemOpen
  \bibinfo {note} {In a previous work, we observed using XMCD-PEEM, that the
  skyrmion profile in Pt/Co/MgO was fitted well by a 360{\textdegree } Bloch
  wall profile \cite {boulleRoomtemperatureChiralMagnetic2016}}\BibitemShut
  {NoStop}%
\bibitem [{\citenamefont {Tomasello}\ \emph {et~al.}(2014)\citenamefont
  {Tomasello}, \citenamefont {Martinez}, \citenamefont {Zivieri}, \citenamefont
  {Torres}, \citenamefont {Carpentieri},\ and\ \citenamefont
  {Finocchio}}]{tomaselloStrategyDesignSkyrmion2014}%
  \BibitemOpen
  \bibfield  {author} {\bibinfo {author} {\bibfnamefont {R.}~\bibnamefont
  {Tomasello}}, \bibinfo {author} {\bibfnamefont {E.}~\bibnamefont {Martinez}},
  \bibinfo {author} {\bibfnamefont {R.}~\bibnamefont {Zivieri}}, \bibinfo
  {author} {\bibfnamefont {L.}~\bibnamefont {Torres}}, \bibinfo {author}
  {\bibfnamefont {M.}~\bibnamefont {Carpentieri}}, \ and\ \bibinfo {author}
  {\bibfnamefont {G.}~\bibnamefont {Finocchio}},\ }\bibfield  {title} {\enquote
  {\bibinfo {title} {A strategy for the design of skyrmion racetrack
  memories},}\ }\href {\doibase 10.1038/srep06784} {\bibfield  {journal}
  {\bibinfo  {journal} {Sci. Rep.}\ }\textbf {\bibinfo {volume} {4}},\ \bibinfo
  {pages} {6784} (\bibinfo {year} {2014})}\BibitemShut {NoStop}%
\bibitem [{\citenamefont {Miron}\ \emph {et~al.}(2011)\citenamefont {Miron},
  \citenamefont {Moore}, \citenamefont {Szambolics}, \citenamefont
  {{Buda-Prejbeanu}}, \citenamefont {Auffret}, \citenamefont {Rodmacq},
  \citenamefont {Pizzini}, \citenamefont {Vogel}, \citenamefont {Bonfim},
  \citenamefont {Schuhl},\ and\ \citenamefont
  {Gaudin}}]{mironFastCurrentinducedDomainwall2011}%
  \BibitemOpen
  \bibfield  {author} {\bibinfo {author} {\bibfnamefont {I.~M.}\ \bibnamefont
  {Miron}}, \bibinfo {author} {\bibfnamefont {T.}~\bibnamefont {Moore}},
  \bibinfo {author} {\bibfnamefont {H.}~\bibnamefont {Szambolics}}, \bibinfo
  {author} {\bibfnamefont {L.~D.}\ \bibnamefont {{Buda-Prejbeanu}}}, \bibinfo
  {author} {\bibfnamefont {S.}~\bibnamefont {Auffret}}, \bibinfo {author}
  {\bibfnamefont {B.}~\bibnamefont {Rodmacq}}, \bibinfo {author} {\bibfnamefont
  {S.}~\bibnamefont {Pizzini}}, \bibinfo {author} {\bibfnamefont
  {J.}~\bibnamefont {Vogel}}, \bibinfo {author} {\bibfnamefont
  {M.}~\bibnamefont {Bonfim}}, \bibinfo {author} {\bibfnamefont
  {A.}~\bibnamefont {Schuhl}}, \ and\ \bibinfo {author} {\bibfnamefont
  {G.}~\bibnamefont {Gaudin}},\ }\bibfield  {title} {\enquote {\bibinfo {title}
  {Fast current-induced domain-wall motion controlled by the {{Rashba}}
  effect},}\ }\href {\doibase 10.1038/nmat3020} {\bibfield  {journal} {\bibinfo
   {journal} {Nat. Mater.}\ }\textbf {\bibinfo {volume} {10}},\ \bibinfo
  {pages} {419} (\bibinfo {year} {2011})}\BibitemShut {NoStop}%
\bibitem [{\citenamefont {Reichhardt}\ \emph
  {et~al.}(2015{\natexlab{a}})\citenamefont {Reichhardt}, \citenamefont {Ray},\
  and\ \citenamefont {Reichhardt}}]{reichhardtQuantizedTransportSkyrmion2015}%
  \BibitemOpen
  \bibfield  {author} {\bibinfo {author} {\bibfnamefont {C.}~\bibnamefont
  {Reichhardt}}, \bibinfo {author} {\bibfnamefont {D.}~\bibnamefont {Ray}}, \
  and\ \bibinfo {author} {\bibfnamefont {C.~J.~O.}\ \bibnamefont
  {Reichhardt}},\ }\bibfield  {title} {\enquote {\bibinfo {title} {Quantized
  transport for a skyrmion moving on a two-dimensional periodic substrate},}\
  }\href {\doibase 10.1103/PhysRevB.91.104426} {\bibfield  {journal} {\bibinfo
  {journal} {Phys. Rev. B}\ }\textbf {\bibinfo {volume} {91}},\ \bibinfo
  {pages} {104426} (\bibinfo {year} {2015}{\natexlab{a}})}\BibitemShut
  {NoStop}%
\bibitem [{\citenamefont {Reichhardt}\ \emph
  {et~al.}(2015{\natexlab{b}})\citenamefont {Reichhardt}, \citenamefont {Ray},\
  and\ \citenamefont
  {Reichhardt}}]{reichhardtCollectiveTransportProperties2015}%
  \BibitemOpen
  \bibfield  {author} {\bibinfo {author} {\bibfnamefont {C.}~\bibnamefont
  {Reichhardt}}, \bibinfo {author} {\bibfnamefont {D.}~\bibnamefont {Ray}}, \
  and\ \bibinfo {author} {\bibfnamefont {C.~J.~O.}\ \bibnamefont
  {Reichhardt}},\ }\bibfield  {title} {\enquote {\bibinfo {title} {Collective
  {{Transport Properties}} of {{Driven Skyrmions}} with {{Random Disorder}}},}\
  }\href {\doibase 10.1103/PhysRevLett.114.217202} {\bibfield  {journal}
  {\bibinfo  {journal} {Phys. Rev. Lett.}\ }\textbf {\bibinfo {volume} {114}},\
  \bibinfo {pages} {217202} (\bibinfo {year} {2015}{\natexlab{b}})}\BibitemShut
  {NoStop}%
\bibitem [{\citenamefont {Reichhardt}\ and\ \citenamefont
  {Reichhardt}(2016)}]{reichhardtNoiseFluctuationsDrive2016}%
  \BibitemOpen
  \bibfield  {author} {\bibinfo {author} {\bibfnamefont {C.}~\bibnamefont
  {Reichhardt}}\ and\ \bibinfo {author} {\bibfnamefont {C.~J.~O.}\ \bibnamefont
  {Reichhardt}},\ }\bibfield  {title} {\enquote {\bibinfo {title} {Noise
  fluctuations and drive dependence of the skyrmion {{Hall}} effect in
  disordered systems},}\ }\href {\doibase 10.1088/1367-2630/18/9/095005}
  {\bibfield  {journal} {\bibinfo  {journal} {New J. Phys.}\ }\textbf {\bibinfo
  {volume} {18}},\ \bibinfo {pages} {095005} (\bibinfo {year}
  {2016})}\BibitemShut {NoStop}%
\bibitem [{\citenamefont {Metaxas}\ \emph {et~al.}(2007)\citenamefont
  {Metaxas}, \citenamefont {Jamet}, \citenamefont {Mougin}, \citenamefont
  {Cormier}, \citenamefont {Ferr{\'e}}, \citenamefont {Baltz}, \citenamefont
  {Rodmacq}, \citenamefont {Dieny},\ and\ \citenamefont
  {Stamps}}]{metaxasCreepFlowRegimes2007}%
  \BibitemOpen
  \bibfield  {author} {\bibinfo {author} {\bibfnamefont {P.~J.}\ \bibnamefont
  {Metaxas}}, \bibinfo {author} {\bibfnamefont {J.~P.}\ \bibnamefont {Jamet}},
  \bibinfo {author} {\bibfnamefont {A.}~\bibnamefont {Mougin}}, \bibinfo
  {author} {\bibfnamefont {M.}~\bibnamefont {Cormier}}, \bibinfo {author}
  {\bibfnamefont {J.}~\bibnamefont {Ferr{\'e}}}, \bibinfo {author}
  {\bibfnamefont {V.}~\bibnamefont {Baltz}}, \bibinfo {author} {\bibfnamefont
  {B.}~\bibnamefont {Rodmacq}}, \bibinfo {author} {\bibfnamefont
  {B.}~\bibnamefont {Dieny}}, \ and\ \bibinfo {author} {\bibfnamefont {R.~L.}\
  \bibnamefont {Stamps}},\ }\bibfield  {title} {\enquote {\bibinfo {title}
  {{Creep} and {Flow} {Regimes} of {Magnetic} {Domain}-{Wall} {Motion} in
  {Ultrathin} {Pt}/{Co}/{Pt} {Films} with {Perpendicular} {Anisotropy}},}\
  }\href {\doibase 10.1103/PhysRevLett.99.217208} {\bibfield  {journal}
  {\bibinfo  {journal} {Phys. Rev. Lett.}\ }\textbf {\bibinfo {volume} {99}},\
  \bibinfo {pages} {217208} (\bibinfo {year} {2007})}\BibitemShut {NoStop}%
\bibitem [{\citenamefont {Pham}\ \emph {et~al.}(2016)\citenamefont {Pham},
  \citenamefont {Vogel}, \citenamefont {Sampaio}, \citenamefont {Va{\v n}atka},
  \citenamefont {{Rojas-S{\'a}nchez}}, \citenamefont {Bonfim}, \citenamefont
  {Chaves}, \citenamefont {Choueikani}, \citenamefont {Ohresser}, \citenamefont
  {Otero}, \citenamefont {Thiaville},\ and\ \citenamefont
  {Pizzini}}]{phamVeryLargeDomain2016}%
  \BibitemOpen
  \bibfield  {author} {\bibinfo {author} {\bibfnamefont {T.~H.}\ \bibnamefont
  {Pham}}, \bibinfo {author} {\bibfnamefont {J.}~\bibnamefont {Vogel}},
  \bibinfo {author} {\bibfnamefont {J.}~\bibnamefont {Sampaio}}, \bibinfo
  {author} {\bibfnamefont {M.}~\bibnamefont {Va{\v n}atka}}, \bibinfo {author}
  {\bibfnamefont {J.-C.}\ \bibnamefont {{Rojas-S{\'a}nchez}}}, \bibinfo
  {author} {\bibfnamefont {M.}~\bibnamefont {Bonfim}}, \bibinfo {author}
  {\bibfnamefont {D.~S.}\ \bibnamefont {Chaves}}, \bibinfo {author}
  {\bibfnamefont {F.}~\bibnamefont {Choueikani}}, \bibinfo {author}
  {\bibfnamefont {P.}~\bibnamefont {Ohresser}}, \bibinfo {author}
  {\bibfnamefont {E.}~\bibnamefont {Otero}}, \bibinfo {author} {\bibfnamefont
  {A.}~\bibnamefont {Thiaville}}, \ and\ \bibinfo {author} {\bibfnamefont
  {S.}~\bibnamefont {Pizzini}},\ }\bibfield  {title} {\enquote {\bibinfo
  {title} {Very large domain wall velocities in {{Pt}}/{{Co}}/{{GdOx}} and
  {{Pt}}/{{Co}}/{{Gd}} trilayers with {{Dzyaloshinskii}}-{{Moriya}}
  interaction},}\ }\href {\doibase 10.1209/0295-5075/113/67001} {\bibfield
  {journal} {\bibinfo  {journal} {EPL}\ }\textbf {\bibinfo {volume} {113}},\
  \bibinfo {pages} {67001} (\bibinfo {year} {2016})}\BibitemShut {NoStop}%
\bibitem [{\citenamefont {Min}\ \emph {et~al.}(2010)\citenamefont {Min},
  \citenamefont {McMichael}, \citenamefont {Donahue}, \citenamefont {Miltat},\
  and\ \citenamefont {Stiles}}]{minEffectsDisorderInternal2010}%
  \BibitemOpen
  \bibfield  {author} {\bibinfo {author} {\bibfnamefont {H.}~\bibnamefont
  {Min}}, \bibinfo {author} {\bibfnamefont {R.~D.}\ \bibnamefont {McMichael}},
  \bibinfo {author} {\bibfnamefont {M.~J.}\ \bibnamefont {Donahue}}, \bibinfo
  {author} {\bibfnamefont {J.}~\bibnamefont {Miltat}}, \ and\ \bibinfo {author}
  {\bibfnamefont {M.~D.}\ \bibnamefont {Stiles}},\ }\bibfield  {title}
  {\enquote {\bibinfo {title} {Effects of {{Disorder}} and {{Internal
  Dynamics}} on {{Vortex Wall Propagation}}},}\ }\href {\doibase
  10.1103/PhysRevLett.104.217201} {\bibfield  {journal} {\bibinfo  {journal}
  {Phys. Rev. Lett.}\ }\textbf {\bibinfo {volume} {104}},\ \bibinfo {pages}
  {217201} (\bibinfo {year} {2010})}\BibitemShut {NoStop}%
\bibitem [{\citenamefont {Beaujour}\ \emph {et~al.}(2006)\citenamefont
  {Beaujour}, \citenamefont {Lee}, \citenamefont {Kent}, \citenamefont
  {Krycka},\ and\ \citenamefont
  {Kao}}]{beaujourMagnetizationDampingUltrathin2006}%
  \BibitemOpen
  \bibfield  {author} {\bibinfo {author} {\bibfnamefont {J.-M.~L.}\
  \bibnamefont {Beaujour}}, \bibinfo {author} {\bibfnamefont {J.~H.}\
  \bibnamefont {Lee}}, \bibinfo {author} {\bibfnamefont {A.~D.}\ \bibnamefont
  {Kent}}, \bibinfo {author} {\bibfnamefont {K.}~\bibnamefont {Krycka}}, \ and\
  \bibinfo {author} {\bibfnamefont {C.-C.}\ \bibnamefont {Kao}},\ }\bibfield
  {title} {\enquote {\bibinfo {title} {Magnetization damping in ultrathin
  polycrystalline {{Co}} films: {{Evidence}} for nonlocal effects},}\ }\href
  {\doibase 10.1103/PhysRevB.74.214405} {\bibfield  {journal} {\bibinfo
  {journal} {Phys. Rev. B}\ }\textbf {\bibinfo {volume} {74}},\ \bibinfo
  {pages} {214405} (\bibinfo {year} {2006})}\BibitemShut {NoStop}%
\bibitem [{\citenamefont {Kim}\ \emph {et~al.}(2012)\citenamefont {Kim},
  \citenamefont {Moon}, \citenamefont {Lee},\ and\ \citenamefont
  {Lee}}]{kimPredictionGiantSpin2012}%
  \BibitemOpen
  \bibfield  {author} {\bibinfo {author} {\bibfnamefont {K.-W.}\ \bibnamefont
  {Kim}}, \bibinfo {author} {\bibfnamefont {J.-H.}\ \bibnamefont {Moon}},
  \bibinfo {author} {\bibfnamefont {K.-J.}\ \bibnamefont {Lee}}, \ and\
  \bibinfo {author} {\bibfnamefont {H.-W.}\ \bibnamefont {Lee}},\ }\bibfield
  {title} {\enquote {\bibinfo {title} {Prediction of {{Giant Spin Motive
  Force}} due to {{Rashba Spin}}-{{Orbit Coupling}}},}\ }\href {\doibase
  10.1103/PhysRevLett.108.217202} {\bibfield  {journal} {\bibinfo  {journal}
  {Phys. Rev. Lett.}\ }\textbf {\bibinfo {volume} {108}},\ \bibinfo {pages}
  {217202} (\bibinfo {year} {2012})}\BibitemShut {NoStop}%
\bibitem [{\citenamefont {Kim}(2015)}]{kimRoleNonlinearAnisotropic2015}%
  \BibitemOpen
  \bibfield  {author} {\bibinfo {author} {\bibfnamefont {J.-V.}\ \bibnamefont
  {Kim}},\ }\bibfield  {title} {\enquote {\bibinfo {title} {Role of nonlinear
  anisotropic damping in the magnetization dynamics of topological solitons},}\
  }\href {\doibase 10.1103/PhysRevB.92.014418} {\bibfield  {journal} {\bibinfo
  {journal} {Phys. Rev. B}\ }\textbf {\bibinfo {volume} {92}},\ \bibinfo
  {pages} {014418} (\bibinfo {year} {2015})}\BibitemShut {NoStop}%
\bibitem [{\citenamefont {Khvalkovskiy}\ \emph {et~al.}(2013)\citenamefont
  {Khvalkovskiy}, \citenamefont {Cros}, \citenamefont {Apalkov}, \citenamefont
  {Nikitin}, \citenamefont {Krounbi}, \citenamefont {Zvezdin}, \citenamefont
  {Anane}, \citenamefont {Grollier},\ and\ \citenamefont
  {Fert}}]{khvalkovskiyMatchingDomainwallConfiguration2013}%
  \BibitemOpen
  \bibfield  {author} {\bibinfo {author} {\bibfnamefont {A.~V.}\ \bibnamefont
  {Khvalkovskiy}}, \bibinfo {author} {\bibfnamefont {V.}~\bibnamefont {Cros}},
  \bibinfo {author} {\bibfnamefont {D.}~\bibnamefont {Apalkov}}, \bibinfo
  {author} {\bibfnamefont {V.}~\bibnamefont {Nikitin}}, \bibinfo {author}
  {\bibfnamefont {M.}~\bibnamefont {Krounbi}}, \bibinfo {author} {\bibfnamefont
  {K.~A.}\ \bibnamefont {Zvezdin}}, \bibinfo {author} {\bibfnamefont
  {A.}~\bibnamefont {Anane}}, \bibinfo {author} {\bibfnamefont
  {J.}~\bibnamefont {Grollier}}, \ and\ \bibinfo {author} {\bibfnamefont
  {A.}~\bibnamefont {Fert}},\ }\bibfield  {title} {\enquote {\bibinfo {title}
  {Matching domain-wall configuration and spin-orbit torques for efficient
  domain-wall motion},}\ }\href {\doibase 10.1103/PhysRevB.87.020402}
  {\bibfield  {journal} {\bibinfo  {journal} {Phys. Rev. B}\ }\textbf {\bibinfo
  {volume} {87}},\ \bibinfo {pages} {020402} (\bibinfo {year}
  {2013})}\BibitemShut {NoStop}%
\bibitem [{\citenamefont {Buda}\ \emph {et~al.}(2002)\citenamefont {Buda},
  \citenamefont {Prejbeanu}, \citenamefont {Ebels},\ and\ \citenamefont
  {Ounadjela}}]{budaMicromagneticSimulationsMagnetisation2002}%
  \BibitemOpen
  \bibfield  {author} {\bibinfo {author} {\bibfnamefont {L.~D.}\ \bibnamefont
  {Buda}}, \bibinfo {author} {\bibfnamefont {I.~L.}\ \bibnamefont {Prejbeanu}},
  \bibinfo {author} {\bibfnamefont {U.}~\bibnamefont {Ebels}}, \ and\ \bibinfo
  {author} {\bibfnamefont {K.}~\bibnamefont {Ounadjela}},\ }\bibfield  {title}
  {\enquote {\bibinfo {title} {Micromagnetic simulations of magnetisation in
  circular cobalt dots},}\ }\href {\doibase 10.1016/S0927-0256(02)00184-2}
  {\bibfield  {journal} {\bibinfo  {journal} {Comput. Mater. Sci.}\ }\textbf
  {\bibinfo {volume} {24}},\ \bibinfo {pages} {181} (\bibinfo {year}
  {2002})}\BibitemShut {NoStop}%
\bibitem [{\citenamefont {Tomasello}\ \emph {et~al.}(2018)\citenamefont
  {Tomasello}, \citenamefont {Giordano}, \citenamefont {Chiappini},
  \citenamefont {Zivieri}, \citenamefont {Siracusano}, \citenamefont
  {Puliafito}, \citenamefont {Medlej}, \citenamefont {La~Corte}, \citenamefont
  {Azzerboni}, \citenamefont {Carpentieri}, \citenamefont {Zeng},\ and\
  \citenamefont {Finocchio}}]{tomaselloMicromagneticUnderstandingSkyrmion2018}%
  \BibitemOpen
  \bibfield  {author} {\bibinfo {author} {\bibfnamefont {R.}~\bibnamefont
  {Tomasello}}, \bibinfo {author} {\bibfnamefont {A.}~\bibnamefont {Giordano}},
  \bibinfo {author} {\bibfnamefont {S.}~\bibnamefont {Chiappini}}, \bibinfo
  {author} {\bibfnamefont {R.}~\bibnamefont {Zivieri}}, \bibinfo {author}
  {\bibfnamefont {G.}~\bibnamefont {Siracusano}}, \bibinfo {author}
  {\bibfnamefont {V.}~\bibnamefont {Puliafito}}, \bibinfo {author}
  {\bibfnamefont {I.}~\bibnamefont {Medlej}}, \bibinfo {author} {\bibfnamefont
  {A.}~\bibnamefont {La~Corte}}, \bibinfo {author} {\bibfnamefont
  {B.}~\bibnamefont {Azzerboni}}, \bibinfo {author} {\bibfnamefont
  {M.}~\bibnamefont {Carpentieri}}, \bibinfo {author} {\bibfnamefont
  {Z.}~\bibnamefont {Zeng}}, \ and\ \bibinfo {author} {\bibfnamefont
  {G.}~\bibnamefont {Finocchio}},\ }\bibfield  {title} {\enquote {\bibinfo
  {title} {Micromagnetic understanding of the skyrmion {{Hall}} angle current
  dependence in perpendicularly magnetized ferromagnets},}\ }\href {\doibase
  10.1103/PhysRevB.98.224418} {\bibfield  {journal} {\bibinfo  {journal} {Phys.
  Rev. B}\ }\textbf {\bibinfo {volume} {98}},\ \bibinfo {pages} {224418}
  (\bibinfo {year} {2018})}\BibitemShut {NoStop}%
\bibitem [{\citenamefont {Voto}\ \emph {et~al.}(2016)\citenamefont {Voto},
  \citenamefont {{Lopez-Diaz}},\ and\ \citenamefont
  {Torres}}]{votoEffectsGrainSize2016}%
  \BibitemOpen
  \bibfield  {author} {\bibinfo {author} {\bibfnamefont {M.}~\bibnamefont
  {Voto}}, \bibinfo {author} {\bibfnamefont {L.}~\bibnamefont {{Lopez-Diaz}}},
  \ and\ \bibinfo {author} {\bibfnamefont {L.}~\bibnamefont {Torres}},\
  }\bibfield  {title} {\enquote {\bibinfo {title} {Effects of grain size and
  disorder on domain wall propagation in {{CoFeB}} thin films},}\ }\href
  {\doibase 10.1088/0022-3727/49/18/185001} {\bibfield  {journal} {\bibinfo
  {journal} {J. Phys. D: Appl. Phys.}\ }\textbf {\bibinfo {volume} {49}},\
  \bibinfo {pages} {185001} (\bibinfo {year} {2016})}\BibitemShut {NoStop}%
\bibitem [{\citenamefont {{Garcia-Sanchez}}\ \emph {et~al.}(2016)\citenamefont
  {{Garcia-Sanchez}}, \citenamefont {Sampaio}, \citenamefont {Reyren},
  \citenamefont {Cros},\ and\ \citenamefont
  {Kim}}]{garcia-sanchezSkyrmionbasedSpintorqueNanooscillator2016}%
  \BibitemOpen
  \bibfield  {author} {\bibinfo {author} {\bibfnamefont {F.}~\bibnamefont
  {{Garcia-Sanchez}}}, \bibinfo {author} {\bibfnamefont {J.}~\bibnamefont
  {Sampaio}}, \bibinfo {author} {\bibfnamefont {N.}~\bibnamefont {Reyren}},
  \bibinfo {author} {\bibfnamefont {V.}~\bibnamefont {Cros}}, \ and\ \bibinfo
  {author} {\bibfnamefont {J.-V.}\ \bibnamefont {Kim}},\ }\bibfield  {title}
  {\enquote {\bibinfo {title} {A skyrmion-based spin-torque nano-oscillator},}\
  }\href {\doibase 10.1088/1367-2630/18/7/075011} {\bibfield  {journal}
  {\bibinfo  {journal} {New J. Phys.}\ }\textbf {\bibinfo {volume} {18}},\
  \bibinfo {pages} {075011} (\bibinfo {year} {2016})}\BibitemShut {NoStop}%
\bibitem [{\citenamefont {Iwasaki}\ \emph {et~al.}(2013)\citenamefont
  {Iwasaki}, \citenamefont {Mochizuki},\ and\ \citenamefont
  {Nagaosa}}]{iwasakiUniversalCurrentvelocityRelation2013}%
  \BibitemOpen
  \bibfield  {author} {\bibinfo {author} {\bibfnamefont {J.}~\bibnamefont
  {Iwasaki}}, \bibinfo {author} {\bibfnamefont {M.}~\bibnamefont {Mochizuki}},
  \ and\ \bibinfo {author} {\bibfnamefont {N.}~\bibnamefont {Nagaosa}},\
  }\bibfield  {title} {\enquote {\bibinfo {title} {Universal current-velocity
  relation of skyrmion motion in chiral magnets},}\ }\href {\doibase
  10.1038/ncomms2442} {\bibfield  {journal} {\bibinfo  {journal} {Nat.
  Commun.}\ }\textbf {\bibinfo {volume} {4}},\ \bibinfo {pages} {1463}
  (\bibinfo {year} {2013})}\BibitemShut {NoStop}%
\bibitem [{\citenamefont {Salimath}\ \emph {et~al.}(2019)\citenamefont
  {Salimath}, \citenamefont {Abbout}, \citenamefont {Brataas},\ and\
  \citenamefont {Manchon}}]{salimathCurrentdrivenSkyrmionDepinning2019}%
  \BibitemOpen
  \bibfield  {author} {\bibinfo {author} {\bibfnamefont {A.}~\bibnamefont
  {Salimath}}, \bibinfo {author} {\bibfnamefont {A.}~\bibnamefont {Abbout}},
  \bibinfo {author} {\bibfnamefont {A.}~\bibnamefont {Brataas}}, \ and\
  \bibinfo {author} {\bibfnamefont {A.}~\bibnamefont {Manchon}},\ }\bibfield
  {title} {\enquote {\bibinfo {title} {Current-driven skyrmion depinning in
  magnetic granular films},}\ }\href {\doibase 10.1103/PhysRevB.99.104416}
  {\bibfield  {journal} {\bibinfo  {journal} {Phys. Rev. B}\ }\textbf {\bibinfo
  {volume} {99}},\ \bibinfo {pages} {104416} (\bibinfo {year}
  {2019})}\BibitemShut {NoStop}%
\bibitem [{\citenamefont {Aballe}\ \emph {et~al.}(2015)\citenamefont {Aballe},
  \citenamefont {Foerster}, \citenamefont {Pellegrin}, \citenamefont
  {Nicolas},\ and\ \citenamefont
  {Ferrer}}]{AballeALBAspectroscopicLEEMPEEM2015}%
  \BibitemOpen
  \bibfield  {author} {\bibinfo {author} {\bibfnamefont {L.}~\bibnamefont
  {Aballe}}, \bibinfo {author} {\bibfnamefont {M.}~\bibnamefont {Foerster}},
  \bibinfo {author} {\bibfnamefont {E.}~\bibnamefont {Pellegrin}}, \bibinfo
  {author} {\bibfnamefont {J.}~\bibnamefont {Nicolas}}, \ and\ \bibinfo
  {author} {\bibfnamefont {S.}~\bibnamefont {Ferrer}},\ }\bibfield  {title}
  {\enquote {\bibinfo {title} {The {{ALBA}} spectroscopic {{LEEM}}-{{PEEM}}
  experimental station: Layout and performance},}\ }\href {\doibase
  10.1107/S1600577515003537} {\bibfield  {journal} {\bibinfo  {journal} {J
  Synchrotron Rad.}\ }\textbf {\bibinfo {volume} {22}},\ \bibinfo {pages} {745}
  (\bibinfo {year} {2015})}\BibitemShut {NoStop}%
\bibitem [{\citenamefont {Foerster}\ \emph {et~al.}(2016)\citenamefont
  {Foerster}, \citenamefont {Prat}, \citenamefont {Massana}, \citenamefont
  {Gonzalez}, \citenamefont {Fontsere}, \citenamefont {Molas}, \citenamefont
  {Matilla}, \citenamefont {Pellegrin},\ and\ \citenamefont
  {Aballe}}]{FoersterCustomsampleenvironments2016}%
  \BibitemOpen
  \bibfield  {author} {\bibinfo {author} {\bibfnamefont {M.}~\bibnamefont
  {Foerster}}, \bibinfo {author} {\bibfnamefont {J.}~\bibnamefont {Prat}},
  \bibinfo {author} {\bibfnamefont {V.}~\bibnamefont {Massana}}, \bibinfo
  {author} {\bibfnamefont {N.}~\bibnamefont {Gonzalez}}, \bibinfo {author}
  {\bibfnamefont {A.}~\bibnamefont {Fontsere}}, \bibinfo {author}
  {\bibfnamefont {B.}~\bibnamefont {Molas}}, \bibinfo {author} {\bibfnamefont
  {O.}~\bibnamefont {Matilla}}, \bibinfo {author} {\bibfnamefont
  {E.}~\bibnamefont {Pellegrin}}, \ and\ \bibinfo {author} {\bibfnamefont
  {L.}~\bibnamefont {Aballe}},\ }\bibfield  {title} {\enquote {\bibinfo {title}
  {Custom sample environments at the {{ALBA XPEEM}}},}\ }\href {\doibase
  10.1016/j.ultramic.2016.08.016} {\bibfield  {journal} {\bibinfo  {journal}
  {Ultramicroscopy}\ }\textbf {\bibinfo {volume} {171}},\ \bibinfo {pages} {63}
  (\bibinfo {year} {2016})}\BibitemShut {NoStop}%
\bibitem [{\citenamefont {Mente{\c s}}\ \emph {et~al.}(2014)\citenamefont
  {Mente{\c s}}, \citenamefont {Zamborlini}, \citenamefont {Sala},\ and\
  \citenamefont {Locatelli}}]{mentesCathodeLensSpectromicroscopy2014}%
  \BibitemOpen
  \bibfield  {author} {\bibinfo {author} {\bibfnamefont {T.~O.}\ \bibnamefont
  {Mente{\c s}}}, \bibinfo {author} {\bibfnamefont {G.}~\bibnamefont
  {Zamborlini}}, \bibinfo {author} {\bibfnamefont {A.}~\bibnamefont {Sala}}, \
  and\ \bibinfo {author} {\bibfnamefont {A.}~\bibnamefont {Locatelli}},\
  }\bibfield  {title} {\enquote {\bibinfo {title} {Cathode lens
  spectromicroscopy: Methodology and applications},}\ }\href {\doibase
  10.3762/bjnano.5.198} {\bibfield  {journal} {\bibinfo  {journal} {Beilstein
  J. Nanotechnol.}\ }\textbf {\bibinfo {volume} {5}},\ \bibinfo {pages} {1873}
  (\bibinfo {year} {2014})}\BibitemShut {NoStop}%
\end{thebibliography}%


\newcommand{\noopsort}[1]{}
\begin{thebibliography}{11}%
\makeatletter
\providecommand \@ifxundefined [1]{%
 \@ifx{#1\undefined}
}%
\providecommand \@ifnum [1]{%
 \ifnum #1\expandafter \@firstoftwo
 \else \expandafter \@secondoftwo
 \fi
}%
\providecommand \@ifx [1]{%
 \ifx #1\expandafter \@firstoftwo
 \else \expandafter \@secondoftwo
 \fi
}%
\providecommand \natexlab [1]{#1}%
\providecommand \enquote  [1]{``#1''}%
\providecommand \bibnamefont  [1]{#1}%
\providecommand \bibfnamefont [1]{#1}%
\providecommand \citenamefont [1]{#1}%
\providecommand \href@noop [0]{\@secondoftwo}%
\providecommand \href [0]{\begingroup \@sanitize@url \@href}%
\providecommand \@href[1]{\@@startlink{#1}\@@href}%
\providecommand \@@href[1]{\endgroup#1\@@endlink}%
\providecommand \@sanitize@url [0]{\catcode `\\12\catcode `\$12\catcode
  `\&12\catcode `\#12\catcode `\^12\catcode `\_12\catcode `\%12\relax}%
\providecommand \@@startlink[1]{}%
\providecommand \@@endlink[0]{}%
\providecommand \url  [0]{\begingroup\@sanitize@url \@url }%
\providecommand \@url [1]{\endgroup\@href {#1}{\urlprefix }}%
\providecommand \urlprefix  [0]{URL }%
\providecommand \Eprint [0]{\href }%
\providecommand \doibase [0]{http://dx.doi.org/}%
\providecommand \selectlanguage [0]{\@gobble}%
\providecommand \bibinfo  [0]{\@secondoftwo}%
\providecommand \bibfield  [0]{\@secondoftwo}%
\providecommand \translation [1]{[#1]}%
\providecommand \BibitemOpen [0]{}%
\providecommand \bibitemStop [0]{}%
\providecommand \bibitemNoStop [0]{.\EOS\space}%
\providecommand \EOS [0]{\spacefactor3000\relax}%
\providecommand \BibitemShut  [1]{\csname bibitem#1\endcsname}%
\let\auto@bib@innerbib\@empty
%</preamble>
\bibitem [{\citenamefont {Bandiera}\ \emph {et~al.}(2011)\citenamefont
  {Bandiera}, \citenamefont {Sousa}, \citenamefont {Rodmacq},\ and\
  \citenamefont {Dieny}}]{bandieraAsymmetricInterfacialPerpendicular2011}%
  \BibitemOpen
  \bibfield  {author} {\bibinfo {author} {\bibfnamefont {S.}~\bibnamefont
  {Bandiera}}, \bibinfo {author} {\bibfnamefont {R.~C.}\ \bibnamefont {Sousa}},
  \bibinfo {author} {\bibfnamefont {B.}~\bibnamefont {Rodmacq}}, \ and\
  \bibinfo {author} {\bibfnamefont {B.}~\bibnamefont {Dieny}},\ }\bibfield
  {title} {\enquote {\bibinfo {title} {Asymmetric {{Interfacial Perpendicular
  Magnetic Anisotropy}} in {{Pt}}/{{Co}}/{{Pt Trilayers}}},}\ }\href {\doibase
  10.1109/LMAG.2011.2174032} {\bibfield  {journal} {\bibinfo  {journal} {IEEE
  Magn. Lett.}\ }\textbf {\bibinfo {volume} {2}},\ \bibinfo {pages} {3000504}
  (\bibinfo {year} {2011})}\BibitemShut {NoStop}%
\bibitem [{\citenamefont {Belmeguenai}\ \emph {et~al.}(2015)\citenamefont
  {Belmeguenai}, \citenamefont {Adam}, \citenamefont {Roussign{\'e}},
  \citenamefont {Eimer}, \citenamefont {Devolder}, \citenamefont {Kim},
  \citenamefont {Cherif}, \citenamefont {Stashkevich},\ and\ \citenamefont
  {Thiaville}}]{belmeguenaiInterfacialDzyaloshinskiiMoriyaInteraction2015}%
  \BibitemOpen
  \bibfield  {author} {\bibinfo {author} {\bibfnamefont {M.}~\bibnamefont
  {Belmeguenai}}, \bibinfo {author} {\bibfnamefont {J.-P.}\ \bibnamefont
  {Adam}}, \bibinfo {author} {\bibfnamefont {Y.}~\bibnamefont {Roussign{\'e}}},
  \bibinfo {author} {\bibfnamefont {S.}~\bibnamefont {Eimer}}, \bibinfo
  {author} {\bibfnamefont {T.}~\bibnamefont {Devolder}}, \bibinfo {author}
  {\bibfnamefont {J.-V.}\ \bibnamefont {Kim}}, \bibinfo {author} {\bibfnamefont
  {S.~M.}\ \bibnamefont {Cherif}}, \bibinfo {author} {\bibfnamefont
  {A.}~\bibnamefont {Stashkevich}}, \ and\ \bibinfo {author} {\bibfnamefont
  {A.}~\bibnamefont {Thiaville}},\ }\bibfield  {title} {\enquote {\bibinfo
  {title} {Interfacial {{Dzyaloshinskii}}-{{Moriya}} interaction in
  perpendicularly magnetized {Pt}/{Co}/{AlO}$_x$ ultrathin films measured by
  {{Brillouin}} light spectroscopy},}\ }\href {\doibase
  10.1103/PhysRevB.91.180405} {\bibfield  {journal} {\bibinfo  {journal} {Phys.
  Rev. B}\ }\textbf {\bibinfo {volume} {91}},\ \bibinfo {pages} {180405}
  (\bibinfo {year} {2015})}\BibitemShut {NoStop}%
\bibitem [{\citenamefont {Boulle}\ \emph {et~al.}(2016)\citenamefont {Boulle},
  \citenamefont {Vogel}, \citenamefont {Yang}, \citenamefont {Pizzini},
  \citenamefont {{\noopsort{souza chaves}}{de Souza Chaves}}, \citenamefont
  {Locatelli}, \citenamefont {Mente{\c s}}, \citenamefont {Sala}, \citenamefont
  {{Buda-Prejbeanu}}, \citenamefont {Klein}, \citenamefont {Belmeguenai},
  \citenamefont {Roussign{\'e}}, \citenamefont {Stashkevich}, \citenamefont
  {Ch{\'e}rif}, \citenamefont {Aballe}, \citenamefont {Foerster}, \citenamefont
  {Chshiev}, \citenamefont {Auffret}, \citenamefont {Miron},\ and\
  \citenamefont {Gaudin}}]{boulleRoomtemperatureChiralMagnetic2016}%
  \BibitemOpen
  \bibfield  {author} {\bibinfo {author} {\bibfnamefont {O.}~\bibnamefont
  {Boulle}}, \bibinfo {author} {\bibfnamefont {J.}~\bibnamefont {Vogel}},
  \bibinfo {author} {\bibfnamefont {H.}~\bibnamefont {Yang}}, \bibinfo {author}
  {\bibfnamefont {S.}~\bibnamefont {Pizzini}}, \bibinfo {author} {\bibfnamefont
  {D.}~\bibnamefont {{\noopsort{souza chaves}}{de Souza Chaves}}}, \bibinfo
  {author} {\bibfnamefont {A.}~\bibnamefont {Locatelli}}, \bibinfo {author}
  {\bibfnamefont {T.~O.}\ \bibnamefont {Mente{\c s}}}, \bibinfo {author}
  {\bibfnamefont {A.}~\bibnamefont {Sala}}, \bibinfo {author} {\bibfnamefont
  {L.~D.}\ \bibnamefont {{Buda-Prejbeanu}}}, \bibinfo {author} {\bibfnamefont
  {O.}~\bibnamefont {Klein}}, \bibinfo {author} {\bibfnamefont
  {M.}~\bibnamefont {Belmeguenai}}, \bibinfo {author} {\bibfnamefont
  {Y.}~\bibnamefont {Roussign{\'e}}}, \bibinfo {author} {\bibfnamefont
  {A.}~\bibnamefont {Stashkevich}}, \bibinfo {author} {\bibfnamefont {S.~M.}\
  \bibnamefont {Ch{\'e}rif}}, \bibinfo {author} {\bibfnamefont
  {L.}~\bibnamefont {Aballe}}, \bibinfo {author} {\bibfnamefont
  {M.}~\bibnamefont {Foerster}}, \bibinfo {author} {\bibfnamefont
  {M.}~\bibnamefont {Chshiev}}, \bibinfo {author} {\bibfnamefont
  {S.}~\bibnamefont {Auffret}}, \bibinfo {author} {\bibfnamefont {I.~M.}\
  \bibnamefont {Miron}}, \ and\ \bibinfo {author} {\bibfnamefont
  {G.}~\bibnamefont {Gaudin}},\ }\bibfield  {title} {\enquote {\bibinfo {title}
  {Room-temperature chiral magnetic skyrmions in ultrathin magnetic
  nanostructures},}\ }\href {\doibase 10.1038/nnano.2015.315} {\bibfield
  {journal} {\bibinfo  {journal} {Nat. Nanotech.}\ }\textbf {\bibinfo {volume}
  {11}},\ \bibinfo {pages} {449} (\bibinfo {year} {2016})}\BibitemShut
  {NoStop}%
\bibitem [{\citenamefont {Vansteenkiste}\ \emph {et~al.}(2014)\citenamefont
  {Vansteenkiste}, \citenamefont {Leliaert}, \citenamefont {Dvornik},
  \citenamefont {Helsen}, \citenamefont {{Garcia-Sanchez}},\ and\ \citenamefont
  {Waeyenberge}}]{vansteenkisteDesignVerificationMuMax32014}%
  \BibitemOpen
  \bibfield  {author} {\bibinfo {author} {\bibfnamefont {A.}~\bibnamefont
  {Vansteenkiste}}, \bibinfo {author} {\bibfnamefont {J.}~\bibnamefont
  {Leliaert}}, \bibinfo {author} {\bibfnamefont {M.}~\bibnamefont {Dvornik}},
  \bibinfo {author} {\bibfnamefont {M.}~\bibnamefont {Helsen}}, \bibinfo
  {author} {\bibfnamefont {F.}~\bibnamefont {{Garcia-Sanchez}}}, \ and\
  \bibinfo {author} {\bibfnamefont {B.~V.}\ \bibnamefont {Waeyenberge}},\
  }\bibfield  {title} {\enquote {\bibinfo {title} {The design and verification
  of {{MuMax3}}},}\ }\href {\doibase 10.1063/1.4899186} {\bibfield  {journal}
  {\bibinfo  {journal} {AIP Adv.}\ }\textbf {\bibinfo {volume} {4}},\ \bibinfo
  {pages} {107133} (\bibinfo {year} {2014})}\BibitemShut {NoStop}%
\bibitem [{\citenamefont {Metaxas}\ \emph {et~al.}(2007)\citenamefont
  {Metaxas}, \citenamefont {Jamet}, \citenamefont {Mougin}, \citenamefont
  {Cormier}, \citenamefont {Ferr{\'e}}, \citenamefont {Baltz}, \citenamefont
  {Rodmacq}, \citenamefont {Dieny},\ and\ \citenamefont
  {Stamps}}]{metaxasCreepFlowRegimes2007}%
  \BibitemOpen
  \bibfield  {author} {\bibinfo {author} {\bibfnamefont {P.~J.}\ \bibnamefont
  {Metaxas}}, \bibinfo {author} {\bibfnamefont {J.~P.}\ \bibnamefont {Jamet}},
  \bibinfo {author} {\bibfnamefont {A.}~\bibnamefont {Mougin}}, \bibinfo
  {author} {\bibfnamefont {M.}~\bibnamefont {Cormier}}, \bibinfo {author}
  {\bibfnamefont {J.}~\bibnamefont {Ferr{\'e}}}, \bibinfo {author}
  {\bibfnamefont {V.}~\bibnamefont {Baltz}}, \bibinfo {author} {\bibfnamefont
  {B.}~\bibnamefont {Rodmacq}}, \bibinfo {author} {\bibfnamefont
  {B.}~\bibnamefont {Dieny}}, \ and\ \bibinfo {author} {\bibfnamefont {R.~L.}\
  \bibnamefont {Stamps}},\ }\bibfield  {title} {\enquote {\bibinfo {title}
  {{Creep} and {Flow} {Regimes} of {Magnetic} {Domain}-{Wall} {Motion} in
  {Ultrathin} {Pt}/{Co}/{Pt} {Films} with {Perpendicular} {Anisotropy}},}\
  }\href {\doibase 10.1103/PhysRevLett.99.217208} {\bibfield  {journal}
  {\bibinfo  {journal} {Phys. Rev. Lett.}\ }\textbf {\bibinfo {volume} {99}},\
  \bibinfo {pages} {217208} (\bibinfo {year} {2007})}\BibitemShut {NoStop}%
\bibitem [{\citenamefont {Shepley}\ \emph {et~al.}(2018)\citenamefont
  {Shepley}, \citenamefont {Tunnicliffe}, \citenamefont {Shahbazi},
  \citenamefont {Burnell},\ and\ \citenamefont
  {Moore}}]{shepleyMagneticPropertiesDomainwall2018}%
  \BibitemOpen
  \bibfield  {author} {\bibinfo {author} {\bibfnamefont {P.~M.}\ \bibnamefont
  {Shepley}}, \bibinfo {author} {\bibfnamefont {H.}~\bibnamefont
  {Tunnicliffe}}, \bibinfo {author} {\bibfnamefont {K.}~\bibnamefont
  {Shahbazi}}, \bibinfo {author} {\bibfnamefont {G.}~\bibnamefont {Burnell}}, \
  and\ \bibinfo {author} {\bibfnamefont {T.~A.}\ \bibnamefont {Moore}},\
  }\bibfield  {title} {\enquote {\bibinfo {title} {Magnetic properties,
  domain-wall creep motion, and the {{Dzyaloshinskii}}-{{Moriya}} interaction
  in {{Pt}}/{{Co}}/{{Ir}} thin films},}\ }\href {\doibase
  10.1103/PhysRevB.97.134417} {\bibfield  {journal} {\bibinfo  {journal} {Phys.
  Rev. B}\ }\textbf {\bibinfo {volume} {97}},\ \bibinfo {pages} {134417}
  (\bibinfo {year} {2018})}\BibitemShut {NoStop}%
\bibitem [{\citenamefont {Juge}\ \emph {et~al.}(2018)\citenamefont {Juge},
  \citenamefont {Je}, \citenamefont {{\noopsort{souza chaves}}{de Souza
  Chaves}}, \citenamefont {Pizzini}, \citenamefont {{Buda-Prejbeanu}},
  \citenamefont {Aballe}, \citenamefont {Foerster}, \citenamefont {Locatelli},
  \citenamefont {Mente{\c s}}, \citenamefont {Sala}, \citenamefont
  {Maccherozzi}, \citenamefont {Dhesi}, \citenamefont {Auffret}, \citenamefont
  {Gautier}, \citenamefont {Gaudin}, \citenamefont {Vogel},\ and\ \citenamefont
  {Boulle}}]{jugeMagneticSkyrmionsConfined2018}%
  \BibitemOpen
  \bibfield  {author} {\bibinfo {author} {\bibfnamefont {R.}~\bibnamefont
  {Juge}}, \bibinfo {author} {\bibfnamefont {S.-G.}\ \bibnamefont {Je}},
  \bibinfo {author} {\bibfnamefont {D.}~\bibnamefont {{\noopsort{souza
  chaves}}{de Souza Chaves}}}, \bibinfo {author} {\bibfnamefont
  {S.}~\bibnamefont {Pizzini}}, \bibinfo {author} {\bibfnamefont {L.~D.}\
  \bibnamefont {{Buda-Prejbeanu}}}, \bibinfo {author} {\bibfnamefont
  {L.}~\bibnamefont {Aballe}}, \bibinfo {author} {\bibfnamefont
  {M.}~\bibnamefont {Foerster}}, \bibinfo {author} {\bibfnamefont
  {A.}~\bibnamefont {Locatelli}}, \bibinfo {author} {\bibfnamefont {T.~O.}\
  \bibnamefont {Mente{\c s}}}, \bibinfo {author} {\bibfnamefont
  {A.}~\bibnamefont {Sala}}, \bibinfo {author} {\bibfnamefont {F.}~\bibnamefont
  {Maccherozzi}}, \bibinfo {author} {\bibfnamefont {S.~S.}\ \bibnamefont
  {Dhesi}}, \bibinfo {author} {\bibfnamefont {S.}~\bibnamefont {Auffret}},
  \bibinfo {author} {\bibfnamefont {E.}~\bibnamefont {Gautier}}, \bibinfo
  {author} {\bibfnamefont {G.}~\bibnamefont {Gaudin}}, \bibinfo {author}
  {\bibfnamefont {J.}~\bibnamefont {Vogel}}, \ and\ \bibinfo {author}
  {\bibfnamefont {O.}~\bibnamefont {Boulle}},\ }\bibfield  {title} {\enquote
  {\bibinfo {title} {Magnetic skyrmions in confined geometries: {{Effect}} of
  the magnetic field and the disorder},}\ }\href {\doibase
  10.1016/j.jmmm.2017.10.030} {\bibfield  {journal} {\bibinfo  {journal} {J.
  Magn. Magn. Mater.}\ }\textbf {\bibinfo {volume} {455}},\ \bibinfo {pages}
  {3} (\bibinfo {year} {2018})}\BibitemShut {NoStop}%
\bibitem [{\citenamefont {Gross}\ \emph {et~al.}(2018)\citenamefont {Gross},
  \citenamefont {Akhtar}, \citenamefont {Hrabec}, \citenamefont {Sampaio},
  \citenamefont {Mart{\'i}nez}, \citenamefont {Chouaieb}, \citenamefont
  {Shields}, \citenamefont {Maletinsky}, \citenamefont {Thiaville},
  \citenamefont {Rohart},\ and\ \citenamefont
  {Jacques}}]{grossSkyrmionMorphologyUltrathin2018}%
  \BibitemOpen
  \bibfield  {author} {\bibinfo {author} {\bibfnamefont {I.}~\bibnamefont
  {Gross}}, \bibinfo {author} {\bibfnamefont {W.}~\bibnamefont {Akhtar}},
  \bibinfo {author} {\bibfnamefont {A.}~\bibnamefont {Hrabec}}, \bibinfo
  {author} {\bibfnamefont {J.}~\bibnamefont {Sampaio}}, \bibinfo {author}
  {\bibfnamefont {L.~J.}\ \bibnamefont {Mart{\'i}nez}}, \bibinfo {author}
  {\bibfnamefont {S.}~\bibnamefont {Chouaieb}}, \bibinfo {author}
  {\bibfnamefont {B.~J.}\ \bibnamefont {Shields}}, \bibinfo {author}
  {\bibfnamefont {P.}~\bibnamefont {Maletinsky}}, \bibinfo {author}
  {\bibfnamefont {A.}~\bibnamefont {Thiaville}}, \bibinfo {author}
  {\bibfnamefont {S.}~\bibnamefont {Rohart}}, \ and\ \bibinfo {author}
  {\bibfnamefont {V.}~\bibnamefont {Jacques}},\ }\bibfield  {title} {\enquote
  {\bibinfo {title} {Skyrmion morphology in ultrathin magnetic films},}\ }\href
  {\doibase 10.1103/PhysRevMaterials.2.024406} {\bibfield  {journal} {\bibinfo
  {journal} {Phys. Rev. Mater.}\ }\textbf {\bibinfo {volume} {2}},\ \bibinfo
  {pages} {024406} (\bibinfo {year} {2018})}\BibitemShut {NoStop}%
\bibitem [{\citenamefont {Thiaville}\ \emph {et~al.}(2012)\citenamefont
  {Thiaville}, \citenamefont {Rohart}, \citenamefont {Ju{\'e}}, \citenamefont
  {Cros},\ and\ \citenamefont
  {Fert}}]{thiavilleDynamicsDzyaloshinskiiDomain2012}%
  \BibitemOpen
  \bibfield  {author} {\bibinfo {author} {\bibfnamefont {A.}~\bibnamefont
  {Thiaville}}, \bibinfo {author} {\bibfnamefont {S.}~\bibnamefont {Rohart}},
  \bibinfo {author} {\bibfnamefont {{\'E}.}~\bibnamefont {Ju{\'e}}}, \bibinfo
  {author} {\bibfnamefont {V.}~\bibnamefont {Cros}}, \ and\ \bibinfo {author}
  {\bibfnamefont {A.}~\bibnamefont {Fert}},\ }\bibfield  {title} {\enquote
  {\bibinfo {title} {Dynamics of {{Dzyaloshinskii}} domain walls in ultrathin
  magnetic films},}\ }\href {\doibase 10.1209/0295-5075/100/57002} {\bibfield
  {journal} {\bibinfo  {journal} {Europhys. Lett.}\ }\textbf {\bibinfo {volume}
  {100}},\ \bibinfo {pages} {57002} (\bibinfo {year} {2012})}\BibitemShut
  {NoStop}%
\bibitem [{\citenamefont {Garello}\ \emph {et~al.}(2013)\citenamefont
  {Garello}, \citenamefont {Miron}, \citenamefont {Avci}, \citenamefont
  {Freimuth}, \citenamefont {Mokrousov}, \citenamefont {Bl{\"u}gel},
  \citenamefont {Auffret}, \citenamefont {Boulle}, \citenamefont {Gaudin},\
  and\ \citenamefont {Gambardella}}]{garelloSymmetryMagnitudeSpinorbit2013}%
  \BibitemOpen
  \bibfield  {author} {\bibinfo {author} {\bibfnamefont {K.}~\bibnamefont
  {Garello}}, \bibinfo {author} {\bibfnamefont {I.~M.}\ \bibnamefont {Miron}},
  \bibinfo {author} {\bibfnamefont {C.~O.}\ \bibnamefont {Avci}}, \bibinfo
  {author} {\bibfnamefont {F.}~\bibnamefont {Freimuth}}, \bibinfo {author}
  {\bibfnamefont {Y.}~\bibnamefont {Mokrousov}}, \bibinfo {author}
  {\bibfnamefont {S.}~\bibnamefont {Bl{\"u}gel}}, \bibinfo {author}
  {\bibfnamefont {S.}~\bibnamefont {Auffret}}, \bibinfo {author} {\bibfnamefont
  {O.}~\bibnamefont {Boulle}}, \bibinfo {author} {\bibfnamefont
  {G.}~\bibnamefont {Gaudin}}, \ and\ \bibinfo {author} {\bibfnamefont
  {P.}~\bibnamefont {Gambardella}},\ }\bibfield  {title} {\enquote {\bibinfo
  {title} {Symmetry and magnitude of spin-orbit torques in ferromagnetic
  heterostructures},}\ }\href {\doibase 10.1038/nnano.2013.145} {\bibfield
  {journal} {\bibinfo  {journal} {Nat. Nanotech.}\ }\textbf {\bibinfo {volume}
  {8}},\ \bibinfo {pages} {587} (\bibinfo {year} {2013})}\BibitemShut {NoStop}%
\bibitem [{\citenamefont {Avci}\ \emph {et~al.}(2014)\citenamefont {Avci},
  \citenamefont {Garello}, \citenamefont {Gabureac}, \citenamefont {Ghosh},
  \citenamefont {Fuhrer}, \citenamefont {Alvarado},\ and\ \citenamefont
  {Gambardella}}]{avciInterplaySpinorbitTorque2014}%
  \BibitemOpen
  \bibfield  {author} {\bibinfo {author} {\bibfnamefont {C.~O.}\ \bibnamefont
  {Avci}}, \bibinfo {author} {\bibfnamefont {K.}~\bibnamefont {Garello}},
  \bibinfo {author} {\bibfnamefont {M.}~\bibnamefont {Gabureac}}, \bibinfo
  {author} {\bibfnamefont {A.}~\bibnamefont {Ghosh}}, \bibinfo {author}
  {\bibfnamefont {A.}~\bibnamefont {Fuhrer}}, \bibinfo {author} {\bibfnamefont
  {S.~F.}\ \bibnamefont {Alvarado}}, \ and\ \bibinfo {author} {\bibfnamefont
  {P.}~\bibnamefont {Gambardella}},\ }\bibfield  {title} {\enquote {\bibinfo
  {title} {Interplay of spin-orbit torque and thermoelectric effects in
  ferromagnet/normal-metal bilayers},}\ }\href {\doibase
  10.1103/PhysRevB.90.224427} {\bibfield  {journal} {\bibinfo  {journal} {Phys.
  Rev. B}\ }\textbf {\bibinfo {volume} {90}},\ \bibinfo {pages} {224427}
  (\bibinfo {year} {2014})}\BibitemShut {NoStop}%
\end{thebibliography}%

\end{document}

\title{\Large Supplementary Information}

\maketitle

\section{Sample characterisation}

This section presents the measurement of the different parameters used in the analytical calculation of the skyrmion velocity and Skyrmion Hall Angle (SkHA) and in the micromagnetic simulations. All the measurements, except for the saturation magnetisation measurements, were performed on a Ta(3)/Pt(3)/Co(0.6-1.1)/MgO(0.9)/Ta(2) (thicknesses in nm) non-patterned reference sample (hereafter referred to as Pt/Co/MgO) which was deposited together with the wafer in which the skyrmion dynamics was studied. 

\subsection{Hysteresis loops}
\label{section_hyst_loops}

\begin{figure}[h]
\includegraphics[width=\textwidth]{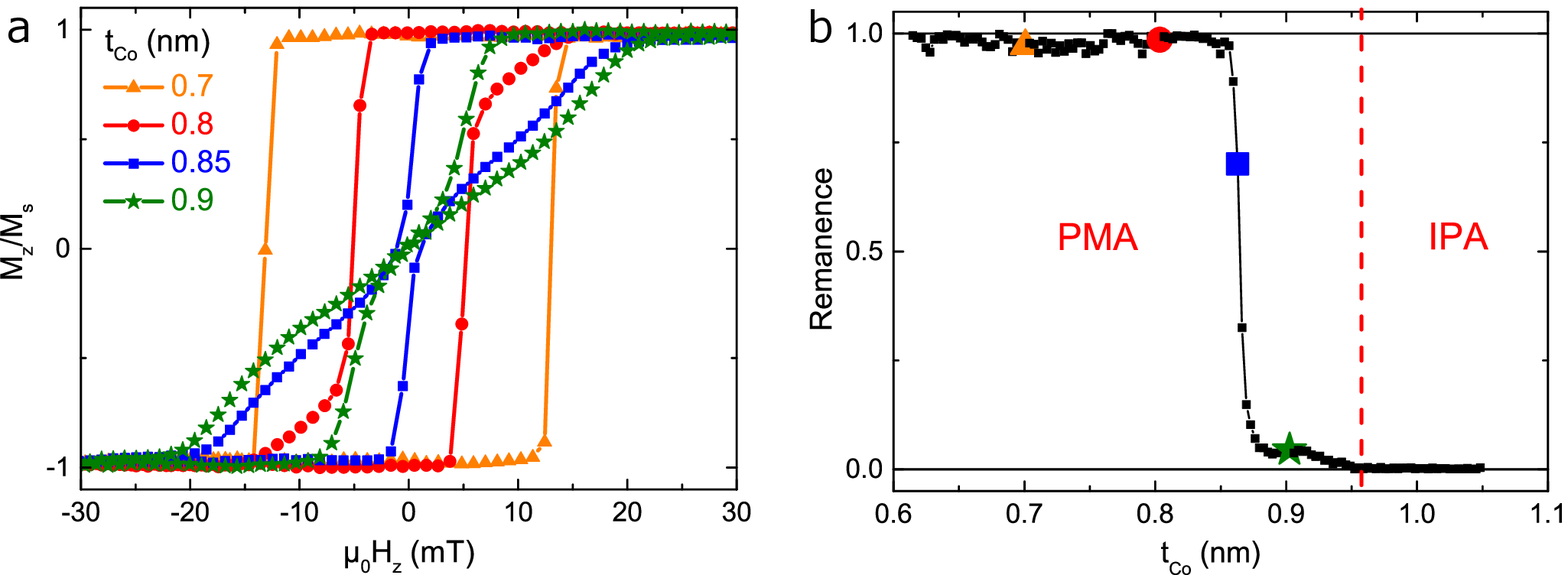}
\centering
\caption{Polar Magneto-Optical Kerr Effect (pMOKE) measurements. The sample consists of a non-patterned Pt/Co(0.6-1.1nm)/MgO and the magnetic field is applied perpendicularly to the sample plane. \textbf{a.} Hysteresis loops measured at different locations on the wedge corresponding to different Co thicknesses $t_{Co}$. \textbf{b.} pMOKE amplitude measured at zero field (remanence) as a function of the Co thickness. The hysteresis loops in \textbf{a.} are indicated by the corresponding symbols in \textbf{b.}}
\label{remanence_hyst}
\end{figure}

The Pt/Co(0.6-1.1)/MgO film was deposited by DC magnetron sputtering at room temperature on a 100-mm high-resistivity Si wafer and annealed at 250°C under vacuum for 10 min. The Co layer was deposited as a wedge. Fig. \ref{remanence_hyst}.a. displays the hysteresis loops measured at different locations on the wedge using polar Magneto-Optical Kerr Effect (pMOKE). The magnetic field is applied perpendicularly to the film plane. 

\FloatBarrier

At $t_{Co}=0.7$ nm, the large Perpendicular Magnetic Anisotropy (PMA) leads to a square hysteresis loop, indicating that the sample is fully magnetised at remanence. Upon increasing the Co thickness, the decrease in the PMA combined with the larger demagnetising field lead to a bended loop characteristic of a reversal involving multi-domain states. A decrease in the magnetisation at remanence ($H_z=0$) is observed around $t_{Co}=0.85$ nm, indicating the presence of stripe domains at zero field. Fig. \ref{remanence_hyst}.b. displays the normalised remanence, obtained from the zero-field pMOKE amplitude, as a function of the Co thickness. It shows the decrease in the anisotropy for increasing $t_{Co}$ and the transition between perpendicular and in-plane magnetic anisotropy (PMA and IPA) for $t_{Co}\approx{}0.95$ nm. The current-driven skyrmion motion experiments are performed for $t_{Co}=0.9$ nm.

\subsection{Saturation magnetisation, effective anisotropy and Dzyaloshinskii-Moriya Interaction measurements}

\begin{figure}[h]
\includegraphics[width=0.5\textwidth]{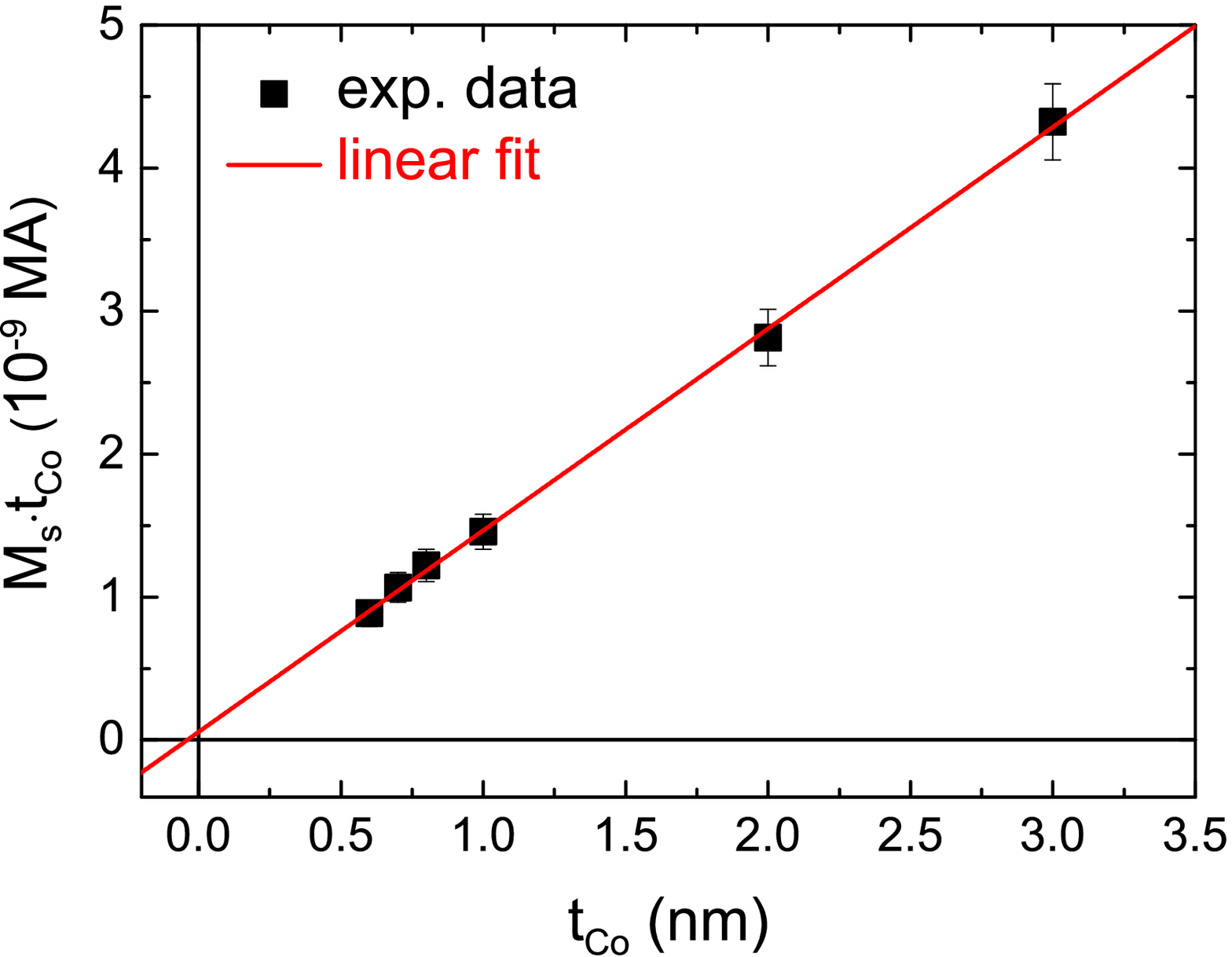}
\centering
\caption{Saturation magnetisation measurements. Magnetic moment per unit area measured on different Pt/Co($t_{Co}$)/MgO samples using SQUID magnetometry at room temperature.}
\label{Ms}
\end{figure}

To measure the saturation magnetisation $M_s$, different (non-wedged) Pt/Co($t_{Co}$)/MgO samples were deposited with $t_{Co}$ ranging from 0.6 nm to 3 nm. The total magnetic moment was measured from the IP and OOP hysteresis loops at room temperature on $\approx{}5\times5$ mm$^2$ samples using a Superconducting QUantum Interference Device (SQUID). The area of each sample was precisely measured from scanning electron microscopy images. Fig. \ref{Ms} displays the measured magnetic moment per unit area $M_s\cdot{}t_{Co}$ as a function of the Co thickness $t_{Co}$. The saturation magnetisation is extracted from a linear fit and a value $M_s=1.42\pm0.05$ MA m$^{-1}$ is found. The intercept of the fit with the $x$-axis is found very close to $t_{Co}=0$, indicating the absence of a magnetically dead layer, consistently with a previous experimental study \cite{bandieraAsymmetricInterfacialPerpendicular2011}. The effective anisotropy was measured using SQUID on a Pt/Co(0.9nm)/MgO film, identical to the sample in which we studied the skyrmion dynamics (same position on the wedge). From the hard-axis hysteresis loop, we measure the anisotropy field (in-plane saturation field) from which we deduce the effective anisotropy constant $K_{eff}=(7\pm2)\times{}10^{4}$ J m$^{-3}$. The uniaxial anisotropy constant $K_u$ can be estimated using the relation $K_u=K_{eff}+\mu{}_0M_s^2/2$ ; we find $K_u=(1.34\pm0.12)\times{}10^{6}$ J m$^{-3}$.

\FloatBarrier

\begin{figure}[h]
\includegraphics[width=\textwidth]{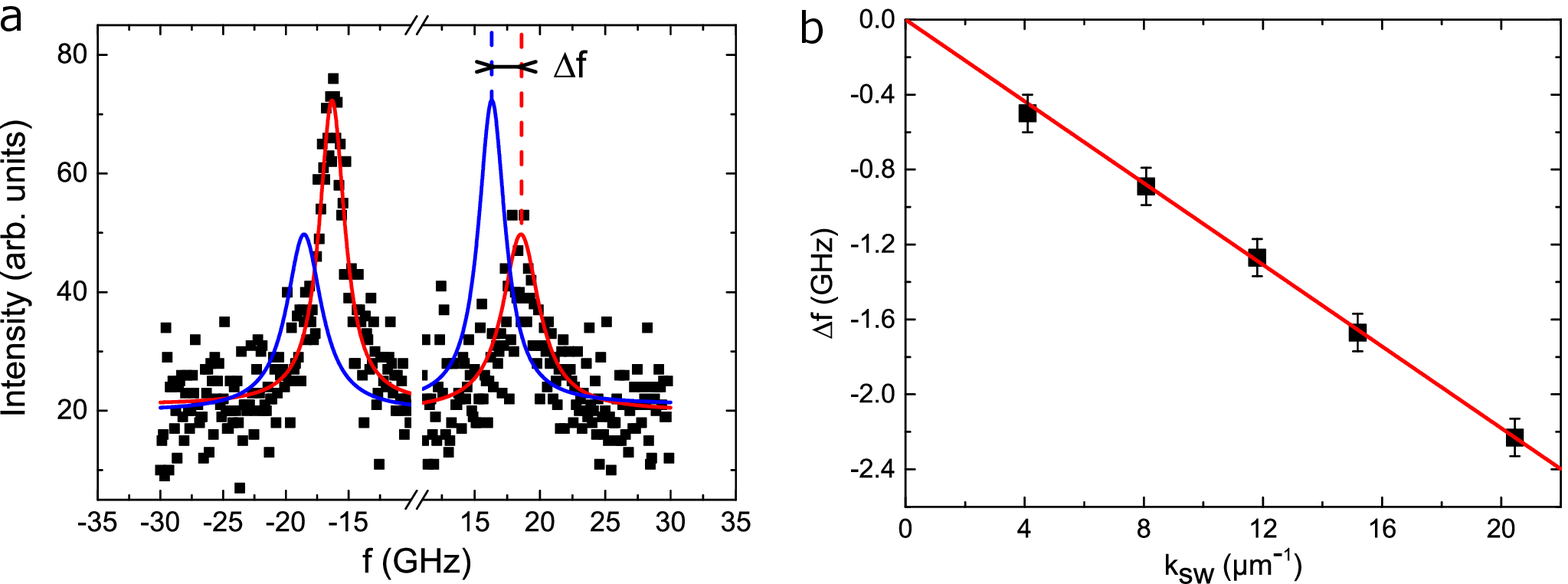}
\centering
\caption{Brillouin Light Scattering (BLS) experiments. The sample is a non-patterned Pt/Co(0.91nm)/MgO film (thicknesses in nm). \textbf{a.} BLS spectrum showing the intensity as a function of the frequency $f$  measured for an in-plane applied magnetic field $\mu_0H=0.6$ T and for $k_{SW}$ = 20.45 \textmu{}m$^{-1}$. The black squares refer to the experimental data and the red line to a Lorentzian fit. The blue line is the Lorentzian fit inverted with respect to $f=0$. $\Delta{}f$ is the shift between the Stokes and anti-Stokes peak frequencies. \textbf{b.} Frequency shift $\Delta{}f$ as a function of the wave vector $k_{SW}$ (black squares). The red line is a linear fit.}
\label{BLS_DMI}
\end{figure}

In order to measure the Dzyaloshinskii-Moriya Interaction (DMI) parameter $D$, we performed Brillouin Light Scattering (BLS) experiments on a Pt/Co(0.91nm)/MgO sample. The principle of the technique is the following: the magnetisation is saturated in the film plane by an external magnetic field and spin waves (SW) propagating along the direction perpendicular to this field are probed by a laser with a well-defined wave vector $k_{SW}$. The DMI introduces a preferred handedness an therefore leads to an energy difference between two SW propagating with opposite wave vectors. This energy difference corresponds to a shift in frequency : $\Delta{}f\left(k_{SW}\right)=f_S\left(k_{SW}\right)-f_{AS}\left(k_{SW}\right)$ where $f_{S}$ and $f_{AS}$ are the Stokes (a SW is created) and Anti-Stokes (a SW was absorbed) frequencies respectively. This frequency shift is directly related to the DMI by the following relation : $\Delta{}f\left(k_{SW}\right)=2\gamma{}k_{SW}D/\left(\pi{}M_s\right)$ with $\gamma$ the gyromagnetic ratio \cite{belmeguenaiInterfacialDzyaloshinskiiMoriyaInteraction2015}. Fig. \ref{BLS_DMI}.a. displays a BLS spectrum measured for an in-plane external field $\mu_0H=0.6$ T and for $k_{SW}$ = 20.45 \textmu{}m$^{-1}$. It shows the S and AS peaks with $f_{S}>0$ and $f_{AS}<0$ respectively. The frequency shift $\Delta{}f$ is obtained from a Lorentzian fit (red line) of the experimental data. These measurements are performed for several wave vectors (see Fig. \ref{BLS_DMI}.b, black squares). The DMI is then extracted from the linear fit (Fig. \ref{BLS_DMI}.b, red line). We find $D=-1.27\pm0.04$ mJ m$^{-2}$ using $M_s=1.42$ MA m$^{-1}$ and $\gamma{}/(2\pi{})=31$ GHz T$^{-1}$ extracted from ferromagnetic resonance experiments. In these BLS experiments, a negative value for $D$ indicates a left-handed chirality, consistently with our previous work \cite{boulleRoomtemperatureChiralMagnetic2016}. Note that using BLS, it is not required to know the DW width (and therefore the exchange constant and effective anisotropy) to extract the DMI parameter, unlike other methods based on asymmetric DW propagation. 

\FloatBarrier

\subsection{Estimation of the exchange constant}
\label{section_measure_exchage}

\begin{figure}[h]
\includegraphics[width=\textwidth]{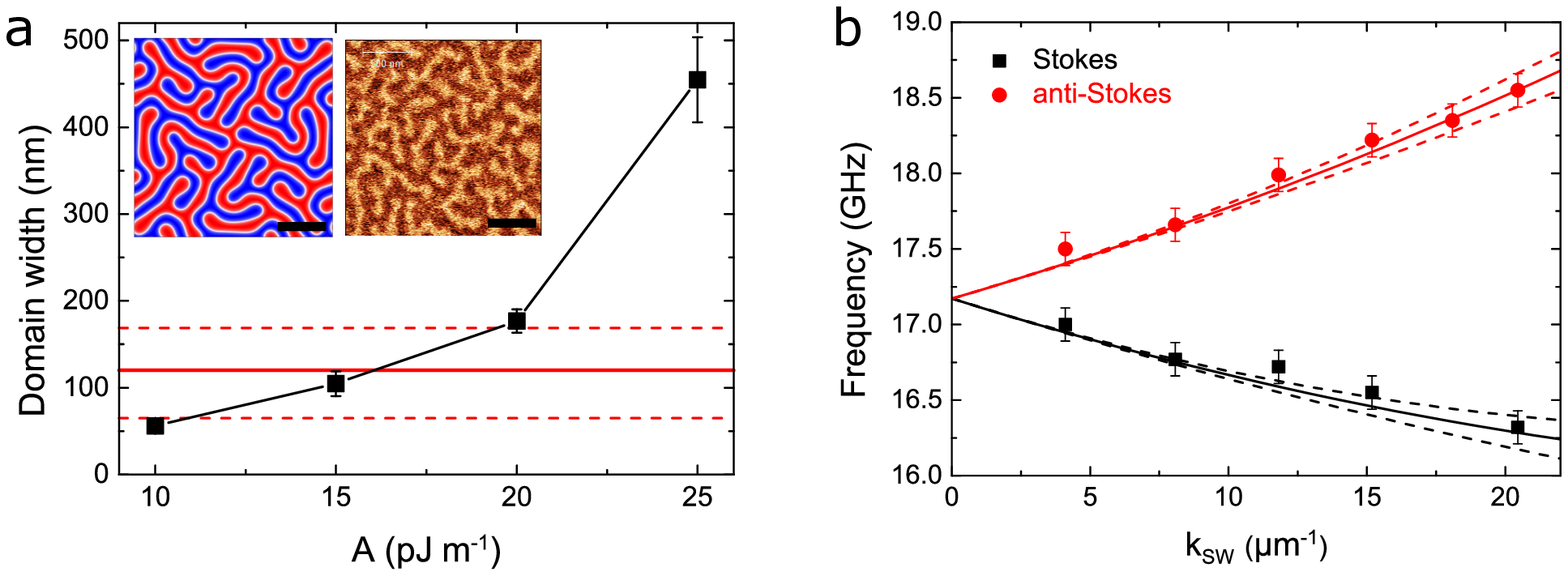}
\centering
\caption{Estimation of the exchange constant $A$. \textbf{a.} Micromagnetic simulations : average domain width as a function of $A$. The error bars denote the standard deviation and the solid black line is guide for the eye. The red solid line is the average domain sizes measured from MFM images at zero applied magnetic field. The insets display the demagnetised domain configuration (zero applied field) obtained from MFM images and from micromagnetic simulations with $A=15$ pJ m$^{-1}$ (scale bar is 500 nm). \textbf{b.} BLS measurements: measured Stokes (black squares) and Anti-Stokes (red dots) frequencies as a function of the wave vector $k_{SW}$. The lines represent the model described in ref. \cite{belmeguenaiInterfacialDzyaloshinskiiMoriyaInteraction2015} with $A=16\pm{}6$ pJ m$^{-1}$.}
\label{estimationA}
\end{figure}

In order to determine the exchange constant, we compared the domain size at remanence measured on our Pt/Co(0.91nm)/MgO stack (same sample used for BLS measurements) using Magnetic Force Microscopy (MFM) and the one extracted from micromagnetic simulations. The simulations were performed with Mumax3 \cite{vansteenkisteDesignVerificationMuMax32014} using the aforementioned micromagnetic parameters and for different values of the exchange constant $A$. Fig. \ref{estimationA}.a displays the calculated domain size as a function of $A$. Fig. \ref{estimationA}.b and c show respectively the typical domain configuration as measured using MFM under zero external field and the one obtained from the zero-field simulations with $A = 15$ pJ m$^{-1}$. In the simulations, the initial state, a 200-nm-diameter bubble, is relaxed at zero external field, resulting in this particular domain pattern. Note that the equilibrium domain size does not depend on the initial state. In addition, periodic boundary conditions are considered in both directions to mimic an infinite film. The red solid in Fig. \ref{estimationA}.a indicates the average domain size extracted from MFM images : $120\pm{}49$ nm.  \\

\FloatBarrier

We find $A=16\pm{}6$ pJ m$^{-1}$, a value in line with those found in the literature for ultrathin Co films \cite{metaxasCreepFlowRegimes2007, shepleyMagneticPropertiesDomainwall2018}. This value has been chosen for the micromagnetic simulations. We note the imprecision on the estimation of $A$ given the uncertainty on the domain width measurement. To support these findings, we note that the exchange constant can also be extracted from BLS measurements (section S.1.2), more precisely from the dependence of the Stokes and Anti-Stokes frequencies ($f_{S}$ and $f_{AS}$) on the SW wave vector $k_{SW}$. Fig. \ref{estimationA}.b displays the measured $f_{S}$ and $f_{AS}$. Using the model given in ref. \cite{belmeguenaiInterfacialDzyaloshinskiiMoriyaInteraction2015}, we plot the analytical values of the two frequencies (solid lines) corresponding to $A=16$ pJ m$^{-1}$. The dashed lines illustrate for each frequency the uncertainty on the exchange constant. A good agreement is found in the range of $A$ values extracted from domain width measurements and the best fit is found for $A\approx18$ pJ m$^{-1}$. This confirms the relevance of using the value $A=16$ pJ m$^{-1}$ in the micromagnetic simulations. Note that for the modelling of the skyrmion dynamics, the value of $A$ impacts mainly the DW width $\Delta$. Since $\Delta{}=\sqrt{A/K_{eff}}$, a small change of $A$ would have a negligible influence on the calculated skyrmion velocity and SkHA. 

\newpage

\subsection{Experimental skyrmion size}

\begin{figure}[h]
\includegraphics[width=0.5\textwidth]{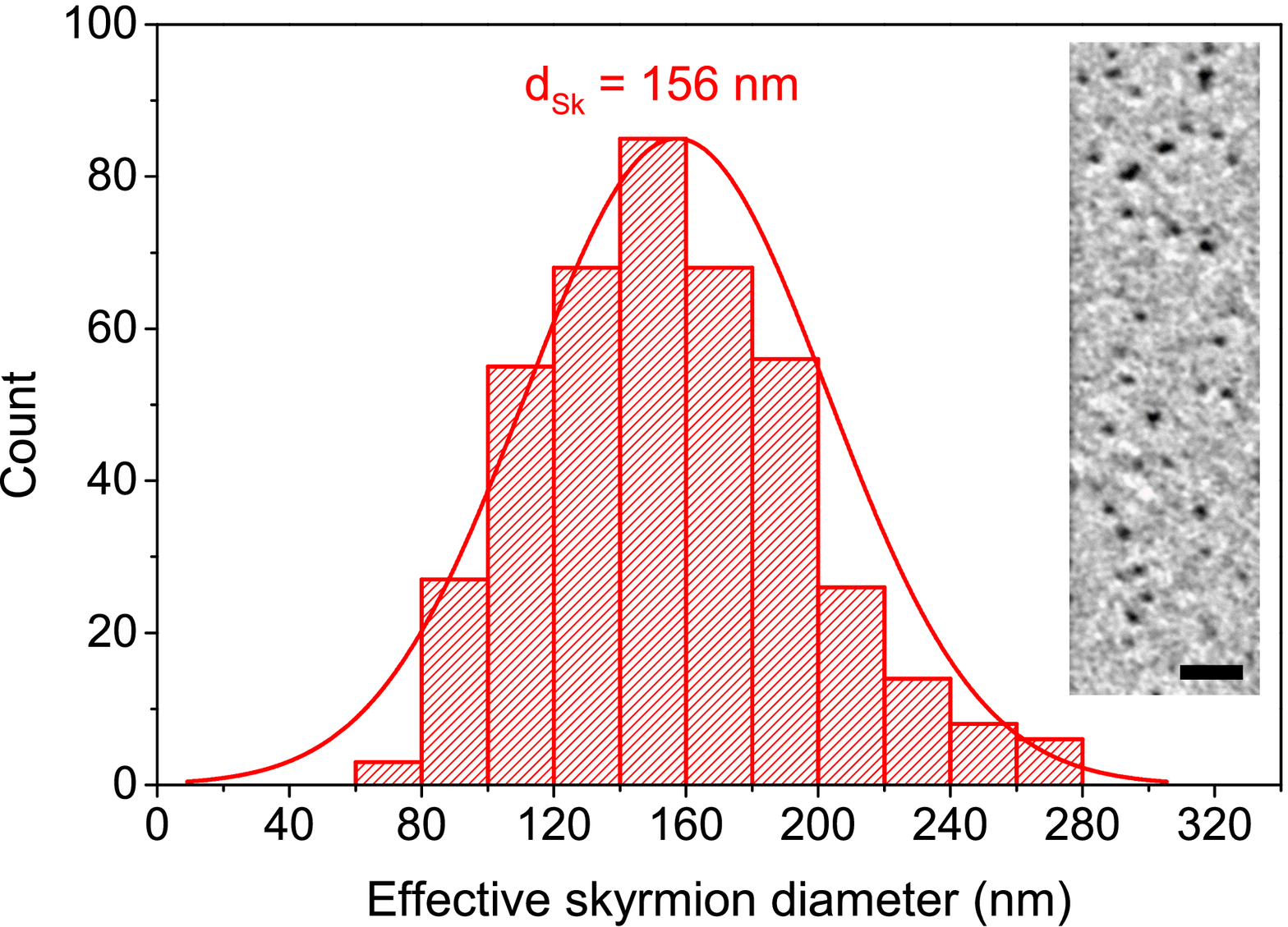}
\centering
\caption{Skyrmion size measured in Pt/Co/MgO. Distribution of the measured effective skyrmion diameter under an external field $\mu_0H_z\approx{}-5$ mT. The solid line is a fit with a Gaussian distribution with average 156 nm and standard deviation 45 nm. The inset shows a XMCD-PEEM image of a magnetic track at $\mu_0H_z\approx{}-5$ mT (scale bar is 1 \textmu{}m).}
\label{SkyrmionSize}
\end{figure}

From XMCD-PEEM images, we can extract an average diameter for the skyrmions. For deformed skyrmion, we calculate an effective diameter assuming an elliptical shape. Fig. \ref{SkyrmionSize} displays a histogram of the effective skyrmion diameter measured for a large number of skyrmions under an out-of-plane external field $\mu_0H_z\approx{}-5$ mT. From a Gaussian fit (solid line), we extract an average diameter $d_{Sk}=156\pm{}45$ nm, where the error is the standard deviation. This dispersion is a consequence of the disorder and inhomogeneities inherent to granular films that affect the skyrmion size and morphology \cite{jugeMagneticSkyrmionsConfined2018, grossSkyrmionMorphologyUltrathin2018}. 

\FloatBarrier

\subsection{Damping measurements : field-induced domain wall dynamics}

\begin{figure}[h]
\includegraphics[width=0.5\textwidth]{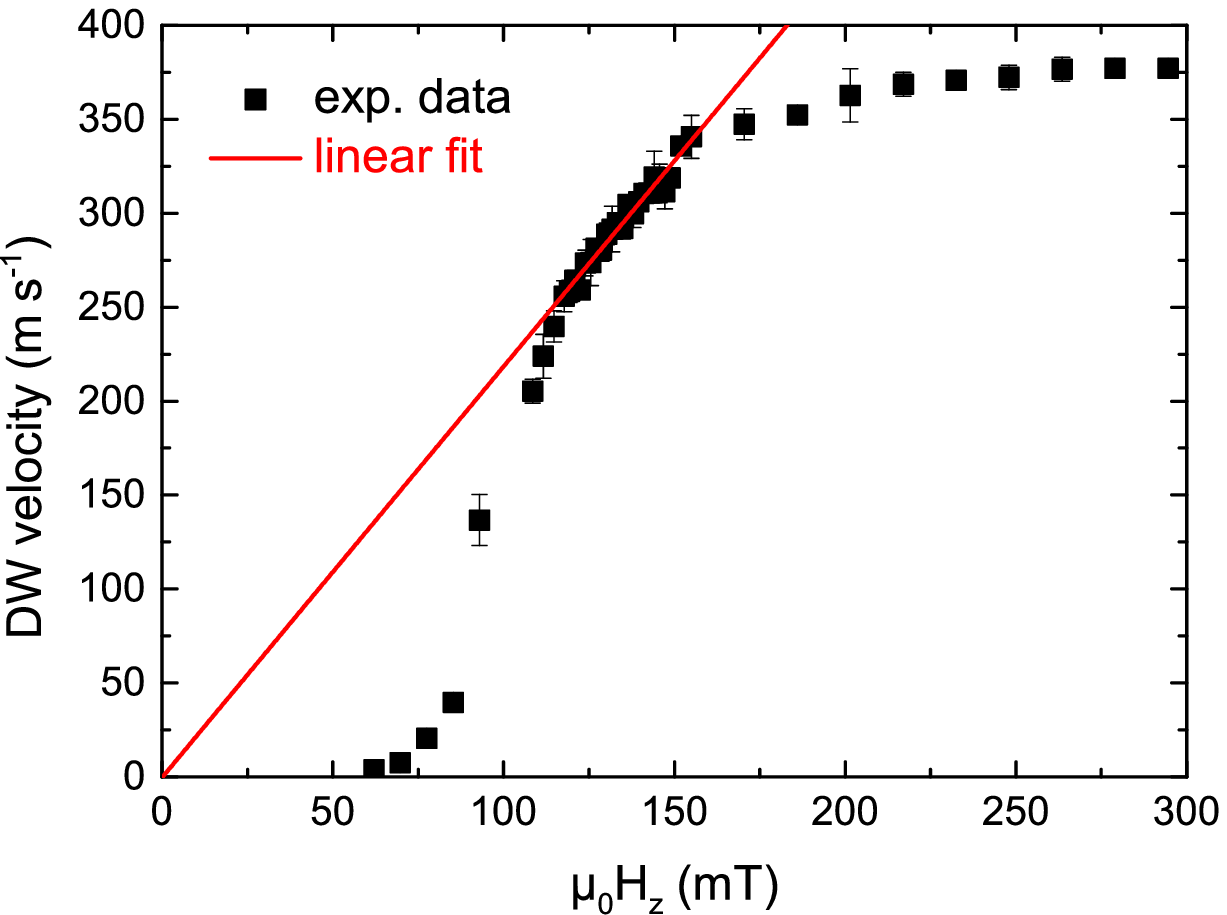}
\centering
\caption{Field-induced Domain Wall (DW) dynamics experiments. The DW velocity, obtained from polar Magneto-Optical Kerr Effect (pMOKE) microscopy as a function of the applied out-of-plane magnetic field measured for a Pt/Co(0.63nm)/MgO film at room-temperature. The black squares denote the experimental data and the red line is a linear fit in the flow regime.}
\label{DWdynamics}
\end{figure}

We performed magnetic field-induced Domain Wall (DW) dynamics measurements using polar Magneto-Optical Kerr Effect (pMOKE) microscopy on the same Pt/Co($t_{Co}$)/MgO stack, at a position on the wedge corresponding to $t_{Co}=0.63$ nm, for which larger domains are observed due to the higher effective anisotropy. The magnetic field is applied along the easy axis, that is perpendicular to the film plane. Fig. \ref{DWdynamics} shows the measured DW velocity as a function of the applied field. In the steady-state regime, one can express the DW velocity as $v_{DW}=\mu_{DW}\mu{}_0H_z$ where $\mu_{DW}$ is the DW mobility. The mobility takes different values in the flow and precessional regimes (\textit{i.e.} below or above the Walker field), respectively $\mu_{DW}=\gamma{\Delta}/{\alpha}$ and $\mu_{DW}=\gamma{\Delta}/{\left(\alpha+\alpha{}^{-1}\right)}$ with $\gamma$ the gyromagnetic ratio, $\Delta=\sqrt{A/K_{eff}}$ the DW width and $A$ the damping parameter \cite{metaxasCreepFlowRegimes2007}. Experimentally, we find $\mu_{DW}=2.2$ m s$^{-1}$ mT$^{-1}$. We extract $\Delta=5$ nm from our material parameters using $\gamma/(2\pi{})=31$ GHz T$^{-1}$, $A=16$ pJ m$^{-1}$ and the measured effective anisotropy $K_{eff}=6.3\times{}10^{5}$ J m$^{-3}$ for this sample. A fit in the flow regime yields $\alpha=0.43\pm0.08$ when taking into account the uncertainty on A. However, using the expression in the precessional regime would lead to an non-physical complex value for the damping. This implies that the DW dynamics in the observed linear regime is in the flow regime. This is consistent with the large DMI measured (section S.1.2) since the Walker field increases with increasing DMI \cite{thiavilleDynamicsDzyaloshinskiiDomain2012}.

\FloatBarrier

\subsection{Spin-orbit torques measurements}
\label{SOTmeasurements}

To determine the Damping-Like (DL) and Field-Like (FL) Spin-Orbit Torques (SOTs), we performed quasi-static harmonic Hall voltage measurements on a Pt/Co(0.9)/MgO film at room-temperature. The film is patterned into a device consisting of a 5-\textmu{}m-wide cross (see Fig. \ref{SOTmeasurements}.a). An AC current of frequency $\omega{}/\left(2\pi\right)=10$ Hz is injected along $\boldsymbol{\hat{x}}$ and the transverse resistance is measured (along $\boldsymbol{\hat{y}}$). The DL-SOT and FL-SOT acting on the magnetisation are given by $\boldsymbol{T}_{DL(FL)} = -\gamma{}_0\left(\boldsymbol{m}\times\boldsymbol{H}_{DL(FL)}\right)$ where $\boldsymbol{m}$ is the reduced magnetisation vector and $\boldsymbol{H}_{DL}$ and $\boldsymbol{H}_{FL}$ are the effective magnetic fields associated with the SOTs. These effective fields can be expressed as follows : $\mu_0\boldsymbol{H}_{DL}=C_{DL}J[(\boldsymbol{\hat{z}}\times\boldsymbol{\hat{j}})\times\boldsymbol{m}]=C_{DL}J\left(\boldsymbol{\hat{y}}\times\boldsymbol{m}\right)$ and $\mu_0\textbf{H}_{FL}=C_{FL}J(\boldsymbol{\hat{z}}\times\boldsymbol{\hat{j}})=C_{FL}J\boldsymbol{\hat{y}}$ where $\boldsymbol{\hat{j}}=\boldsymbol{\hat{x}}$ is te unit vector in the current direction and $\boldsymbol{\hat{y}}$ the unit vector perpendicular to the current direction (see Fig. \ref{SOTmeasurements}.a) \cite{garelloSymmetryMagnitudeSpinorbit2013}. \\

\begin{figure}[h]
\includegraphics[width=\textwidth]{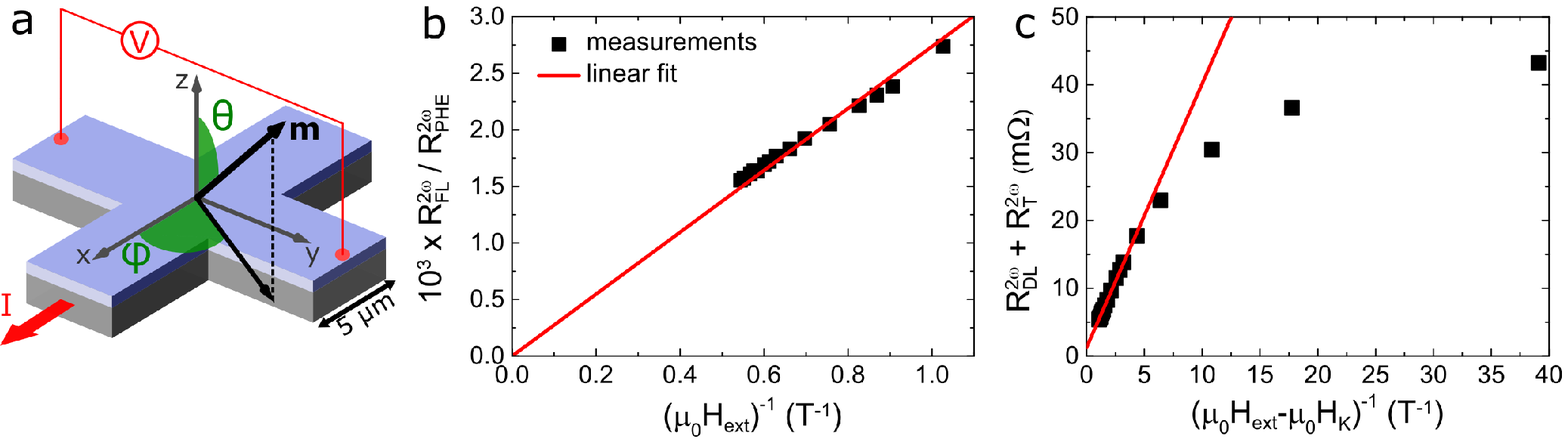}
\centering
\caption{Harmonic Hall voltage measurements. \textbf{a.} Schematics of the device consisting of a 5-\textmu{}m-wide cross. An AC current of frequency $f=10$ Hz is injected along $\boldsymbol{\hat{x}}$ and the transverse resistance (along $\boldsymbol{\hat{y}}$) is measured. \textbf{b.} Second harmonic Field-Like (FL) component as a function of the external applied magnetic field. The red line is a linear fit with a zero intercept since the FL-SOT effective field is zero for an infinitely large external field. \textbf{c.} Second harmonic Damping-Like (DL) and thermal components as a function of the effective magnetic field acting on it. The red line is a linear fit in the regime for which $H_{ext}>H_K$. The DL-SOT effective field is obtained from the slope while the intercept provides the thermal component.}
\label{SOTmeasurements}
\end{figure}
\FloatBarrier

The principle of the measurement is the following : the amplitude of these effective fields is compared to that of an external field. The external field $\boldsymbol{H}_{ext}=\cos\left(\varphi\right)\boldsymbol{\hat{x}}+\sin\left(\varphi\right)\boldsymbol{\hat{y}}$ is rotated in the sample plane and the transverse harmonic resistances are measured for $0\leq\varphi\leq{}2\pi$. Using a spherical coordinate system as defined in Fig. \ref{SOTmeasurements}.a, the first harmonic expression of the Hall resistance can be written as $R_H^{\omega}=R_{AHE}^{\omega}\cos\left(\theta\right) + R_{PHE}^{\omega}\sin{}^2\left(\theta\right)\sin\left(2\varphi\right)$ \cite{avciInterplaySpinorbitTorque2014}. This expression provides the equilibrium position of the magnetisation, which depends on the first harmonic anomalous and planar hall coefficients, $R_{AHE}^{\omega}$ and $R_{PHE}^{\omega}$  respectively. The second harmonic resistance $R_{H}^{2\omega}$ consists of the FL, DL and thermal components $R_{FL}^{2\omega}$, $R_{DL}^{2\omega}$ and $R_{\nabla{}T}^{2\omega}$ respectively, such that $R_H^{2\omega}=R_{FL}^{2\omega}+R_{DL}^{2\omega}+R_{\nabla{}T}^{2\omega}$. The components $R_{FL}^{2\omega}$ and $R_{DL}^{2\omega}+R_{\nabla{}T}^{2\omega}$ can be separated based on their distinct $\varphi$-dependence. The FL-SOT and DL-SOT effective fields can be further obtained from these two terms by considering their dependence on the total magnetic field acting on the magnetisation. One one hand, the action of the FL-SOT on the magnetisation is equivalent to that of an in-plane magnetic field and one can show that $R_{FL}^{2\omega}\propto{}1/$H$_{ext}$. On the other hand, the DL-SOT effective field depends on the magnetisation orientation and since the Co exhibits perpendicular magnetic anisotropy, the action of both external and anisotropy fields needs to be taken into account to extract the DL-SOT. Then, one can show that $R_{DL}^{2\omega}+R_{\nabla{}T}^{2\omega}\propto{}1/\left(H_{ext}-H_K\right)$. Fig. \ref{SOTmeasurements}.b and c display the dependence of these two terms.

The DL and FL components are then extracted from the slope in the linear regime (red lines) and found to be $C_{DL}=2.1$ mT per 10$^{11}$A m$^{-2}$ and $C_{FL}=0.9$ mT per 10$^{11}$A m$^{-2}$ respectively. These experiments provide a direct measurement of the SOTs. Therefore, the value of the DL-SOT effective field can be used to calculate the skyrmion velocity predicted by the model (equation (2) in the main text) and the measured ratio $C_{FL}/C_{DL}$ can be implemented directly in the micromagnetic code. This way, it is not required to estimate the spin Hall angle, which would constitute an additional source of error.

\newpage
\clearpage

\section{Micromagnetic simulations and analytical model}

\subsection{Initial skyrmion state}
\label{skyrmion_profile_simus}

\begin{figure}[h]
\includegraphics[width=0.8\textwidth]{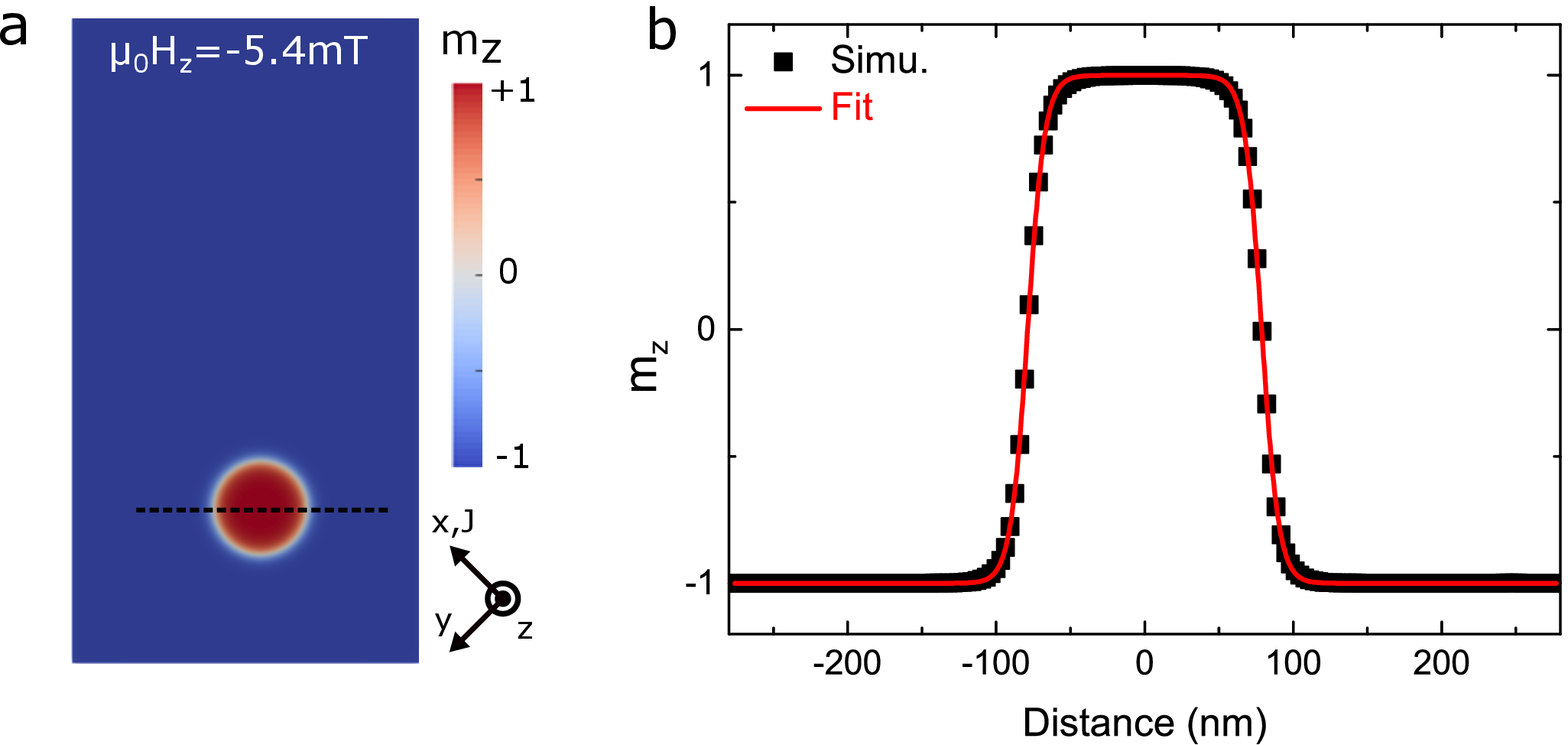}
\centering
\caption{Micromagnetic simulations : initial skyrmion state. \textbf{a.} Initial state of the micromagnetic simulations of the skyrmion dynamics. The applied field is $\mu_0H_z=-5.4$ mT. \textbf{b.} Simulated skyrmion profile (indicated by the dashed black line in \textbf{a.}). The red solid line is a fit using two 180{\textdegree} Bloch DW profiles. From this fit, we extract the  DW width parameter $\Delta=11.5$ nm and the skyrmion radius at rest $R=78.8$ nm.}
\label{skyrmionprofile}
\end{figure}

We now have all the parameters required to run the micromagnetic simulations. Using all the aforementioned parameters, we obtain the initial state displayed in Fig. \ref{skyrmionprofile}.a. The magnetic field was tuned to obtain the same skyrmion size at rest as in the experiments : for $\mu_0H_z=-5.4$ mT, we find a radius of 78.8 nm in quantitative agreement with the observations (section S.1.4). In Fig. \ref{skyrmionprofile}.b is plotted the out-of-plane magnetisation profile ($m_z$) across the skyrmion as indicated by the black dashed line in Fig. \ref{skyrmionprofile}.a. The radius $R$ as well as the DW width parameter $\Delta$ are extracted from a fit (red solid line) using the expression of a 360{\textdegree} Bloch wall profile. We find $\Delta=11.5$ nm. This $\Delta$ value is used in the analytical model to calculate the SkHA and the skyrmion velocity.

\FloatBarrier

\subsection{Influence of the Field-Like Spin-Orbit Torque (FL-SOT)}

\begin{figure}[h]
\includegraphics[width=\textwidth]{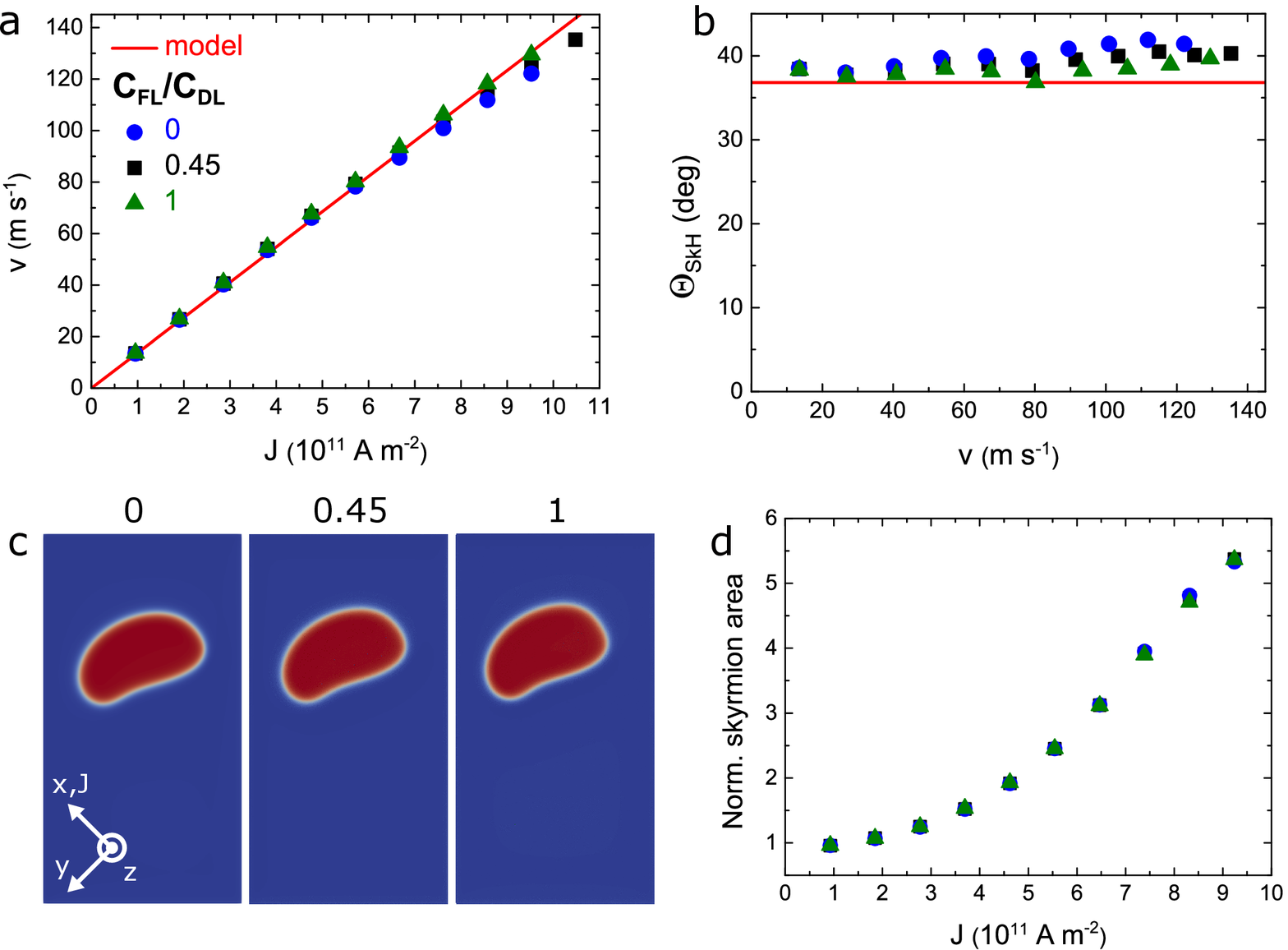}
\centering
\caption{Effect of the Field-Like Spin-Orbit Torque (FL-SOT). \textbf{a.} Skyrmion velocity and \textbf{b.} SkHA $\Theta_{SkH}$ obtained from micromagnetic simulations in a disorder-free film for different values of the ratio $C_{FL}/C_{DL}$. \textbf{c.} Snapshot of a skyrmion after 5 ns computation time for $J=6.5\times{}10^{11}$ A m$^{-2}$ and for different $C_{FL}/C_{DL}$ values. \textbf{d.} Evolution of the skyrmion size with the current denisty (at 5 ns). The size is defined as the total area for which $m_z>0$ (in red in \textbf{c.}) normalised by the skyrmion area at rest.}
\label{simus_FL_DL_SOT}
\end{figure}

Here, we discuss the influence of the FL-SOT on the disorder-free skyrmion dynamics. We considered different values of the ratio $C_{FL}/C_{DL}$ : 0, 0.45 (the measured value) and 1 (see section S1.6 for their definition). The corresponding skyrmion velocity and SkHA obtained from micromagnetic simulations are plotted in Fig. \ref{simus_FL_DL_SOT}.a and b respectively. It shows that, even for FL-SOT amplitudes two times larger than the value measured in our system, its effect on the skyrmion dynamics is negligible in the range of current density considered. Therefore, the FL-SOT cannot account for the velocity-dependence of the SkHA observed experimentally. Furthermore, as highlighted in Fig. \ref{simus_FL_DL_SOT}.c and d, the FL-SOT has very little influence on the dynamical deformation/expansion mentioned in the main text. This dynamical effect is thus largely due to the DL-SOT. These results confirm that the effect of the FL-SOT on the current-driven skyrmion dynamics in Pt/Co/MgO can be neglected.

\FloatBarrier

\newpage

\nocite{apsrev41Control}
\bibliographystyle{apsrev4-1}
\bibliography{supp_bbl}